\documentclass[a4paper,11pt]{article}
\pdfoutput=1 

\usepackage{jheppub} 

\usepackage[T1]{fontenc} 
\usepackage{amsmath}
\usepackage{mathtools}
    \usepackage{amssymb}
    \usepackage{amsthm}
    \usepackage{amsfonts}
    \usepackage{braket}
    \usepackage{slashed}
    \usepackage{graphicx}  
\usepackage{bbm}   
\usepackage[export]{adjustbox}
\usepackage{subcaption}
\usepackage{xcolor}
\usepackage{footnote}  
\makesavenoteenv{tabular} 
\makesavenoteenv{table} 
\usepackage{booktabs}
\usepackage{cancel} 
\usepackage{float}
\usepackage[utf8]{inputenc}
\usepackage[british]{babel} 
\usepackage{lscape} 
\newcommand\mysimeqq{\stackrel{\mathclap{\normalfont\mbox{\text{(\ref{DeltaMB})}}}}{\simeq}}

\title{\boldmath Flavor and $CP$ Violation from a QCD-like Hidden Sector}

 \author{Wafia Bensalem}
 \author{and Daniel Stolarski}
 \affiliation{Ottawa-Carleton Institute for Physics, Carleton University,\\
1125 Colonel By Drive, Ottawa, ON K1S 5B6, Canada}
 \emailAdd{wafiabensalem@cunet.carleton.ca}
 \emailAdd{stolar@physics.carleton.ca}

\abstract{Confining hidden sectors at the GeV scale are well motivated by asymmetric dark matter and naturalness considerations and can also give interesting collider signatures. Here we study such sectors connected to the Standard Model by a TeV scale mediator charged under both QCD and the dark force. Such a mediator admits a Yukawa coupling between quarks and dark quarks which is generically flavour and $CP$ violating. We show that in contrast to expectation, electric dipole moments do not place a strong constraint on this scenario even with $O(1)$ $CP$-violating phases.
We also quantitatively explore constraints from $\Delta F=1,2$ processes as a function of the number of dark quark flavours. Finally, we describe the reach of upcoming measurements at Belle-II and KOTO, and we propose new $CP$-odd observables in rare meson decays that may be sensitive to the $CP$-violating nature of the dark sector. }

\begin{document}

\maketitle

\section{Introduction}
\label{sec:intro}

The nature of dark matter is one of the most important open questions in fundamental physics. The asymmetric dark matter paradigm~\cite{Nussinov:1985xr,Barr:1990ca,Barr:1991qn,Dodelson:1991iv,Kaplan:1991ah,Kuzmin:1996he,Foot:2003jt,Foot:2004pq,Hooper:2004dc,Kitano:2004sv,Gudnason:2006ug,Kaplan:2009ag} (for reviews see~\cite{Davoudiasl:2012uw,PetrakiEtal,Zurek:2013wia}) is a particularly attractive solution to the dark matter problem because it explains the abundance of dark matter in the same way that the abundance of baryons in the universe arises, via an asymmetry of (dark) matter over (dark) anti-matter. This paradigm can explain why the abundances of matter and dark matter in the universe only differ by a factor of five. 

While asymmetric dark matter models can easily explain why the number density of dark matter is similar to that of matter, the abundance of (dark) matter is set by both the number density and the mass. The mass of the proton is controlled by the confinement scale of QCD. This suggests that the asymmetric dark matter should also be a bound state of a new dark strong force. This was realized concretely in the model of Bai and Schwaller~\cite{BaiSchwaller014} (BS) which has an $SU(N_c)$ dark force whose running gauge coupling is related to that of QCD and thus naturally predicts a dark confinement scale at around a GeV.  This model also predicts the existence of a scalar $X$ charged under both QCD and ``dark QCD'' around a TeV scale, and this mediator has a Yukawa coupling to quarks and dark quarks. 

Models with a dark confining force around the GeV scale have received tremendous attention lately. In addition to being motivated by asymmetric dark matter, they can also arise in neutral naturalness models~\cite{Craig:2015pha,Curtin:2015fna,Cheng:2016uqk,Kilic:2018sew}, and provide interesting benchmark for exotic searches at colliders~\cite{Strassler:2006im,Strassler:2006ri,Han:2007ae,EmergingJets,Cohen:2015toa,Csaki:2015fba,Knapen:2016hky,Pierce:2017taw,Burdman:2018ehe,Cheng:2019yai,Linthorne:2021oiz,Knapen:2021eip} (for a review, see chapter 7 of~\cite{Alimena:2019zri}). There is now a direct collider search for this BS model~\cite{Sirunyan:2018njd} using the proposal of~\cite{EmergingJets}, and significant bounds can be placed from collider searches and direct detection~\cite{Mies:2020mzw}.

In this work we focus on the constraints that arise from Yukawa couplings of the $X$ scalar in the BS model. The couplings of this mediator to SM quarks breaks some of the SM accidental flavour symmetries and generically gives rise to flavour and $CP$-violating processes. Therefore, dark QCD models are variants of the Flavoured Dark Matter paradigm~\cite{Kile:2011mn,Kamenik:2011nb,Batell:2011tc,Agrawal:2011ze,Batell:2013zwa,Lopez-Honorez:2013wla,ABG014,Bishara:2014gwa,Hamze:2014wca,Calibbi:2015sfa,Agrawal:2015tfa,Bishara:2015mha,Bhattacharya:2015xha,Chen:2015jkt,Agrawal:2015kje,Blanke:2017fum}. In order to study flavour observables, the effects of the mediator can be parameterized within an effective field theory of quarks and leptons. Models with different types of mediators can thus still be constrained using the analysis of this work. While many models are constructed not to be in conflict with flavour observables (see~\cite{Knapen:2021eip} for a classification), this is not always the case. So while our work uses the BS model for concreteness, it is more broadly applicable. 

Many of the flavour and $CP$-violating processes we study are extremely well measured and all agree\footnote{There are of course several few $\sigma$ anomalies that we do not attempt to explain here.} with the SM predictions, so these processes can be used to place constraints on the parameters of the dark QCD models. The flavour constraints on these models were studied in the work of Renner and Schwaller (RS)~\cite{RennerSchwaller018} under the assumptions of real Yukawa couplings. In this paper we extend the work of RS to the more general case of complex couplings that can potentially violate $CP$. Naively, the strongest bounds on $CP$-violating models come from null results in searches for electric dipole moments (EDMs) of various systems. In this work, we show that the models we are considering, even with generic $\mathcal{O}(1)$ phases, 
are not constrained by current EDM measurements. There may, however, be sensitivity with future experiments. 

The BS model requires the existence of $n_f$ light flavours charged under dark QCD whose mass is at or below the confinement scale, and the dark matter is the lightest baryonic bound state of those light flavours. The simplest case is $n_f=3$ which gives a parallel structure to QCD which also has three light flavours, and that is the focus of the work of~\cite{RennerSchwaller018}. Here we generalize to arbitrary $n_f$, analyzing all possibilities. 

In addition to constraints from flavour and $CP$ observables, we also consider future prospects for discovery, focusing in particular on rare decays with dark quarks or hadrons in the final state. We explore how $CP$-violating phases change predictions of the model. We also propose speculative new measurements that can directly measure $CP$ violation in the dark sector at future experiments. While most of the techniques we explore are difficult with current technology and experiments, we hope this motivates improvements that allow such measurements in the future.

The remainder of this work is structured as follows: in Sec.~\ref{sec:Model} we outline the details of the model and give constraints from the running of the QCD coupling. In Sec.~\ref{sec:EDM}, we explore bounds from EDMs, in Sec.~\ref{Constraints}, we analyze bounds from $\Delta F = 2$ processes, and in Sec.~\ref{Pheno} we compute constraints and discovery prospects from $\Delta F = 1$ processes. We compare the $\Delta F=2$ and $\Delta F = 1$ constraints in Sec.~\ref{Constraints2}, and we conclude in Sec.~\ref{sec:conclusions}. In the appendices we give a brief review of the flavour and $CP$ observables used in this work, and give various technical details on how we compute bounds.

\section{Dark QCD Model}
\label{sec:Model}

We begin with a DQCD model which is an $SU(N_c)$ gauge group with $n_f$ vectorlike fundamental flavors of dark quarks with masses near or below the confinement scale. Cofinement will give rise to a zoo of dark hadrons that get their masses from DQCD effects. For simplicity, we will consider the dark quark masses to be approximately equal. The dark quarks are all neutral under the SM gauge forces. 
Motivated by asymmetric dark matter models and particularly the setup in~\cite{BaiSchwaller014}, we take the confinement scale of the dark QCD to be around the GeV scale. The lightest baryon can serve as dark matter if baryon number is a good symmetry.

The asymmetry generation mechanism of~\cite{BaiSchwaller014} requires a bifundamental scalar portal, $X$, which is a scalar charged under both QCD and dark QCD. This scalar can then be a mediator between the dark sector and the Standard Model via a Yukawa coupling:
\begin{equation}
 \label{IntLag}
 \mathcal{L}_{SM-DM}^{int}= -\lambda_{ij}\bar{d}_{iR}Q_jX + \, {\rm h.c}\ ,
\end{equation}
where $d_{iR}$ are SM right-handed (RH) down-type quarks and we have chosen the hypercharge of the $X$ so that this coupling is allowed. Here $i$ is an SM flavor index running from 1 to 3 over down, strange, and bottom, $Q_j$ are the dark quarks with $j$ a dark flavor index running from 1 to $n_f$. Therefore $\lambda$ is a Yukawa coupling matrix in the SM-DM flavor space.

Coupling to down-type quarks is the scenario considered in~\cite{BaiSchwaller014} and  the simplest scenario that contributes to $K-\bar{K}$ mixing, the quark system that gives rise to the strongest constraints. One could consider couplings to RH up-type quarks, or couplings to left-handed quark doublets which we will explore in the future. These two cases have been studied in the Flavoured Dark Matter context~\cite{BlankeKastRHupQuarks,JubbEtAllRHupQuarks,BlankeEtAlLHquarks}, in models without dark gauge interactions. In this work we also consider generic $CP$ violating effects for the first time. 


Before analyzing the bounds from flavor and $CP$ violation, we briefly establish bounds that do not depend on the Yukawa matrix $\lambda$. Because $X$ carries a QCD charge, it will contribute to the running of $\alpha_{\text{s}}$. Using the analysis of~\cite{KaplanSchwartz} on event shape data at LEP, we find a limit of $M_X >  30$ GeV for $N_c = 3$ and one flavor of $X$. In the case of three degenerate flavors of scalar mediators and $N_c = 8$, we find $M_X > 55$ GeV. 
Future colliders could also place similar bounds, with the study of~\cite{AlvesEWco}, we find a potential lower bound of $100$ GeV from TLEP constraints using the  running of electroweak couplings.

Direct production of $X$ at colliders can also be used to place bounds. The quantum numbers of the $X$ are similar to that of a sbottom in supersymmetry, and LEP has placed bounds on those states of $\mathcal{O}(100)$ GeV~\cite{LEP}. Translating these searches to precise bounds on the $X$ which decays to exotic dark sector states is non-trivial and beyond the scope of this work. At the LHC, CMS has placed direct limits on the $X$~\cite{Sirunyan:2018njd} ruling out parts of the parameter space, but that search did not analyze $X$ masses below 400 GeV. A theoretical recast exploring various other limits was done in~\cite{Mies:2020mzw}.


\section{Electric Dipole Moment}
\label{sec:EDM}
We will consider an arbitrary coupling matrix $\lambda_{ij}$, which means that this model will in general have large $CP$ violation. Typically, the strongest bounds on $CP$ violating new physics come from electric dipole moments (EDMs). In this section we analyze those bounds for the DQCD framework and show that bounds from EDMs are actually quite weak and do not place significant constraints on this model. The experimental limit on the neutron electric dipole moment (nEDM) is $|d_n| <  3.0 \times 10^{-26} \ e\cdot cm$~\cite{nEDMexp}. There are also bounds on EDMs of nuclei, atoms, and molecules ~\cite{EDMatoms} that can be translated into limits 
on the EDMs of quarks which are $1.27\times 10^{-24} \ e\cdot cm$ for the up quark and $1.17\times 10^{-24} \ e\cdot cm$  for the down quark, at the scale of $4$ GeV$^2$~\cite{QuarksEDM}. A limit on the electron electric dipole moment (eEDM) of  $1.1\times 10^{-29} \ e\cdot cm$ can also be placed~\cite{eEDMexp}.

Since the neutron and all concerned nuclei, atoms, and molecules contain hadronic states, their observable EDMs are in general sensitive to any $CP-$odd effects, including gluonic ones~\cite{LeDallRitz2013}. 
The effective $CP-$odd flavor-diagonal Lagrangian up to dimention six, normalized to $1$ GeV is~\cite{LeDallRitz2013,PospelovRitz2005}
\begin{equation}
\label{LCPV}
\begin{aligned}
\mathcal{L}_{eff}^{\cancel{{CP}}}(1\text{ GeV})= & \frac{\bar{\theta}g_s^2}{32\pi^2}G_{\mu\nu}^a\widetilde{G}^{\mu\nu ,a}
-\frac{i}{2}d_i\bar{\psi}_i\sigma_{\mu\nu}\gamma_5\psi_iF^{\mu\nu}
-\frac{i}{2}\widetilde{d}_ig_s\bar{\psi}_it_a\sigma_{\mu\nu}\gamma_5\psi_iG_a^{\mu\nu} \\
& +\frac{1}{3}\omega f^{abc}G_{\mu\nu}^a\widetilde{G}^{\nu\beta ,b}G_\beta^{\mu ,c}
+ C_{ij}(\bar{\psi}_i\psi_i)(\bar{\psi}_ji\gamma_5\psi_j).
\end{aligned}
\end{equation}
where $i$ and $j$ are summed over light flavors, and $G^{\mu\nu}$ ($F^{\mu\nu}$) is the field strength tensor of the gluon (photon). The first term in equation~(\ref{LCPV}) is the QCD $\theta$-term; we assume an axion~\cite{HookStrongCP} or some other mechanism is at work to relax $\bar{\theta} \rightarrow 0$. 
The second and third terms contain respectively the EDM and the chromo-EDM (CEDM) operators. The fourth term is the Weinberg operator~\cite{Weinberg1989}, and the last term has a four quark operator which is very suppressed compared to the other operators~\cite{HamzaouiPospelov1995,PospelovRitz2005}, so it can be neglected in the calculations of nEDM. We are therefore left with the Weinberg and the (C)EDM terms.
We will show in this section, that even if the dark quarks are not mass-degenerate,\footnote{In subsequent sections, we will consider the dark quarks to be approximately mass-degenerate, but in this section we consider the more general possibility.} the contribution of our model to EDMs is negligible.

In order to contribute to the $CP$-odd EDM, diagrams must have an imaginary component. In general, the amplitude of any loop diagram is a sum of operators, whose coefficients can be written in the form: $C=gL$, where $L$ represents the loop integration, and $g$ represents the effective coupling, which is the product of couplings from all the vertices of the diagram. In general, $g$ and $L$ can both be complex, but  
diagrams with heavy $X$ running in the loop and zero external momentum cannot have the $X$ go on-shell. Therefore, the optical theorem\footnote{See for example Section 7.3 of \cite{PeskinQFT}.} implies that $\text{Im}(L) = 0$ and only $g$ can be complex.

\begin{figure}[hbpt]
\centering
\includegraphics[width=50mm,scale=0.5]{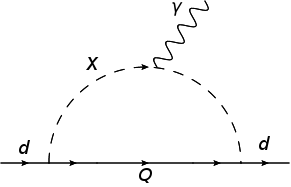}
\caption{One loop contribution of our model to the (C)EDM of the $d$ quark. $Q$ is a dark quark, and $X$ is the bifundamental scalar mediator. The CEDM diagram is obtained when the photon is replaced by a gluon.}
\label{EDM1loop}
\end{figure}
\begin{figure}[hbpt]
\centering
\begin{subfigure}{1.0\linewidth}
 \centering
 \includegraphics[width=\textwidth]{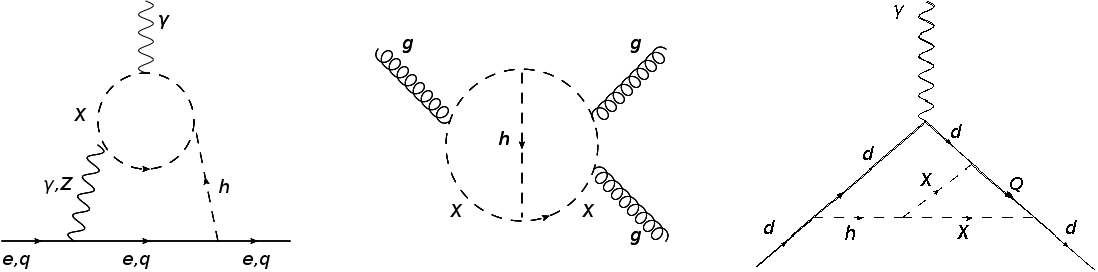}
\end{subfigure}
\hfill
\vspace{1cm}
\begin{subfigure}{1.0\linewidth}
 \centering
 \includegraphics[width=\textwidth]{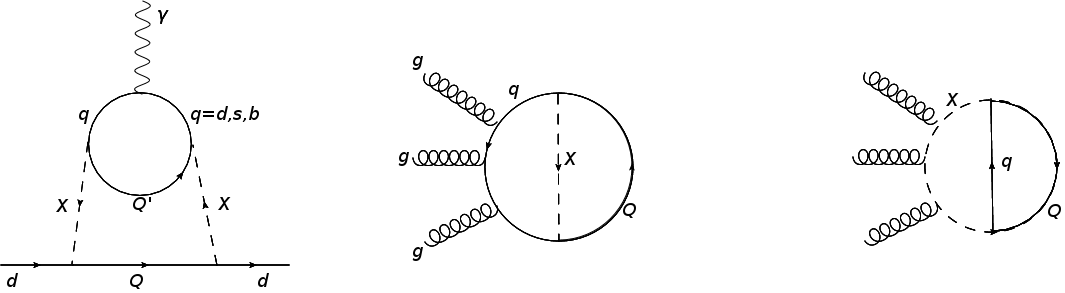}
\end{subfigure}
%
\caption{Contributions of our model to Barr-Zee (left column), Weinberg (central column and lower right), and kite (upper right) diagrams that could participate to nEDM or eEDM. We have  included one diagram per topology, but there are additional diagrams where the photon (gluon) couples to other intermediate particles, or where the Higgs scalar propagator changes place. The CEDM diagrams are obtained from the Barr-Zee and kite ones, when the photon is replaced by a gluon. $Q$ and $Q'$ are dark quarks, and $X$ is the bifundamental scalar mediator. These diagrams, having real effective couplings, do not participate to EDMs.}
\label{BarrZeeWeinberg}
\end{figure}

At one loop there is one that could contribute to the quark EDM, shown in Figure~\ref{EDM1loop}. 
The effective couplings of this diagram, $\lambda_{dQ}\lambda^\ast_{dQ}$,  is real, and thus gives no contributions to EDMs. 
At two loops, there are contributions to the three gluon Weinberg operator~\cite{Weinberg1989}, and  Barr-Zee-like~\cite{BarrZee} and kite diagrams contributing to the quark and electron EDM. These are all shown in Figure~\ref{BarrZeeWeinberg}. The diagrams of the first line of  Figure~\ref{BarrZeeWeinberg} give zero EDMs because their effective couplings are real. Indeed, the coupling between the SM Higgs and $X$ is real, since the only renormalizable gauge invariant interaction between the SM Higgs and $X$ is given by $X^\dagger X H^\dagger H$ which is Hermitian and thus always $CP$ conserving. The $Q$-$X$-$d$ vertices of the kite diagram give the real coupling $\lambda^\ast_{dQ}\lambda_{dQ}$.
The diagrams with fermion loops of the second line give real effective couplings as well. This is because they are of the form $\lambda^\ast_{qQ}\lambda_{qQ}$, for the Weinberg-like diagrams, and $\lambda^\ast_{dQ}\lambda_{dQ}\lambda^\ast_{qQ'}\lambda_{qQ'}$, for the Barr-Zee-like diagram.
All diagrams of Figure~\ref{BarrZeeWeinberg} could have a non-zero EDM if there is a more complicated Higgs sector and/or if there are multiple $X$ scalars.

\begin{figure}[hbpt]
\centering
 \includegraphics[width=\textwidth]{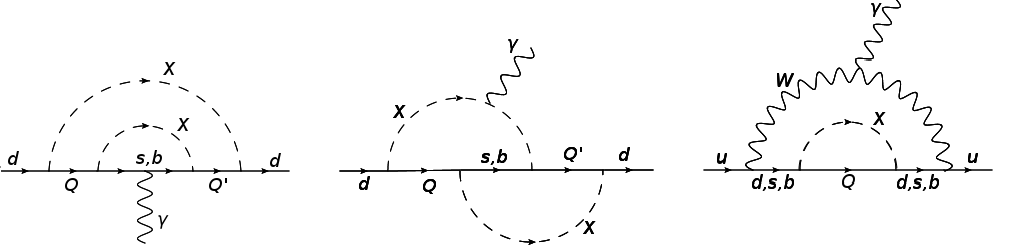}
\caption{Rainbow diagrams, which are other two loop contributions of our model to nEDM. $Q$ and $Q'$ are dark quarks, and $X$ is the bifundamental scalar mediator. Here again, We have included one diagram per topology, but there are additional diagrams where the photon couples to other intermediate particles. The CEDM diagrams are obtained when the photon is replaced by a gluon. The first two diagrams are contributions to the (C)EDM of the $d$ quark, and the third diagram shows contributions to the (C)EDM of the $u$ quark}
\label{DoubleSunSets}
\end{figure}

Another type of two loop diagrams, shown in Figures~\ref{DoubleSunSets}, may contribute to the nEDM via the (C)EDMs of the $d$ and $u$ quarks. 
The three types of rainbow diagrams, when taken individually, with the two internal (dark) quarks being different and not mass-degenerate, will have complex effective couplings. However, because of the symmetry of these diagrams, their contributions to the (C)EDM operators vanish when we sum over all the internal (dark) quarks,
even if the individual contributions are nonzero. 
%
The first and second diagrams of Figure~\ref{DoubleSunSets} will give a sum of operators whose coefficients are of the form: 
\begin{equation}
 C_{(1,2)}=\sum\limits_{q=b,s}\ \sum\limits_{Q,Q'=1}^{n_f} \lambda_{dQ'}^\ast \lambda_{qQ'} \lambda_{qQ}^\ast \lambda_{dQ} L_{QQ'},
 \end{equation}
 where $L_{QQ'}$ is the loop factor  when $Q$ ($Q'$) is the left (right) internal dark quark, with each type of operator having its own specific factor $L_{QQ'}$. From the symmetry of the diagrams, we have $L_{Q'Q}=L_{QQ'}$. Indeed, if we apply a time reversal to one diagram having a left (right) internal quark $Q$ ($Q'$), then we actually change $L_{QQ'}$ to $L_{Q'Q}$. So $L_{Q'Q}$ is the time reversal transform of $L_{QQ'}$, and since $L_{QQ'}$ is real as we explained above, it is then $T-$conserving,\footnote{$L_{QQ'}$ is $CP-$conserving since it is real, which implies it is $T-$conserving, assuming $CPT$ invariance.}  that is, it is equal to $L_{Q'Q}$, its $T-$transform. We can show the equality in a different way for the first diagram of Figure~\ref{DoubleSunSets}. In the loop integral, we have exactly the same integrand when we interchange $Q$ and $Q'$ because, first, $Q$ and $Q'$ have the same momentum, and second, the terms proportional to their masses (which could give different integrands if the masses are different) vanish because they come with the null product $P_RP_L$. Finally we obtain: 
\begin{equation} 
C_{(1,2)} = \sum\limits_{q=b,s} \left[  \sum\limits_Q |\lambda_{dQ}|^2 |\lambda_{qQ}|^2 L_{QQ} + \sum\limits_{Q\neq Q'} 2\text{Re}(\lambda_{dQ'}^\ast \lambda_{qQ'} \lambda_{qQ}^\ast \lambda_{dQ}) L_{QQ'} \right],
\end{equation}
which is real.
With a similar reasoning, we find that the sum over the internal quarks of the third diagram of Figure~\ref{DoubleSunSets} gives a real coefficient as well. Indeed, it takes the form: 
\begin{equation}
C_{(3)} = \sum\limits_Q \left[  \sum\limits_{q=d,s,b} | \lambda_{qQ} |^2 | V_{uq} |^2 L_{qqQ} +  (1/2)\sum\limits_{q\neq q'}  2\text{Re}(\lambda_{qQ}^\ast \lambda_{q'Q} V_{uq'}^\ast V_{uq}) L_{qq'Q}] \right],
\end{equation}
 where $L_{qq'Q}$ is the loop factor when the internal quark $q$ ($q'$) is on the left (right), and $V_{uq}$ are the elements of the CKM matrix. Again, by $T-$symmetry, $L_{qq'Q}=L_{q'qQ}$. One could think that this loop factor symmetry does not apply when the photon connects to one of the inner SM quarks in the right diagram of Figure~\ref{DoubleSunSets}, but the reversal argument still applies to this diagram when the photon is moved from one inner quark to the other.
 In ~\cite{Fujiwara:2021vam} the authors show that the contribution vanishes as well after summing over internal quarks. This work shows, in a model independent way, that the  diagrams of the same type as the left and right ones of Figure~\ref{DoubleSunSets} give zero EDM. The analysis of~\cite{Fujiwara:2020unw} confirms a vanishing EDM in the middle diagram type of Figure~\ref{DoubleSunSets}, since a non-zero contribution requires two non-degenerate scalars, while our model contains only a single scalar.
 Therefore, all the rainbow diagrams give zero (C)EDMs. If there are multiple $X$ scalars, there could be non-zero EDMs and non-trivial bounds.
 
There may be contributions to the nEDM at the three loop level, which we estimate to be of order $\sim e \lambda_{ij}^4(1/16\pi^2)^3 (m_d/M_X^2) \lesssim 10^{-29} \ e\cdot cm$ for a TeV scale mediator. This is far below the experimental limit of order $10^{-26} \ e\cdot cm$, but there may be a non-trivial constraint for light mediators, $M_X \lesssim 100$ GeV.

While the dark sector does not couple directly to electrons, there could also be contributions to the electron electric dipole moment (eEDM) starting at two-loop order. The upper left diagram of Figure~\ref{BarrZeeWeinberg}, is the only one that contributes, but as argued above, it is purely real. 
At three loops, one can take the original Barr-Zee  diagram and insert an $X$ propagator inside the quark loop, but this diagram also has a real effective coupling. Therefore, current EDM bounds are insensitive to $CP$ violation in the minimal model with only one bi-fundamental and one Higgs, but there may be prospects for detection with future experimental improvements.


\section{\boldmath $\Delta F=2$ Constraints}
\label{Constraints}
\begin{figure}[hbpt]
 \centering
 \includegraphics[width=150mm,scale=1.0]{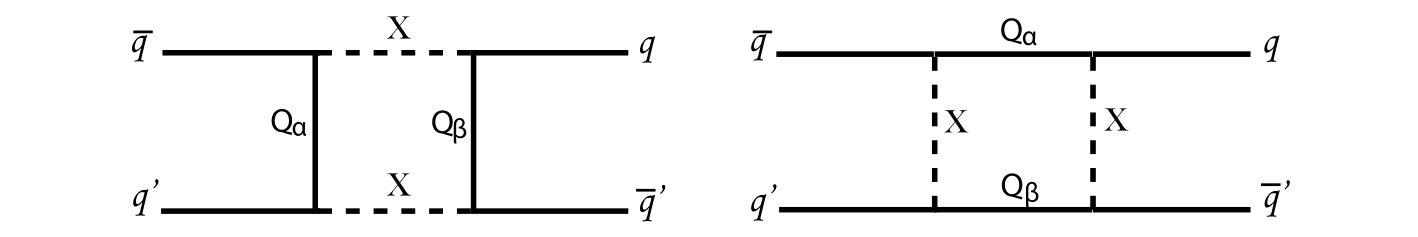}
 \caption{Contributions of the new particles to $\Delta F=2$ meson mixing processes.}
 \label{fig:Boxes}
\end{figure}

We now turn to $\Delta F = 2$ bounds on the DQCD framework. Contributions to $\Delta F=2$ processes arise first at the one loop level where two box diagrams contribute to the $B_d$, $B_s$ or $K$  meson mixing shown in Figure~\ref{fig:Boxes}. These diagrams were computed in~\cite{ABG014} without a dark gauge symmetry, and we have recomputed and confirmed their results, up to the addition of an $N_c$ factor for dark color. 

We take the scenario where there are $n_f$ approximately mass-degenerate dark flavors, $Q_1, \ Q_2, \ .....Q_{n_f}$, and we also take the dark quarks to be parameterically lighter than the scalar mediator, $m_{Q_i}\ll M_X$, where $M_X$ is the mediator mass. 
In this case, the loop function for the diagrams in Figure~\ref{fig:Boxes} is close to unity, and the effects of those diagrams can be parameterized in terms of an effective $\Delta F=2$ Hamiltonian:
\begin{equation} \label{Heff}  
H_{\rm eff}^{\rm NP}(\Delta F=2) = \frac{N_c \  {\xi}_M^2}{128{\pi}^2M_X^2} \left[ (\bar{q}\gamma_{\mu}P_Rq')(\bar{q}\gamma^{\mu}P_Rq')\right]  + {\rm h.c.}
\end{equation}
\begin{equation}
\label{xi}
{\xi}_M = \sum_{Q=Q_1}^{Q_{n_f}} \lambda_{qQ} \lambda_{q'Q}^{\ast} \ , \ \ M =  B_d , B_s \ or \ K.
\end{equation}
Here $N_c$ is the number of dark colors, $q$ and $q'$ are $d$ and $s$ for the $K-$system, or $d$ and $b$ for the $B_d-$system, or $s$ and $b$ for the $B_{s}-$system.  Note that the effective Hamiltonian of $H_{\rm eff}^{\rm NP}$ can also be thought of as a four fermion operator times a Wilson coefficient in an effective field theory with the Wilson coefficient given by
$c_{qq'}^{\rm NP} = N_c \ {\xi}_M^2 /(128{\pi}^2 M_X^2)$. 

We will impose constraints from observables of the $K$, $B_d$, and $B_s$ systems on the $\xi_M$ at the $2\sigma$ level. We take $M_X = 1$ TeV as our main benchmark, but all bounds scale with $1/M_X^2$. Since dark quark confinement occurs at a scale comparable to the mass of the SM mesons considered, we allow for a $\pm 50$ \% uncertainty on the new physics amplitude to account for the non-perturbative DQCD uncertainty~\cite{RennerSchwaller018}, as well as for perturbative QCD and RG corrections. Details of the mapping from experimental data onto flavor observables are reviewed in Appendix~\ref{sec:A}, where all the $\Delta F=2$ observables used in what follows are defined. 

Table~\ref{tab:2} summarizes all the constraints we have found, which are conditions on the $[\xi^\ast_M]^2$ terms, that are proportional to the Wilson coefficients of the effective NP Hamiltonian in equation~(\ref{Heff}). The details of the calculations of these constraints can be found in Appendix~\ref{sec:B}. The $\Delta M$ constraints in the first three rows of Table~\ref{tab:2} are $CP$-conserving constraints which were also explored in~\cite{RennerSchwaller018}, while the last three lines are constraints from $CP$-violating observables.
The $CP$-violating constraints are stronger than the $CP$-conserving ones, but the $CP$-violating constraints are on the phases of the Yukawa couplings, so if the Yukawa matrix $\lambda$ is real, then the $CP$-conserving constraints are the only ones to consider.  Among the different mesons, the $K-$system has the strongest constraints, and the $B_s$-system has the weakest ones. Finally, the constraints get stronger with increasing number of dark colors.

We can now use these constraints to explore specific scenarios where we choose the number of dark flavors and explicitly compute the bounds on the Yukawa matrix. We will present our results both in terms of parameter scans and also in terms of specific slices of the parameter space. For each case, we will first consider purely real couplings, then generalize to the complex case.

\begin{center}
\begin{table}
\centering
\begin{tabular}{|| c | c ||}
\hline \hline
\rule{0pt}{5ex} 
Observable &  Constraint  \\
   &   
\\ \hline\hline
 \rule{0pt}{5ex}  
 $\Delta M_K$  & $-6.93 \times 10^{-4} \leq 
N_c \text{Re}\left[( {\xi}_K^*)^2 \right] \left(\frac{1\text{ TeV}}{M_X} \right)^2 \leq  3.77 \times 10^{-4}$ \\
   & 
\\ \hline
\rule{0pt}{5ex}  
 $\Delta M_d$  & $N_c\mid{\xi}_{B_d}^2\mid \left( \frac{1\text{ TeV}}{M_X} \right)^2 \leq 6.55 \times 10^{-4}$ \\
   & 
 \\ \hline
 \rule{0pt}{5ex}  
 $\Delta M_s$  &  $N_c\mid{\xi}_{B_s}^2\mid \left( \frac{1\text{ TeV}}{M_X} \right)^2 \leq 13.15 \times 10^{-3}$ \\
   &  
 \\ \hline
 \rule{0pt}{5ex}  
 $|\epsilon_K|$ &  $N_c\mid \text{Im}\left[ ( {\xi}_K^*)^2 \right] \mid\left( \frac{1\text{ TeV}}{M_X} \right) ^2   \leq 1.64 \times 10^{-6}$ \\
   & 
 \\ \hline 
 \rule{0pt}{5ex}  
 $S_{\psi K_{\text{S}}}$  & $-0.86\times 10^{-4} \leq N_c\text{Im}\left[ ( {\xi}_{B_d}^*)^2 \right] \left( \frac{1\text{ TeV}}{M_X} \right) ^2   \leq 3.12 \times 10^{-4}$ \\
   & 
  \\ \hline
  \rule{0pt}{5ex}  
 $S_{\psi \phi}$ &  $-3.54\times 10^{-3} \leq N_c\text{Im}\left[ ( {\xi}_{B_s}^*)^2 \right] \left( \frac{1\text{ TeV}}{M_X} \right) ^2   \leq  3.88 \times 10^{-3}$ \\
   & 
 \\ \hline \hline
\end{tabular}
\caption{Summary of the $\Delta F=2$ constraints on our model. The parameter $\xi$ is defined in 
equation~\eqref{xi}, $N_c$ is the number of dark colours, and $M_X$ is the mediator mass.}
\label{tab:2}
\end{table}
\end{center}

%

%
%

\subsection{One Dark Flavor}
\label{sec:1DF}
We consider the simple case where there is only one flavor of dark quarks, which we will call $Q$. In this case, the coupling matrix $\lambda$ between $Q$ and the right handed SM quarks $d_R$, $s_R$, and $b_R$ is a three component one column matrix. It has three real parameters that we call $\lambda_d$, $\lambda_s$, and $\lambda_b$; and three complex phases. We can absorb one of the phases by making a phase change to $Q$, so we are left with two $CP$ violating phases, which we call $\delta$ and $\delta'$.\footnote{The phases of the SM quark fields have already been  chosen to reduce the number of phases of the CKM matrix to one. 
} We will use the following parametrization for $\lambda$
\begin{equation}
 \label{Deflambda1DF}
 \lambda=\begin{pmatrix}
              \lambda_d \\
              \lambda_s \ e^{i\delta} \\
              \lambda_b \ e^{i\delta'}
              \end{pmatrix}.
\end{equation}
The DQCD contribution to the effective Hamiltonian of meson mixing is given by equation~(\ref{Heff}) with $\xi_M$ given by equation~(\ref{xi}), but without the summation over $Q$. The approximation of negligible $(m_Q/M_X)^2$ still holds. Let us define
\begin{equation}
\label{lambdaTilde}
\tilde{\lambda}=N_c^{1/4}\lambda  \ .
\end{equation}
%
    If the couplings are all real ($\delta$ and $\delta'$ are both close to $0$ or $\pi$), only $CP$ conserving constraints apply. In the more general case of complex couplings,  all the constraints of Table~\ref{tab:2} apply.  


\subsubsection{Real Couplings}
\label{1DFreal}
\begin{figure}[hbpt]
\centering
   \begin{subfigure}{0.4\textwidth}
    \includegraphics[width=\textwidth]{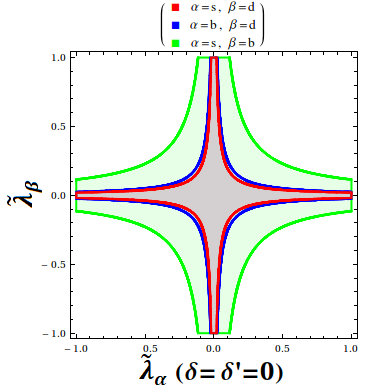} 
   \end{subfigure}
\hspace{1cm}
   \begin{subfigure}{0.4\textwidth}
    \includegraphics[width=\textwidth]{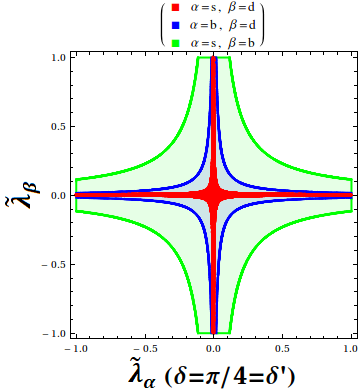} 
   \end{subfigure}
\\
\vspace{1cm}
 \begin{subfigure}{0.4\textwidth}
     \includegraphics[width=\textwidth]{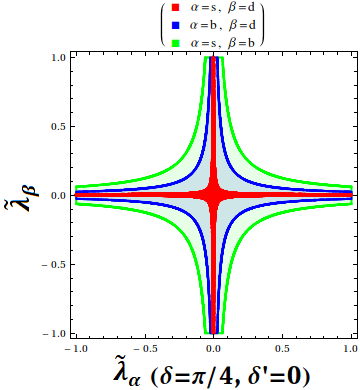}
  \end{subfigure}
\hspace{1cm}
 \begin{subfigure}{0.4\textwidth}
     \includegraphics[width=\textwidth]{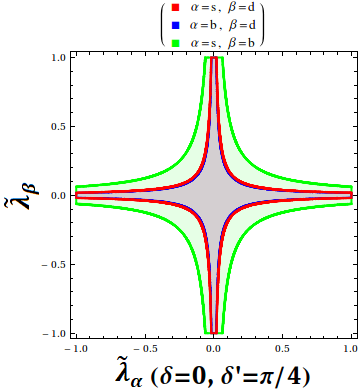}
  \end{subfigure}
\caption{Constraints on the one dark flavour Yukawa couplings defined in equation~\eqref{Deflambda1DF} with $M_X=1$~TeV. The (red, blue, green) lines are constraints from the ($K$, $B$, and $B_s$) systems. These regions are allowed at 95\% CL~from measurements shown in Table~\ref{tab:2}. The upper left figure shows the allowed regions for the couplings between SM and dark quarks when they are all real ($CP$ conservation). 
The upper right plot shows the scenario where the phases of $(\xi_K^\ast)^2$  and $(\xi_{B_d}^\ast)^2$ are close to $\pi /2$ and the phase of $(\xi_{B_s}^\ast)^2$ is close to zero. The lower left plot shows the scenario where the phases of $(\xi_K^\ast)^2$  and $(\xi_{B_s}^\ast)^2$ are respectively close to $\pi /2$ and $-\pi /2$, and the phase of $(\xi_{B_d}^\ast)^2$ is close to zero. Finally the lower right plot shows the scenario where the phases of $(\xi_{B_s}^\ast)^2$  and $(\xi_{B_d}^\ast)^2$ are close to $\pi /2$ and the phase of $(\xi_K^\ast)^2$ is close to zero.}
\label{Fig1DF}
\end{figure}

%
We begin with the case where the phases are set to zero or $\pi$ and the $\lambda$ vector is real. 
In order to guarantee  perturbative unitarity in a conservative manner, we impose the condition $\mid\lambda_\alpha\mid\leq 1$. Applying $CP$ conserving $\Delta F=2$ constraints of Table~\ref{tab:2}, we find: 
\begin{equation}
\begin{aligned}
(\tilde{\lambda}_d\tilde{\lambda}_s)^2\left(\frac{1\text{ TeV}}{M_X} \right)^2 & \leq  3.77 \times 10^{-4} \ , \\
(\tilde{\lambda}_d\tilde{\lambda}_b)^2 \left( \frac{1\text{ TeV}}{M_X} \right)^2 & \leq 6.55 \times 10^{-4} \ , \\
(\tilde{\lambda}_s\tilde{\lambda}_b)^2 \left( \frac{1\text{ TeV}}{M_X} \right)^2 & \leq 13.15 \times 10^{-3}.
 \end{aligned} \label{1DFconstraints}
 \end{equation}
These constraints are shown in the top left panel of Figure~\ref{Fig1DF}, where we took $M_X=1$ TeV. Generic $\mathcal{O}(1)$ values for all three Yukawa couplings are excluded, and the bounds become stronger for larger values of $N_c$, but weaker for a heavier mediator. We also conclude that if two of the $\lambda_{\alpha}$ are close to zero, then the third one can be generic. This is because in that scenario the SM flavour symmetry is preserved and the dark quark cannot mediate flavour violation.

The second and third relations of equation~(\ref{1DFconstraints}) are true for any $\delta$ and $\delta'$, because the $B-$system $CP$ conserving  constraints apply to the magnitude of the $\xi_B^2$ terms. The first constraint will change in the presence of $CP$-violating phases as discussed below.


\subsubsection{Generic Phases}
\label{1DFCPV}
In general, the coupling matrix $\lambda$ can be parameterized by equation~(\ref{Deflambda1DF}). The $\Delta F=2$ constraints of Table~\ref{tab:2} give, in addition to the second and third constraints of equation~(\ref{1DFconstraints}),  the following conditions on the couplings:
\begin{equation}
\begin{aligned}
-6.93 \times 10^{-4}\leq (\tilde{\lambda}_d\tilde{\lambda}_s)^2\cos  (2\delta)\left(\frac{1\text{ TeV}}{M_X} \right)^2  &  \leq  3.77 \times 10^{-4} \ ,
\\
 (\tilde{\lambda}_d\tilde{\lambda}_s)^2\mid\sin (2\delta)\mid\left(\frac{1\text{ TeV}}{M_X} \right)^2 & \leq 1.64\times 10^{-6} \ ,
\\
 -0.86\times 10^{-4} \leq (\tilde{\lambda}_d\tilde{\lambda}_b)^2\sin (2\delta')\left(\frac{1\text{ TeV}}{M_X} \right)^2  & \leq 3.12\times 10^{-4} \ , 
\\
 -3.54\times 10^{-3} \leq (\tilde{\lambda}_s\tilde{\lambda}_b)^2\sin  [2(\delta'-\delta)]  & \leq 3.88\times 10^{-3}.
\label{1DFconstraintsCPV}
\end{aligned}
\end{equation}
The first line is the $CP$-conserving constraint for the Kaon system, while the last three come from $CP$ violation. Note that if $\delta = \pi/2$, the bound on $\tilde{\lambda}_d\tilde{\lambda}_s$ from $\Delta M_K$ is about a factor of two weaker than than if $\delta=0,\pi$.


We now study the special scenarios where CPV effects are maximum, that is, the phases of two of the Wilson coefficients are close to $\pm\pi /2$, and the third one is close to zero.
These scenarios are represented in Figure~\ref{Fig1DF}, where we see that the allowed regions in the planes $(\tilde{\lambda}_d,\tilde{\lambda}_s)$,  $(\tilde{\lambda}_d,\tilde{\lambda}_b)$, and $(\tilde{\lambda}_s,\tilde{\lambda}_b)$, are respectively strongly narrowed when the CPV arguments of $(\xi_K^\ast)^2$, $(\xi_{B_d}^\ast)^2$, and $(\xi_{B_s}^\ast)^2$, are respectively close to $\pm\pi /2$. We also see that $CP$ violation in the kaon system (shown in red) places the strongest constraints. On the other hand, if contritubion to kaon mixing is purely real (bottom right panel), then the constraints are comparable to the $CP$ violating constraints from $B$ mesons (blue). The $B_s$ meson always gives the weakest constraints.


If instead of fixing the phases we fix magnitudes and vary the phases, we see that one phase must be small but not the other. As an example benchmark we take: $(\tilde{\lambda}_d,\tilde{\lambda}_s,\tilde{\lambda}_b)=(0.2,0.15,0.1)$ and find, for $M_X=1$ TeV:
\begin{equation}
\begin{aligned}
& -0.0018 \leq  \text{Arg}(\xi_K^2)\leq 0.0018  \\
 & -0.89 \leq  \text{Arg}(\xi_{B_d}^2)\leq 0.22 
 \end{aligned}
\end{equation}
Recall that $\text{Arg}(\xi_{B_s}^2)$ is the difference between the two arguments above. We see that for this size of $\tilde{\lambda}$, $CP$-violating effects require the phase of the coupling in the Kaon system to be very small, while the constraint on the phases in the $B$ system are much more mild.

%
\subsection{Two Dark Flavors}
\label{sec:2DF}
We now study the case of two dark flavors, $Q$ and $Q'$, with approximately degenerate masses that are small compared to $M_X$. The SM-DM coupling matrix $\lambda$ is then a $3\times 2$ matrix to which we apply the singular value decomposition as follows
\begin{equation}
\label{SVD2flavors}
\lambda = UDV^{\dagger} \ ,
\end{equation}
where $U$ and $V$ are respectively $3\times 3$ and $2\times 2$ unitary matrices, and $D$ is a diagonal $3\times 2$ matrix with positive entries. We parametrize $D$ as follows
\begin{equation}
\label{D2flavors}
D=\begin{pmatrix}
   d_1 & 0  \\
    0 & d_2 \\
    0 & 0
  \end{pmatrix}. 
\end{equation}
%
The matrix $U$, since it is unitary, has nine parameters, three real ones, and six complex phases. This number of phases can be reduced to four thanks to the following symmetry that keeps $\lambda$ invariant
\begin{equation}
\label{symUvsV}
U\longrightarrow U\begin{pmatrix}
    e^{i\alpha_1} & 0 & 0 \\
    0 & e^{i\alpha_2} & 0 \\
    0 & 0 & 1
  \end{pmatrix}
\ , \ \ \ V^\dagger\longrightarrow \begin{pmatrix}
    e^{-i\alpha_1} & 0  \\
    0 & e^{-i\alpha_2}
  \end{pmatrix}V^\dagger.
\end{equation}
We call the physical CPV phases $\delta_{12}$, $\delta_{13}$, $\delta_{23}$, and $\delta$. The matrix $V^\dagger$ can be rotated away due to the dark flavor symmetry $U(2)_Q$.
We now paramaterize $U$ as the product of three rotation matrices:
\begin{equation} \label{Umat2flav}
 U = U_{23}U_{13}U_{12} \ ,
 \end{equation}
\begin{equation} \label{UijMat2flav}
\begin{aligned}
 U_{23} = & \begin{pmatrix}
    e^{-i\delta} & 0 & 0 \\
    0 & \cos\theta_{23} & \sin\theta_{23}e^{-i\delta_{23}} \\
    0 & -\sin\theta_{23}e^{i\delta_{23}} & \cos\theta_{23}
  \end{pmatrix} \ , \\ 
U_{13} = &
  \begin{pmatrix}
    \cos\theta_{13} & 0 & \sin\theta_{13}e^{-i\delta_{13}} \\
    0 & e^{i\delta} & 0 \\
    -\sin\theta_{13}e^{i\delta_{13}} & 0 & \cos\theta_{13}
  \end{pmatrix} \ , \\
U_{12} = &
  \begin{pmatrix}
    \cos\theta_{12}  & \sin\theta_{12}e^{-i\delta_{12}} & 0 \\
    -\sin\theta_{12}e^{i\delta_{12}} & \cos\theta_{12} & 0 \\
     0 & 0 &  e^{i\delta}
  \end{pmatrix}.
  \end{aligned}
\end{equation}

\noindent Therefore the matrix $\lambda$ is given by
\begin{equation}
 \label{lambda2flav}
 \lambda = \begin{pmatrix}
    {\scriptstyle d_1c_{12}c_{13}e^{-i\delta}} &  {\scriptstyle d_2s_{12}c_{13}e^{-i(\delta +\delta_{12})}} \\
   {\scriptstyle -d_1(s_{12}c_{23}e^{i(\delta +\delta_{12})} + c_{12}s_{23}s_{13}e^{i(\delta_{13}-\delta_{23})})}  & {\scriptstyle d_2(c_{12}c_{23}e^{i\delta} - s_{12}s_{23}s_{13}e^{i(\delta_{13}-\delta_{12}-\delta_{23})})} \\
    {\scriptstyle  d_1(s_{12}s_{23}e^{i(\delta +\delta_{12}+\delta_{23})} - c_{12}c_{23}s_{13}e^{i\delta_{13}})} & {\scriptstyle -d_2 (c_{12}s_{23}e^{i(\delta +\delta_{23})} + s_{12}c_{23}s_{13}e^{i(\delta_{13}-\delta_{12})})} 
  \end{pmatrix},
\end{equation}
where we have nine unknown independent parameters, five real ones, $d_1$, $d_2$, $\theta_{12}$, $\theta_{13}$,  and $\theta_{23}$; and four CPV phases $\delta_{12}$, $\delta_{13}$, $\delta_{23}$, and $\delta$. We will take we take $0 \leq \theta_{ij}  \leq \pi/2$, and as before, we define
\begin{equation}
\label{Dtilde}
\tilde{D}= N_c^{1/4}D
\end{equation}
to scale out the dependence on the number of dark colours.

We will see in the next section that in the case of three dark flavors, if the matrix $D$ is proportional to the identity matrix, then $U$ can be rotated away along with $V^\dagger$, and then we land on the flavor universal scenario, $\lambda\propto \mathbbm{1}$. However, in the case of two dark flavors, even if $D$ has similar entries, $d_1\approx d_2$, $U$ cannot be rotated away and there is no flavor universality limit. If $U\propto \mathbbm{1}$, then the dark quarks are aligned with the SM quarks, so there is no flavour violation and thus no constraints. 


\subsubsection{Real Couplings}
\label{2DFreal}
\begin{figure}[hbpt]
 \centering
  \begin{subfigure}{0.48\textwidth}
%
%
    \includegraphics[width=\textwidth]{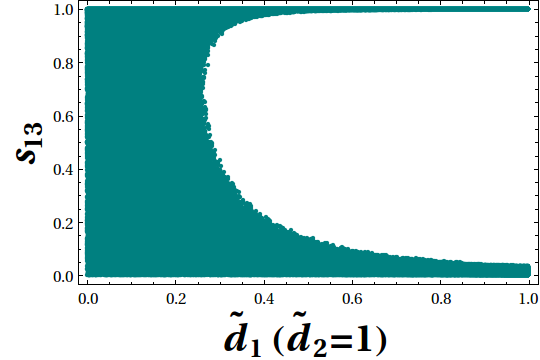}
    \end{subfigure}
\hspace{0.35cm}
%
%
%
%
%
%
%
\begin{subfigure}{0.48\textwidth}
%
%
%
%
    \includegraphics[width=\textwidth]{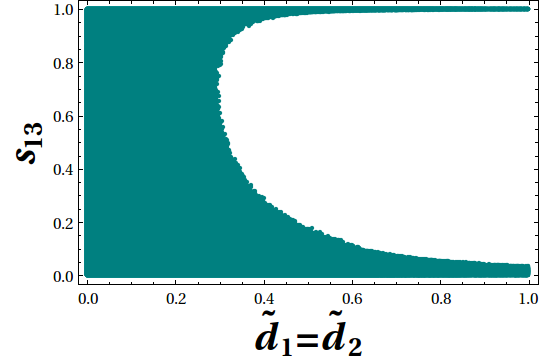}
    \end{subfigure}
%
%
\caption{Case of two dark flavors and a real SM-DM coupling matrix, with $M_X=1$~TeV. Left: Allowed values for $s_{13}$ as a function of $\tilde{d_1}$ when $\tilde{d_2}\approx 1$. Right: $s_{13}$ as a function of $\tilde{d_1}$ when $\tilde{d_1}\approx \tilde{d_2}$. All other mixing angles are randomly scanned. These regions are allowed at 95\% CL from measurements of the mass differences of $K^0-\bar{K}^0$ and $B_{d(s)}^0-\bar{B}_{d(s)}^0$.} 
\label{FIG2DFreal}
\end{figure}

If the CPV phases are all close to $0$ or $\pi$, then $\lambda$ is (approximately) real, and only  the $CP$-conserving constraints of Table~\ref{tab:2} apply.
The singular value decomposition allows us to choose the elements of $D$ to be positive and real. We will consider three cases, first $\tilde{d}_2\approx 1$ and $\tilde{d}_1\leq 1$, second, $\tilde{d}_1\approx 1$ and $\tilde{d}_2\leq 1$, and finally $d_1\approx d_2\leq 1$. As above, we do not consider larger values of $d_i$ to ensure pertubative unitarity.
We randomly scanned all the parameters, and plotted the allowed regions of the mixing angles $\theta_{ij}$  as functions of $\tilde{d}_i$, in the three mentioned scenarios. We find that $s_{12}$ and $s_{23}$ can take any value in the scan, thus we only show the plots of $s_{13}$ in Figure~\ref{FIG2DFreal}.

 We notice the following interesting cases for $M_X = 1$~TeV: 
\begin{itemize}
\item  In the case where $\tilde{d_1}\approx 1$, we obtain the same plot as that of the case where $\tilde{d_2}\approx 1$ (but with $s_{13}$ as a function of $\tilde{d_2}$). 
\item If $\tilde{d}_1\approx \tilde{d}_2\approx 1$, then either $s_{13}\leq 0.05$ or $s_{13}\geq 0.99$. 
\item If $\tilde{d}_{1(2)}\approx 1$ and $\tilde{d}_{2(1)} \leq 0.25$, or $\tilde{d}_1\approx \tilde{d}_2\leq 0.3$ then $s_{13}$ can take on any value. Otherwise, it is constrained to lie in the narrow blue region of the Figures.
\item If $\tilde{d}_1\approx 0$ ($\tilde{d}_2\approx 0$) then all the first (second) column of the matrix $\lambda$ vanishes, so only $Q'$ ($Q$) couples to the SM quarks and we can use the analysis of Section~\ref{sec:1DF}.
\end{itemize}


\subsubsection{Complex Couplings}
\label{2DFreal}
\begin{figure}[hbpt]
 \centering
  \begin{subfigure}{0.3\textwidth}
    \includegraphics[width=\textwidth]{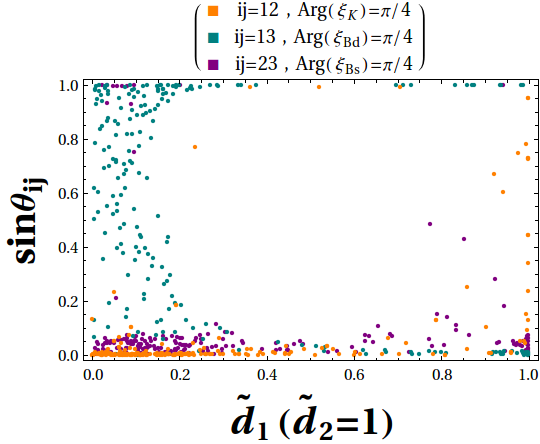}
   \end{subfigure}
\hspace{0.45cm}
\begin{subfigure}{0.3\textwidth}
    \includegraphics[width=\textwidth]{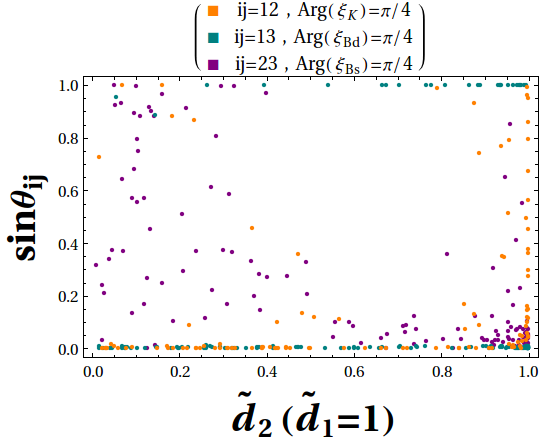}
    \end{subfigure}
\hspace{0.45cm}
\begin{subfigure}{0.3\textwidth}
    \includegraphics[width=\textwidth]{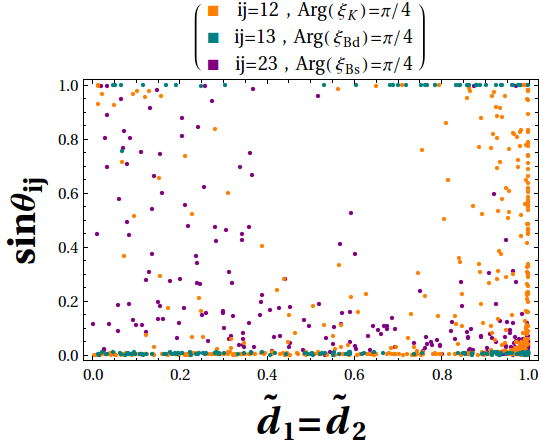}
   \end{subfigure}
\caption{Case of two dark flavors and a complex SM-DM coupling matrix, with $M_X=1$~TeV. From left to right, the first and second plots show the allowed values for the mixing parameters $s_{ij}$ as a function of one element of the matrix $\tilde{D}$, when the other one is close to unity. The third plot shows the case where the two elements of the matrix $\tilde{D}$ are equal.  All other parameters are randomly scanned. 
}
\label{FIGsupMAXCPV2F}
\end{figure}

If the phases are generic and the couplings are complex, we have to apply all the $\Delta F=2$ constraints of Table~\ref{tab:2}.
We consider the cases of maximum $CP$-violation, where we take respectively,  $\text{Arg}(\xi_K^2)=\pi /2$, $\text{Arg}(\xi_{B_D}^2)=\pi /2$, and $\text{Arg}(\xi_{B_S}^2)=\pi /2$; and find the allowed values for $s_{12}$, $s_{13}$, and $s_{23}$ respectively. We have considered the same three scenarios as in the real case, and plotted the allowed regions in the planes $(\tilde{d}_2,s_{ij})$ when $\tilde{d}_1\approx 1$, $(\tilde{d}_1,s_{ij})$ when $\tilde{d}_2\approx 1$, and $(\tilde{d}_i,s_{ij})$ when $\tilde{d}_1\approx \tilde{d}_2$. The plots are shown in Figure~\ref{FIGsupMAXCPV2F}, where we scanned all the parameters while imposing the flavor constraints in addition to $\text{Arg}(\xi_M^2)=\pi /2$, $M=K,B_d,B_s$. The CPV phases were scanned between $0$ and $2\pi$.
We see that the parameter space is highly restricted compared to the real case represented in Figure~\ref{FIG2DFreal}, with only about one in every one million points in the scan allowed by the constraints. This gives a sense of the tuning required in order not to be excluded.
%
%

Rather than fixing one phase and scanning the others, as we did in Figure~\ref{FIGsupMAXCPV2F}, we can also fix all phases of the Wilson coefficients. In the cases where $M_X = 1$ TeV and at least one phase is equal to $\pi /4$, $3\pi /4$, $5\pi /4$, or $7\pi /4$ and the others are zero, a scan of $10^8$ parameter points turns up zero allowed parameter points, indicating that this slice of parameters is either completely excluded or extremely tuned. 


If the elements of the matrix $D$ are equal ($d_1=d_2$), then the dependence on the elements of $U_{12}$ (defined in equation~(\ref{UijMat2flav})) drops from the $\xi_M$ terms, hence letting $s_{12}$ and $\delta_{12}$ unconstrained. This explains the orange dots' broad distribution in the right plot of Figure~\ref{FIGsupMAXCPV2F}.
In this scenario, the $\xi_M$ terms have the following simple expressions\footnote{The subscript ``eq'' represents the case $d_1=d_2$.} 
\begin{equation}
\label{xi2DFdeg}
\begin{aligned}
[\xi_K]_{\text{eq}}=& -d_1^2e^{-i(\delta_{13}-\delta_{23}+\delta)}s_{13}c_{13}s_{23} \\
[\xi_{B_d}]_{\text{eq}}=&  -d_1^2e^{-i(\delta_{13}+\delta)}s_{13}c_{13}c_{23} \\
[\xi_{B_s}]_{\text{eq}}=&  -d_1^2e^{-i\delta_{23}}c_{13}^2s_{23}c_{23} \ .
\end{aligned}
\end{equation}
We notice that all the $\xi_M$ terms vanish and there are no flavour constraints if $c_{13}=0$, so the dark quarks do not couple to the $d$ quark, or if $s_{13}=0=s_{23}$, so the dark quarks do not couple to the $b$ quark. This explains the blue dots' clusters at $s_{13}=0$ and $s_{13}=1$ in the right plot of Figure~\ref{FIGsupMAXCPV2F}.  This case when one SM quark does not participate and $d_1=d_2$ is the two flavour analogue of coupling universality, allowing all the mixing angles to be rotated away so there are no constraints. 

We can also consider scenarios where $d_1\approx d_2$ and only one of the $\xi_M$ is nonzero, thus only one $s_{ij}$ is constrained. First, if $s_{13}=0$ then $\xi_K=0=\xi_{B_d}$, and only $\xi_{B_s}\neq 0$, so the only $CP-$conserving   constraint is $[N_cd_1^4(s_{23}c_{23})^2/M_X^2\leq 0.013]$, and the only CPV phase constrained is $\delta_{23}$. Second, if $s_{23}=0$ then $\xi_K=0=\xi_{B_s}$, and only $\xi_{B_d}\neq 0$, so the only $CP-$conserving   constraint is $[N_cd_1^4(s_{13}c_{13})^2/M_X^2\leq 6.55\times 10^{-4}]$, and the only CPV phases constrained are $\delta_{13}$, and $\delta$. Finally, if $c_{23}=0$ then $\xi_{B_d}=0=\xi_{B_s}$, and only $\xi_K\neq 0$, so  the only $CP-$conserving   constraint is $[N_cd_1^4(s_{13}c_{13})^2/M_X^2\leq 3.77\times 10^{-4}]$, and all the CPV phases except $\delta_{12}$ are constrained.


\subsection{Three Dark Flavors}
\label{sec:3DF}
In this section, the number of dark flavors is $n_f=3$. The $CP$-conserving case was analyzed in detail in~\cite{ABG014}, and we begin by reviewing and updating their results. 
We will label the three dark flavours $d_d$, $s_d$ and $b_d$ and parametrize the Yukawa coupling matrix $\lambda$ using the singular value decomposition, $\lambda = UDV^{\dagger}$, as in the two flavor case. Now $U$, $D$ and $V$ are all $3 \times 3$ matrices. $U$ and $V$ are unitary, and $D$ is diagonal with positive entries. Since we consider that $d_d$, $s_d$ and $b_d$ are approximately mass-degenerate, we can use the $U(3)_{Q_d}$ dark flavor symmetry to rotate $V$ away, so we are left with
\begin{equation} 
\label{lambdaMatrix}
\lambda = UD \ .
\end{equation}
The diagonal matrix $D$ can be parametrized as:
\begin{equation} \label{Dmatrix}
D = \lambda_0 \ \ldotp \ \mathbbm{1} + \text{diag}\left( \lambda_1, \ \lambda_2, \ -(\lambda_1+\lambda_2)\right) \ ,
\end{equation}
and the unitary matrix $U$ can be parametrized in the same way as in the preceding section. Because there is one more quark whose phase can be rotated away, there is one less physical phase than in the $n_f=2$ case.
\begin{equation} \label{Umatrix}
 U = U_{23}U_{13}U_{12} \ ,
 \end{equation}
\begin{equation} 
\label{UijMatrices}
\begin{aligned}
 U_{23} = & \begin{pmatrix}
    1 & 0 & 0 \\
    0 & \cos\theta_{23} & \sin\theta_{23}e^{-i\delta_{23}} \\
    0 & -\sin\theta_{23}e^{i\delta_{23}} & \cos\theta_{23}
  \end{pmatrix} \ , \\ 
U_{13} = &
  \begin{pmatrix}
    \cos\theta_{13} & 0 & \sin\theta_{13}e^{-i\delta_{13}} \\
    0 & 1 & 0 \\
    -\sin\theta_{13}e^{i\delta_{13}} & 0 & \cos\theta_{13}
  \end{pmatrix} \ , \\
U_{12} = &
  \begin{pmatrix}
    \cos\theta_{12}  & \sin\theta_{12}e^{-i\delta_{12}} & 0 \\
    -\sin\theta_{12}e^{i\delta_{12}} & \cos\theta_{12} & 0 \\
     0 & 0 & 1
  \end{pmatrix}.
  \end{aligned}
\end{equation}
Therefore, $\lambda$ has nine unknown independent parameters: $\lambda_0$, $\lambda_1$, $\lambda_2$, $\theta_{12}$, $\theta_{13}$, $\theta_{23}$, $\delta_{12}$, $\delta_{13}$ and $\delta_{23}$. The $\lambda_{ij}$ and $\theta_{ij}$ are real, and the $\delta_{ij}$ are the CPV phases. 

The parameters $\lambda_1$ and $\lambda_2$ measure the deviation from flavor universal couplings where  $\lambda_1\simeq 0\simeq \lambda_2$ and  
$\lambda$ is proportional to the unit matrix, $\lambda = \lambda_0 \ \ldotp \ \mathbbm{1} $. This is also referred to as alignment; the $d$ quark couples only to the dark quark $d_d$, $s$  couples only to $s_d$, and $b$ couples only to $b_d$, all with the same coupling constant $\lambda_0$. In the universal case, the flavor and $CP$ constraints do not apply because an $SU(3)$ flavour symmetry is preserved.
Following~\cite{ABG014}, we define 
\begin{equation}\label{Delta}
\Delta_{ij}=D_{ii}-D_{jj}
\end{equation}
as parameterizing the deviation from universality. We will express our results as functions of $\Delta_{ij}$.

Plugging in our parameterization, the SM-DM coupling matrix has the following expression 
\begin{equation}
\begin{aligned}
 \label{lambdaMatrix2}
 \lambda = & \begin{pmatrix}
    \lambda_{dd_d} \ & \ \lambda_{ds_d} \ & \ \lambda_{db_d}  \\
    \lambda_{sd_d}  \ & \ \lambda_{ss_d} \ & \ \lambda_{sb_d}  \\
     \lambda_{bd_d} \ & \ \lambda_{bs_d} \ & \ \lambda_{bb_d}
  \end{pmatrix} \ , \\
  \lambda_{dd_d} = & \ (\lambda_0+\lambda_1)c_{12}c_{13} \ , \\
  \lambda_{ds_d}= & \ (\lambda_0+\lambda_2)s_{12}c_{13}e^{-i\delta_{12}} \ , \\
  \lambda_{db_d}= & \ (\lambda_0-\lambda_1-\lambda_2)s_{13}e^{-i\delta_{13}} \ , \\ 
  \lambda_{sd_d} = & -(\lambda_0+\lambda_1)(s_{12}c_{23}e^{i\delta_{12}} + c_{12}s_{23}s_{13}e^{i(\delta_{13}-\delta_{23})}) \ , \\
  \lambda_{ss_d}= & \ (\lambda_0+\lambda_2)(c_{12}c_{23} - s_{12}s_{23}s_{13}e^{i(\delta_{13}-\delta_{12}-\delta_{23})}) \ , \\
  \lambda_{sb_d}= & \  (\lambda_0-\lambda_1-\lambda_2)s_{23}c_{13}e^{-i\delta_{23}} \ , \\
   \lambda_{bd_d} = & \ (\lambda_0+\lambda_1)(s_{12}s_{23}e^{i(\delta_{12}+\delta_{23})} - c_{12}c_{23}s_{13}e^{i\delta_{13}}) \ , \\ 
  \lambda_{bs_d}= & -(\lambda_0+\lambda_2) (c_{12}s_{23}e^{i\delta_{23}} + s_{12}c_{23}s_{13}e^{i(\delta_{13}-\delta_{12})}) \ , \\
 \lambda_{bb_d}= & \ (\lambda_0-\lambda_1-\lambda_2)c_{23}c_{13}.
\end{aligned}
\end{equation}
Without loss of generality, we take the matrix $D$ to have positive entries, so $\lambda_0$ is a positive number.  Taking perturbative unitarity into account, it is sufficient to impose the following conditions on the $\lambda_i$ 
\begin{equation}
\begin{aligned}
\label{lambdaCond}
\lambda_0 \ &\approx 1 \ , \\
|\lambda_1| \ & \leq \ 1 \ , \\
|\lambda_2| \ & \leq \ 1 \ , \\
|\lambda_1 +\lambda_2| \ & \leq 1 \ . 
\end{aligned}
\end{equation}
And to avoid double counting, we take $ 0 \leq \theta_{ij}  \leq \pi/4$. Finally we define
\begin{equation}
\label{Deltatilde} \tilde{\lambda}= N_c^{1/4}\lambda \ , \ \ \ \tilde{D}= N_c^{1/4}D \ , \ \ \ \tilde{\Delta}_{ij}= N_c^{1/4}\Delta_{ij}
\end{equation}
to scale out the number of dark colors as in the other cases.

\subsubsection{Real Coupling Matrix}
\label{sec:3DFreal}
\begin{figure}[hbpt]
\centering
   \begin{subfigure}{0.45\textwidth}
    \includegraphics[width=\textwidth]{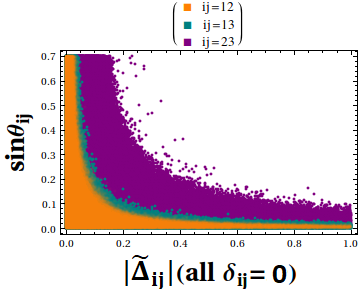} 
   \end{subfigure} \\
\vspace{1cm}
   \begin{subfigure}{0.45\textwidth}
    \includegraphics[width=\textwidth]{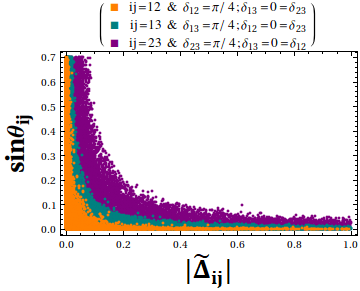} 
   \end{subfigure}
\hspace{1cm}
 \begin{subfigure}{0.45\textwidth}
     \includegraphics[width=\textwidth]{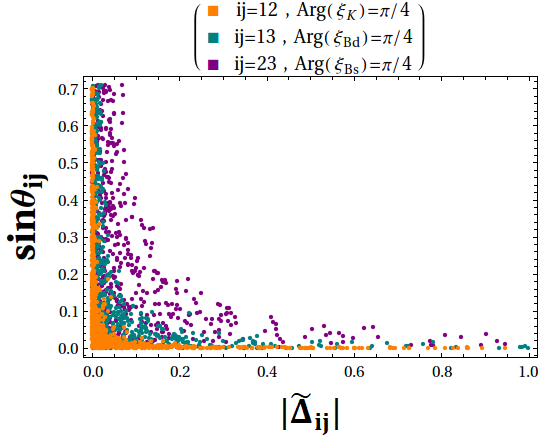}
  \end{subfigure}
\caption{Case of three dark flavors with $M_X=1$~TeV. Allowed values for each mixing angle $s_{ij}$ as a function of $\tilde{\Delta}_{ij}$ defined in equations~(\ref{Delta}) and~(\ref{Deltatilde}). All other mixing angles and $\tilde{\Delta}$ parameters are randomly scanned. The upper plot represents the scenario of a real SM-DM mixing matrix. The lower left plot corresponds to the scenario where one of the CPV phases $\delta_{ij}$ is close to $\pi /4$, and the two others are close to zero. Finally, the lower right plot represents the scenarios where the argument of one  Wilson coefficient of $H_{\rm eff}^{\rm NP}$ is close to $\pi /2$. Note that $|\tilde{\Delta}_{ij}|$ can take values up to 2, we just stopped at 1 for compactness.}
\label{Fig3DF}
\end{figure}

We consider a real matrix  $\lambda$, i.e. all the $\delta_{ij}$ are close to $0$ or $\pi$. From the $CP$-conserving constraints of Table~\ref{tab:2}, we plot in the first row of Figure~\ref{Fig3DF} the allowed regions for the mixing angles $\theta_{ij}$  as functions of $\tilde{\Delta}_{ij}$. To do so, we randomly generated all the parameters and subjected each point of the parameter space to the flavor constraints. Our plots differ slightly from~\cite{ABG014} and we have confirmed that when using the same input data as~\cite{ABG014} and setting $N_c=1$ we are able to precisely reproduce their results.
Therefore, our Figure is an update to Figure~5 of \cite{ABG014} using data summarized in Table~\ref{tab:1}. We also studied the specific scenarios shown in Figure~5 of~\cite{RennerSchwaller018} with $N_c=3$, and our results are similar. 

\subsubsection{Complex Coupling Matrix}
\label{sec:3DFcomplex}
When the matrix $\lambda$ is complex, all the $\Delta F=2$ constraints of Table~\ref{tab:2} apply. To find the parameter space, we randomly scan all the parameters and impose the flavor constraints on each generated point. The CPV phases were scanned between $0$ and $2\pi$. While the $CP$-violating constraints are stronger for generic values of the phases, the $CP$-conserving ones are important when the phases are accidentally small. We see that if $\tilde{\Delta}_{ij}$ and $\sin\theta_{ij}$ are both $\mathcal{O}(1)$ then the scenario is always excluded regardless of the other parameters. 

In the scenarios where one $\delta_{ij}$ is close to $\pi /2$ or $3\pi /2$ and the two others are close to zero the allowed region is slightly smaller than the real case. 
In the scenarios, shown in the lower left of Figure~\ref{Fig3DF}, where one $\delta_{ij}$ is close to $\pi /4$,  $3\pi /4$, $5\pi /4$, or $7\pi /4$ and the two others are close to zero, we notice that for given $k,l $, when $\delta_{kl}=\pi /4$, only the corresponding $s_{kl}$ is more constrained than it is in the real case.  

There is large $CP$-violation when the argument of only one  $\xi^\ast_M$ approaches $\pi /4$, $3\pi /4$, $5\pi /4$, or $7\pi /4$. These phases are: $\text{Arg}(\xi^\ast_M) = -(1/2) \text{Arg(Wilson coefficient)}$ of the effective Hamiltonian~(\ref{Heff}), and they are combinations of the CPV phases $\delta_{ij}$. So these scenarios correspond to the cases where the Wilson coefficients are purely imaginary. In each of these cases, the $CP$-violating constraint on $(\xi^\ast_M)^2$ from Table~\ref{tab:2} is the strongest. The lower right plot of Figure~\ref{Fig3DF} shows these scenarios, where we took the argument of one $\xi^\ast_M$ term close to $\pi /4$. 
 To produce these plots, we scanned all the parameters, including the CPV phases $\delta_{ij}$, and we imposed to the generated points the value of one Wilson coefficient's argument, in addition to the flavor constraints. We find the narrowest allowed region for ($s_{12}$, $s_{13}$, $s_{23}$) when ($\text{Arg}(\xi^\ast_K)=\pi /4$,   $\text{Arg}(\xi^\ast_{B_d})=\pi /4$, $\text{Arg}(\xi^\ast_{B_s})=\pi /4$), and among them, the constraints on $s_{12}$ are the strongest. Only about one in every 300 thousand  $(|\tilde{\Delta}_{23}|,s_{23})$ points in the scan are allowed by the constraints. This gives an idea of the tuning required in order not to be excluded, and explains the scattered purple scan in the lower right plot of Figure~\ref{Fig3DF}.
 
 Like in the two dark flavor case, the case where the argument of at least one $\xi^\ast_M$ (or all of them) approaches $\pi /4$, $3\pi /4$, $5\pi /4$, or $7\pi /4$, with the others near zero, are excluded or extremely tuned because a scan of $10^8$ did not find any allowed points.


We now consider the $ij-$degeneracy scenario, where the $i^{th}$ and $j^{th}$ generations  of dark quarks are quasi-degenerate in their coupling to SM quarks and  $\Delta_{ij}=0$, for some fixed values of $i$ and $j$. In that case, the  $s_{ij}$ mixing parameter  as well as the corresponding CPV phase $\delta_{ij}$ can be rotated away. The allowed regions for the two other mixing parameters are smaller than those of the real non-degenerate case. For example, if $\lambda_1\simeq \lambda_2$, then we are in the $12-$degeneracy scenario, $s_{13}$ and $s_{23}$ are constrained in regions narrower than those of the real $\lambda$ case.

Let us mention that ${\xi}_M  , \  M=K,B_d,B_s$, are independent of $\theta_{12}$ and $\delta_{12}$ in the
$12-$degeneracy case, where we have equal first two elements of the matrix $D$ in equation~(\ref{lambdaMatrix}), which happens when $\lambda_1 \approx \lambda_2$. Then no constraints apply on $s_{12}$ as well as $\delta_{12}$, for any values of the other parameters. This is because the dependence on the elements of $U_{12}$ drops from ${\xi}_M$ in this case. However, the situation is different in the $13-$ ($23-$) degeneracy case, which happens when $\lambda_2 \approx -2\lambda_1$ ($\lambda_2 \approx -\lambda_1/2$). In this case, ${\xi}_M$ terms still depend on the elements of $U_{13}$ ($U_{23}$) but all values $s_{13}$ and $\delta_{13}$ ($s_{23}$ and $\delta_{23}$) can still be allowed when scanning all the other parameters. The special property of the $12-$degeneracy case is due to the parameterization of equation~(\ref{Umatrix}), where $U_{12}$ multiplies $D$ directly in equation~(\ref{lambdaMatrix}). If we change the order of the $U_{ij}$ in $U$, then, the analysis of the constraints on the parameters will differ.

We will now focus on the $12-$degeneracy scenario ($\lambda_1=\lambda_2$). 
The $\xi_M$ terms have the following simple expressions
\begin{equation}
\label{xi3DF12deg}
\begin{aligned}
[\xi_K]_{ 12-\text{deg}}=& \ 3\lambda_1(\lambda_1-2\lambda_0)e^{-i(\delta_{13}-\delta_{23})}s_{13}c_{13}s_{23} \\
[\xi_{B_d}]_{ 12-\text{deg}}=&  \ 3\lambda_1(\lambda_1-2\lambda_0)e^{-i\delta_{13}}s_{13}c_{13}c_{23} \\
[\xi_{B_s}]_{ 12-\text{deg}}=&  \ 3\lambda_1(\lambda_1-2\lambda_0)e^{-i\delta_{23}}c_{13}^2s_{23}c_{23} \ .
\end{aligned}
\end{equation}
In our choice of the mixing angles' domain, $0\leq \theta_{ij}\leq \pi /4$, so $c_{ij} \geq 1/\sqrt{2}$.  Like in the two dark flavor case,  all the $\xi_M$ terms vanish, thus there are constraints, if $s_{13}=0=s_{23}$, so the $b$ quark does couple to the first and second generations of dark quarks, and the third generation of dark quarks does not couple to $d$ and $s$ quarks.

If only $s_{13}=0$, and the $d$ quark does not couple to the third generation of dark quarks, then $\xi_K=0=\xi_{B_d}$, and only $\xi_{B_s}\neq 0$, so only $s_{23}$ and $\delta_{23}$ are constrained. If $s_{23}=0$ so the $s$ does not couple to the third generation of dark quarks , then $\xi_K=0=\xi_{B_s}$, and only $\xi_{B_d}\neq 0$, so only $s_{13}$ and $\delta_{13}$  are constrained.
Finally, let us mention that from the expression in equation~(\ref{lambdaMatrix2}), we see that if $\lambda_1 \approx -\lambda_0$, then all the elements of the first column of the matrix $\lambda$ are zero and $d_d$ does not couple to SM particles, so we can use the two flavour analysis. The same scenario happens to $s_d$ if $\lambda_2 \approx -\lambda_0$, and to $b_d$ if $\lambda_1 +\lambda_2 \approx \lambda_0$.


\subsection{What if \boldmath{$n_f > 3$} ?}
\label{sec:nfDF}
Let us now consider the case of more than three flavors of dark quarks, $Q_1, Q_2, Q_3, Q_4, ... ,Q_{n_f}$, and compute the number of unknown parameters. The SM-DM mixing $\lambda$ is a $3\times n_f$ matrix. We apply to it the singular value decomposition:
\begin{equation}
\label{SVDnflavors}
\lambda = UDV^{\dagger} \ ,
\end{equation}
where $U$ and $V$ are respectively $3\times 3$ and $n_f\times n_f$ unitary matrices, and $D$ is a diagonal $3\times n_f$ matrix with positive entries. We use the following parametrization for $D$:
\begin{equation}
\label{Dnflavors}
D=\begin{pmatrix}
    D_1 & 0 & 0 & 0 & . . . . & 0 \\
    0 & D_2 & 0 & 0 & . . . . & 0\\
    0 & 0 & D_3 & 0 & . . . . & 0
  \end{pmatrix}, 
\end{equation}
where there are $n_f$ columns. The matrix $U$ has three real parameters, and six complex phases. We can reduce these phases to three thanks to the following symmetry that keeps $\lambda$ invariant
\begin{equation}
\label{symUvsVnfDF}
U\longrightarrow U\begin{pmatrix}
    e^{i\alpha_1} & 0 & 0 \\
    0 & e^{i\alpha_2} & 0 \\
    0 & 0 & e^{i\alpha_3}
  \end{pmatrix}
\ , \ \ \ V^\dagger\longrightarrow \begin{pmatrix}
    e^{-i\alpha_1} & 0 & 0 & 0 & . . . . & 0  \\
    0 & e^{-i\alpha_2} & 0 & 0 & . . . . & 0 \\
    0 & 0 & e^{-i\alpha_3} & 0 & . . . . & 0 \\
    0 & 0 & 0 & 1 & 0 . . . & 0 \\
    ... \\
    0 & 0 & 0 & 0 & 0 . . . & 1 
  \end{pmatrix}V^\dagger.
\end{equation}
Then provided that the $n_f$ dark quarks are mass degenerate, $V^\dagger$ can be rotated away using the $U(n_f)$ dark flavor symmetry. Therefore, we are left with $\lambda = UD$, and we can take for $U$ the parametrization of equation~(\ref{Umatrix}, \ref{UijMatrices}) since it is the same number of real and complex parameters. In summary, as long as the quarks are approximately degenerate in mass, the flavour constraints on the $n_f > 3$ case are equivalent to those from the $n_f=3$ case.

There is, however, a remnant of the $U(n_f)$ flavour symmetry that rotates the dark pions among themselves that is left unbroken if $n_f > 3$. In particular, the Yukawa couplings of equation~\eqref{IntLag} explicitly break $U(n_f) \rightarrow U(n_f - 3)$, so the dark pions charged under the remnant symmetry will be stable. This could lead to new phenomenology and constraints from missing energy searches. Small quark mass splittings and/or higher dimensional operators could destabilize the dark pions, but if they are very long lived or stable, there will also be constraints from cosmology.

\section{\boldmath{$\Delta F =1$} Phenomenology and \boldmath{$CP-$}Violation}
\label{Pheno}
The high energy collider signatures of this class of dark QCD models are explored in~\cite{EmergingJets,RennerSchwaller018,Mies:2020mzw,Linthorne:2021oiz}, and direct detection constrains are also discussed in~\cite{BaiSchwaller014,RennerSchwaller018}. In this section we will consider some novel phenomenology, focusing on rare meson decays to dark sector states and the study of large $CP$-violating phases. Our main results are summarized in Tables~\ref{tab:3} and \ref{tab:4} at the end of this section.   


If the number of dark flavours  $n_f>1$, then the dark pions can be considered as pseudo Nambu-Goldstone bosons (pNGBs) resulting from the breaking of the chiral symmetry, $SU(n_f)_L \times SU(n_f)_R$ in the dark sector~\cite{BaiSchwaller014,RennerSchwaller018} analogous to SM pions. Therefore, their mass is generically smaller than the dark baryon masses and is proportional to the dark quark masses. As we have done throughout, we will work in mass degenerate limit for the dark quarks, so there is a multiplet of $n_f^2-1$ degenerate dark pions.
 In order for dark pions to appear in meson decays, we consider the parameter space $M_D \lesssim 4$ GeV where $M_D$ is a generic dark pion mass. 
Decays of SM mesons to dark sector states are thus dominated by dark pions in the final state because they are generically the lightest dark sector hadrons. Furthermore, heavier mesons will typically decay rapidly down to dark pions, analogous to $\rho \rightarrow \pi\pi$ in the SM.  We will consider the case of a single dark quark flavor below, after discussing the multi-flavor scenario.


To better understand the phenomenology, we now estimate the lifetime of the dark pions and compare it to a typical detector size. If decays to kaons are kinematically allowed, $M_D \gtrsim 2 M_K$, which means that we are in the mass range 1 GeV$\lesssim M_D \lesssim 4$ GeV,
the lifetime of such dark pions (composed of dark quark-antiquark pairs such as $Q\bar{Q'}$) can be estimated by computing decays into SM quark-antiquark pairs. For this mass range, $\pi_{\text{D}} \rightarrow s\bar{s}$, and the decay rate is
\begin{equation}
\label{PiDssbar}
\Gamma (\pi_{\text{D}} \rightarrow s\bar{s}) \simeq \frac{3|\lambda_{sQ}\lambda_{sQ'}^\ast |^2}{128\pi M_X^4}F_D^2M_Dm_s^2 \ : \;\;
M_D \gtrsim 2 M_K,
\end{equation}
where $F_D$ is the decay constant of $\pi_{\text{D}}$. 
The dark pion's lifetime is then
\begin{equation}
\label{HeavyPiDlife}
\tau(\pi_{\text{D}} \rightarrow s\bar{s})\simeq  10^{-4}\text{ s}\left(\frac{M_X}{1\text{ TeV}}\right)^4\left(\frac{0.01}{|\lambda_{sQ}\lambda_{sQ'}^\ast |}\right)^2\left(\frac{1\text{ GeV}}{F_D}\right)^2\left(\frac{1\text{ GeV}}{M_D}\right) \ ,   
\end{equation}
which translates to a flight distance on the kilometer scale, much longer than typical detectors for this choice of parameters. 

If the dark pions are in the mass range 300 MeV $\lesssim M_D \lesssim 1$ GeV, then they can decay to different final states such as, 2 or 3 pions, and 1 or 2 pions with one kaon. These decays have either ($\pi_{\text{D}} \rightarrow d\bar{d}$) or ($\pi_{\text{D}} \rightarrow d\bar{s}$) as quark level modes. We can estimate $\Gamma(\pi_{\text{D}} \rightarrow d\bar{d})$  from equation~(\ref{PiDssbar}) by making the appropriate changes, and for ($\pi_{\text{D}} \rightarrow d\bar{s}$), we have
\begin{equation}
\label{PiDds}
\Gamma (\pi_{\text{D}} \rightarrow d\bar{s}) \simeq \frac{3|\lambda_{dQ}\lambda_{sQ'}^\ast |^2}{256\pi M_X^4}F_D^2M_D\left(m_s^2-\frac{m_s^4}{M_D^2}\right) \ : \;\;
300 \text{ MeV} \lesssim M_D \lesssim 1 \text{ GeV} \ . 
\end{equation}
The corresponding lifetimes are then:
\begin{equation}
\begin{aligned}
\label{MediumPiDlife}
\tau(\pi_{\text{D}} \rightarrow d\bar{d})\simeq & \; 1\text{ s}\left(\frac{M_X}{1\text{ TeV}}\right)^4\left(\frac{0.01}{|\lambda_{dQ}\lambda_{dQ'}^\ast |}\right)^2\left(\frac{350\text{ MeV}}{F_D}\right)^2\left(\frac{350\text{ MeV}}{M_D}\right) \ , \\
\tau(\pi_{\text{D}} \rightarrow d\bar{s})\simeq & \; 4 \times 10^{-4}\text{ s}\left(\frac{M_X}{1\text{ TeV}}\right)^4\left(\frac{0.01}{|\lambda_{dQ}\lambda_{sQ'}^\ast |}\right)^2\left(\frac{800\text{ MeV}}{F_D}\right)^2\left(\frac{800\text{ MeV}}{M_D}\right) \ . 
\end{aligned}
\end{equation}
 Therefore, a typical flight distance for a final state that can decay to $d\bar{d}$ ($d\bar{s}$) is on the $10^5$~kilometer (kilometer) scale. 
%
If decays to kaons or pions are not kinematically allowed, $M_D \lesssim 300$ MeV, then the leading decay will be a loop decay to two photons. 
 We estimate the decay rate of the mode $(\pi_{\text{D}}\rightarrow \gamma\gamma)$ via a $\pi_{\text{D}}-\pi^0$ mixing followed by the SM decay $(\pi^0\rightarrow \gamma\gamma)$. We find\footnote{This expression assumes $\frac{|\lambda_{dQ}\lambda_{dQ'}^\ast |F_DM^3_D}{M^2_X}\ll |M^2_D-M^2_{\pi}|$.  If the two states are very nearly degenerate, then the mixing angle is approximately $1/\sqrt{2}$.}
\begin{equation}
\label{PiD2gamma}
\Gamma(\pi_{\text{D}}\rightarrow \gamma\gamma)\approx \frac{\alpha^2_{em}}{4096\pi^3}\frac{|\lambda_{dQ}\lambda_{dQ'}^\ast |^2}{M^4_X}\frac{F^2_DM^7_D}{(M^2_D-M^2_{\pi})^2} \ : \;\;
M_D \lesssim 300 \text{ MeV} \ .
\end{equation}
Then, for $M_D\simeq 150$ MeV, the dark pion's lifetime is 
\begin{equation}
\label{LightPiDlife}
\tau(\pi_{\text{D}} \rightarrow \gamma\gamma)\simeq 0.8\text{ s}\left(\frac{M_X}{1\text{ TeV}}\right)^4\left(\frac{1}{|\lambda_{dQ}\lambda_{dQ'}^\ast |}\right)^2\left(\frac{150\text{ MeV}}{F_D}\right)^2 \ ,   
\end{equation}
and it drops to $0.4$ s, for $M_D\simeq 300$ MeV. Note that this lifetime is for $\mathcal{O}(1)$ couplings, as opposed to the previous estimates that used $\mathcal{O}(0.1)$.

We see for almost all of the parameter space, the lifetimes are significantly larger than the size of detectors, so most dark pions will decay outside of the detector volume and only contribute to missing energy. The case where $n_f > 3$ and some of dark pions are stable can thus be treated the same way. This differs from the assumptions of~\cite{EmergingJets} where the dark pions are at the GeV scale and all the couplings are $\mathcal{O}(1)$, leading to lifetimes in the meter range. 

On the lighter side of the parameter space, the lifetimes can be close to the limit from Big Bang Nucleosynthesis (BBN), which excludes pion lifetimes longer than $\sim 1$ second~\cite{RennerSchwaller018}. Therefore we also restrict ourselves to $M_D \gtrsim 150$ MeV. We have assumed throughout that all the entries of $\lambda$ are the same order of magnitude, but if this is not the case, then dark pions can have vastly different lifetimes, potentially leading to interesting phenomenology~\cite{RennerSchwaller018}.

So far, we have been discussing the lifetimes of the dark pions when $n_f \geq 2$, let us now discuss the $n_f = 1$ scenario. A (spontaneously broken) chiral symmetry as well as PNGBs are absent in this case of a single flavor. A study of a QCD with one flavour found~\cite{1FlavorQCD} that the lightest hadrons are a pseudoscalar meson $\eta$ which is a bound state of a quark and an antiquark, and a corresponding scalar meson $\sigma$ about a factor 1.5 heavier, so decays of the scalar to the pseudo-scalar are kinematically forbidden. We take those two states  to be the only relevant ones for our one flavour dark QCD. These states will decay mostly to $s\bar{s}$, and we estimate their decay rate as
\begin{equation}
\label{EtaDssbar}
\Gamma (\eta (\sigma) \rightarrow s\bar{s}) \simeq \frac{3|\lambda_{sQ}|^4}{128\pi M_X^4}F_{\eta (\sigma)}^2 M_{\eta (\sigma)}\, m_s^2 \ ,
\end{equation}
where $F_{\eta (\sigma)}$ is the decay constant of $\eta\, (\sigma)$, defined by:
\begin{equation}
\begin{aligned}
\label{FetaAndFsigma}
& \bra{0}\bar{Q}\gamma^{\mu}\gamma_5Q\ket{\eta}=iF_{\eta}p^{\mu}  \ ,  \\
& \bra{0}\bar{Q}\gamma^{\mu}Q\ket{\sigma}=iF_{\sigma}p^{\mu} \ ,
\end{aligned}
\end{equation}
$p^{\mu}$ being the 4-momentum of $\eta\, (\sigma)$.
The lifetimes are then
\begin{equation}
\begin{aligned}
\label{EtaDandSigmaDlife}
& \tau(\eta \rightarrow s\bar{s})\simeq  1\times 10^{-5}\text{ s}\left(\frac{M_X}{1\text{ TeV}}\right)^4\left(\frac{0.1}{|\lambda_{sQ}|}\right)^4\left(\frac{2\text{ GeV}}{F_\eta}\right)^2\left(\frac{2\text{ GeV}}{M_\eta}\right) \ ,  \\
& \tau(\sigma \rightarrow s\bar{s})\simeq  4\times 10^{-6}\text{ s}\left(\frac{M_X}{1\text{ TeV}}\right)^4\left(\frac{0.1}{|\lambda_{sQ}|}\right)^4\left(\frac{3\text{ GeV}}{F_\sigma}\right)^2\left(\frac{3\text{ GeV}}{M_\sigma}\right) \ ,
\end{aligned}
\end{equation}
which translate to flight distances of order 1 km. Therefore the dark hadrons in the $n_f=1$ scenario can also be treated as missing energy.

As most the dark hadrons have long lifetimes in most of the parameter space, they can be searched for in invisible or semi-visible decays of SM hadrons.  Below, we consider various such possibilities. 


\subsection{\boldmath{$B_{d(s)} \rightarrow \slashed{E}$} decays}
\label{pheno1}

\begin{figure}[hbpt]
 \centering
 \includegraphics[width=30mm,scale=0.5]{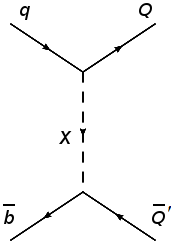}
 \caption{Leading diagram for the modes $B_q\rightarrow Q\bar{Q'}$, as well as the inclusive modes $b\rightarrow qQ\bar{Q'}$. $Q$ and $Q'$ are dark quarks, and $q$ is a $d$ or an $s$ SM quark. If we replace $b$ by $s$ and $q=d$, then the diagram represents the inclusive modes of $K\rightarrow \pi\pi_{\text{D}}$.}
 \label{bsQQbar}
\end{figure}

Our first candidate decay is the fully invisible decay of $B_d$ and $B_s$ mesons.
We will first study the scenario where the DQCD confinement scale is below the $B$ meson masses ($\Lambda_{\text{DQCD}}<M_B$) where we can treat final states in the dark sector in the quark picture ($B_{d(s)} \rightarrow Q\bar{Q'}$). The two quark final state is the dominant NP invisible final state since the four dark quark state is suppressed by $\alpha_{\text{D}}(M_B)$ in addition to a phase space factor. The dark quarks will then hadronize either to dark pions or to heavier dark hadrons which will then decay to dark pions. In the 1 DF scenario, the dark quarks will hadronize to $\eta$ or $\sigma$ mesons or to heavier dark hadrons which will then decay to $\eta$ or $\sigma$. There is also the possibility of production of dark baryons which are absolutely stable, but that does not change the picture.
The leading Feynman diagram for the decays ($B_{d(s)} \rightarrow Q\bar{Q'}$) is the tree-level one shown in Figure~\ref{bsQQbar} with $q\equiv d(s)$.
%
%
%

Searches for invisible $B_d$ decays done by Belle \cite{Belle} and BaBar \cite{BaBar} collaborations have not observed any signals, and the strongest limit is from BaBar, $\mathcal{B}(B_d \rightarrow \slashed{E})_{\text{exp}}<2.4\times 10^{-5}$ at $90\%$ CL  \cite{BaBar}. Belle II is expected to improve the upper limit to $1.5 \times 10^{-6}$ \cite{Belle2}.   
Experimental searches for invisible $B_s$ decays have not been done yet, and Belle II collaboration expects to reach an upper limit of order $10^{-5}$  for $\mathcal{B}(B_s \rightarrow \slashed{E})$  \cite{Belle2}. All of these are many orders of magnitude above the SM predictions~\cite{BadinPetrov, PetrovEtAl},
%
so the observation of  $(B_{d(s)} \rightarrow \slashed{E})$ is a direct manifestation of new physics.

We can estimate the branching ratios (BRs) of the decays of $B_{d(s)}$ to dark quark-antiquark pairs  from the tree diagram of Figure~\ref{bsQQbar}.  Since we can not experimentally discern the different dark flavors, all being invisible, our calculation of the BRs can be done by summing over all possible dark flavours (DFs):
\begin{equation}
 \label{BR-BtoQQbar1}
 \sum\limits_{Q,Q'}\mathcal{B}(B_q \rightarrow Q\bar{Q'}) = N_c\frac{\chi_{qb}}{128\pi M_X^4}\frac{F_{B_q}^2m_Q^2M_{B_q}}{\Gamma_{B_q}}\sqrt{1-\frac{4m_Q^2}{M_{B_q}^2}} \ ,
\end{equation}
where $q=b$ or $s$, and
\begin{equation}
\label{chi}
\chi_{qq'}=\sum\limits_{Q,Q'}|\lambda_{qQ}\lambda_{q'Q'}^\ast|^2 \ . 
\end{equation}
$F_{B_q}$ is the decay constant of the meson, $M_{B_q}$ is its mass, and $\Gamma_{B_q}$ is its total decay width. All the experimental input data we use are given in Table~\ref{tab:1}. The dark sector parameters are $N_c$, the number of dark colors, $m_Q$ the common mass of the dark quarks, and the combination of the Yukawa matrix elements parameterized by $\chi$.
For $m_Q\simeq 1$ GeV, we have
\begin{equation}
 \label{BR-BtoQQbar2}
 \begin{aligned}
 \sum\limits_{Q,Q'}\mathcal{B}(B_d \rightarrow Q\bar{Q'}) \simeq & \ 1.01\times 10^{-3}N_c\left(\frac{1\text{  TeV}}{ M_X}\right)^4 \chi_{db} \ , \\
 \sum\limits_{Q,Q'}\mathcal{B}(B_s \rightarrow Q\bar{Q'}) \simeq & \ 1.50\times 10^{-3}N_c\left(\frac{1\text{  TeV}}{ M_X}\right)^4 \chi_{sb} \ .
\end{aligned}
\end{equation}
Complementary bounds on the elements of the matrix $\lambda$ from $\Delta F = 2$ processes are computed in Section~\ref{Constraints} and summarised in Table~\ref{tab:2}. Those constraints are on the matrix element of the flavour mixing and thus scale as
\begin{equation}
    \frac{N_c}{M_X^2} \left|\sum\limits_Q\lambda_{qQ}\lambda_{q'Q}^\ast\right|^2\,.
\end{equation}
Therefore, the scaling with $N_c$ is the same for $\Delta F = 2$ and $\Delta F = 1$, and predictions for the rare meson decay rates that saturate the $\Delta F = 2$ bounds will be independent of $N_c$. On the other hand, the scaling with $M_X$ differs in the two classes of processes, and predictions for rare meson decays  will scale with $1/M_X^2$. 

For the dependence on the $\lambda_{ij}$ couplings, we see that for $\Delta F = 1$ we must square then sum on two indices over dark flavours, while for $\Delta F = 2$ we sum only on one index, then square. These are identical with one dark flavour, but could be significantly different with more flavours. In particular, there could be destructive interference that makes the $\Delta F = 2$ constraints weak and allows for relatively large rates in the processes considered in this section. 
 
\paragraph{Case of 1 DF:} In this scenario, the dependence on the SM-DM couplings for $\Delta F = 2$ and $\Delta F = 1$ processes are the same. Thus we directly apply the second and third lines of Table~\ref{tab:2} on the $\chi_{qb}$ factors of equation~(\ref{BR-BtoQQbar2}), and get upper bounds on the BRs. For $m_Q\simeq 1$ GeV, we obtain
\begin{equation}
 \label{BR-BtoQQbar1DF}
 \begin{aligned}
 \mathcal{B}(B_d \rightarrow Q\bar{Q}) & \lesssim 6.62\times 10^{-7}\left(\frac{1\text{ TeV}}{M_X} \right)^2 \ , \\
 \mathcal{B}(B_s \rightarrow Q\bar{Q}) & \lesssim 1.97\times 10^{-5}\left(\frac{1\text{ TeV}}{M_X} \right)^2 \ .
 \end{aligned}
\end{equation}
Both $B_d$ and $B_s$ decays to invisible are potentially within the reach of the Belle II experiment if $M_X$ is somewhat less than 1 TeV.

\paragraph{Case of 2 DFs:}
We will consider a few specific slices of parameter space in the two dark flavour scenario. One interesting case described at the end of Section~\ref{sec:2DF}, where $d_1=d_2$ and $c_{13}=0$, has no flavour constraints. 
In this case, $\chi_{db}=0$\footnote{We do not need $d_1=d_2$ to have $\chi_{db}=0$, we just need $c_{13}=0$. Adding $d_1=d_2$ guarantees the absence of constraints.} , but $\chi_{sb}\neq 0$, so we can have non-zero contributions to $\mathcal{B}(B_s \rightarrow Q\bar{Q})$. 
\begin{equation}
\begin{aligned}
 \label{BR-BtoQQbar2DFspecial1}
 \sum\limits_{Q,Q'}\mathcal{B}(B_d \rightarrow Q\bar{Q'})_{(c_{13}=0)} = & \ 0  \ , \\
 \sum\limits_{Q,Q'}\mathcal{B}(B_s \rightarrow Q\bar{Q'})_{(d_1=d_2,\ c_{13}=0)} \simeq & \ 1.50\times 10^{-3}N_c\left(\frac{1\text{  TeV}}{ M_X}\right)^4 \chi_{sb}^{\text{\tiny{Generic}}} \ ,
\end{aligned}
\end{equation}
where $\chi_{sb}$ is defined in equation~\ref{chi}, and ``Generic'' means there are no constraints on it from $\Delta F=2$ processes described in section \ref{Constraints}. In this case, an experimental measurement of the $B_s$ invisible decay would be a measurement of  $N_c \chi_{sb}/M_X^4$. 

Another case with no flavor constraints is when $d_1=d_2$, $s_{13}=0=s_{23}$. This latter (even if $d_1\neq d_2$) gives $\chi_{db}=0=\chi_{sb}$, thus,
\begin{equation}
 \label{BR-BtoQQbar2DFspecial2}
 \sum\limits_{Q,Q'}\mathcal{B}(B_{d(s)} \rightarrow Q\bar{Q'})_{(s_{13}=0=s_{23})} = 0  \ .
\end{equation}
An observation of the $B_d$ invisible decay would impose $c_{13}\neq 0$; and an observation of only the $B_s$ invisible decay would impose $c_{13}=0$ and $s_{13}\neq 0$ or $s_{23}\neq 0$, or the number of DFs is bigger than 2.

If we keep $d_1=d_2$ and now take $s_{13}=0$, then only constraints from $B_s$ apply ($\xi_K=0=\xi_{B_d}$), and we have $\chi_{db}=|\xi_{B_s}|^2/c_{23}^2$ and $\chi_{sb}=|\xi_{B_s}|^2$. 
Applying the third line of Table~\ref{tab:2}, the BRs are then, for $m_Q\simeq 1$ GeV:
\begin{equation}
 \label{BR-BtoQQbar2DF2}
 \begin{aligned}
 \sum\limits_Q\mathcal{B}(B_d \rightarrow Q\bar{Q})_{(d_1=d_2,\ s_{13}=0)} & \lesssim \frac{1.33\times 10^{-5}}{c_{23}^2}\left(\frac{1\text{  TeV}}{ M_X}\right)^2 \ , \\
 \sum\limits_Q\mathcal{B}(B_s \rightarrow Q\bar{Q})_{(d_1=d_2,\ s_{13}=0)} & \lesssim 1.97\times 10^{-5}\left(\frac{1\text{  TeV}}{ M_X}\right)^2 \ .
\end{aligned}
\end{equation}
These bounds  get weaker with decreasing $M_X$, and the first bound could increase considerably if $s_{23}$ approaches $1$. This scenario is also potentially within reach at Belle II.

%
\paragraph{Case of 3 or more DFs:}  If the number of DFs is $n_f\geq 3$, we can, like in the 2 DF case,\footnote{For the 3 DF case, we cannot have $c_{13}=0$, since $0\leq \theta_{ij}\leq \pi /4$.} escape the flavor constraints when the matrix $D$ is $12-$degenerate ($\Delta_{12}=0$) and $s_{13}=0=s_{23}$, giving $\chi_{db}=\chi_{sb}\neq 0$. 
%
\begin{equation}
 \label{BR-BtoQQbar3DF}
 \begin{aligned}
 & \sum\limits_Q\mathcal{B}(B_d \rightarrow Q\bar{Q})_{(12-\text{deg},\ s_{13}=0=s_{23})} \simeq \ 1.01\times 10^{-3}N_c\left(\frac{1\text{  TeV}}{ M_X}\right)^2 \chi_{db}^{\text{\tiny{Generic}}} \ , \\
 & \sum\limits_Q\mathcal{B}(B_s \rightarrow Q\bar{Q})_{(12-\text{deg},\ s_{13}=0=s_{23})} \simeq \ 1.50\times 10^{-3}N_c\left(\frac{1\text{  TeV}}{ M_X}\right)^2 \chi_{sb}^{\text{\tiny{Generic}}} \ , \\
\end{aligned}
\end{equation}
%
Here $\chi_{db}^{\rm Generic}$ is uncostrained from $\Delta F=2$ constraints but it is constrained by the BaBar constraint of $\mathcal{B}(B_d \rightarrow \slashed{E})_{\text{exp}}<2.4\times 10^{-5}$~\cite{BaBar}.
In general, a measurement of the $B_d$ ($B_s$) invisible decay would induce a measurement of the parameter $\chi_{db}$ ($\chi_{sb}$) for fixed $N_c$ and $M_X$. If these latter are found equal, then we can conclude $s_{13}=0=s_{23}$ and $\lambda_1=\lambda_2$.
%
 
We now turn to the scenario where the DQCD scale is above the $B$ meson masses ($\Lambda_{\text{DQCD}}\geq M_B$). In this case, dark pions are the only kinematically allowed final states in $B_q$ decays, since they are PNGBs. It should be noted that such decays are kinematically forbidden in the 1 DF case, since the lightest dark hadrons have mass $\simeq \Lambda_\text{DQCD}$. For 2 or more DF scenarios, the dominant invisible modes are the decays of $B_q$ to 2 dark pions (not necessarily flavor diagonal). Decays to more than 2 dark pions will be phase space suppressed. At the dark quark level, the decay $(B_q\rightarrow \pi_{\text{D}}\pi_{\text{D}})$ corresponds to $(B\rightarrow Q\bar{Q'})$ studied above, and since the decay rate of the inclusive mode is always bigger than the rate of the corresponding exclusive mode, then the upper bounds of equations~(\ref{BR-BtoQQbar2DFspecial1}-\ref{BR-BtoQQbar3DF}) still apply to $(B_q\rightarrow \pi_{\text{D}}\pi_{\text{D}})$.
\paragraph{\boldmath{$CP-$}Asymmetries in \boldmath{$B_{d(s)} \rightarrow \slashed{E}$} decays:}

We will now study the time-dependent mixing induced $CP-$asymmetries (defined in Appendix~\ref{sec:CPasym}) in invisible $B$ decays. Although such asymmetries are currently only measurable in fully visible decays of the $B$ mesons, we hope that, in the future, there will be experiments that have the capability to measure the asymmetries in the kinds of invisible decays discussed here. 

We will focus on the scenario ($\Lambda_{\text{DQCD}}<M_B$) and 1 DF. So we restrict ourselves to the decays ($B_q \rightarrow Q\bar{Q}$) in which the product is a $CP-$odd eigenstate.\footnote{A bound state of a particle $A$ and its antiparticle $\bar{A}$ is a $CP-$eigenstate. The CP-parity of the system, whether the particle $A$ is a boson or a fermion, is: $(P_A) (P_{\bar{A}}) (-1)^S=(-1)^{S+1}$, where $P_A \ (P_{\bar{A}})$ is the parity of $A$ ($\bar{A}$) and $S$ is the total spin of the system. Since $B_q$ has $J=0$, then the system $Q\bar{Q}$ must have either $(S=0 \ \text{and}\  L=0)$ or (S=1\ \text{and}\ L=1). The first case,  being the ground state, is dominant  over the second one (an excited state); thus the $CP-$parity of $Q\bar{Q}$ is $\zeta_{Q\bar{Q}}=-1$.} The weak phase of decay from the dominant tree diagram of Figure~\ref{bsQQbar} is, $\phi_D=\text{Arg}(\lambda_{qQ}\lambda_{bQ}^\ast)$ which we call $\delta_{qbQ} \ (q=d \ \text{or} \ s)$. Therefore, we are in the special case where the $CP-$Asymmetry described in Appendix~\ref{sec:CPasym} is ``clean'' from hadronic uncertainties and takes the simple form of equation~(\ref{CPasym3}), that is,
\begin{equation}
\label{mixCPasym}
\mathcal{A}_{CP}(t)= \zeta_f \sin(2\phi_D-2\phi_M)\sin (\Delta M_qt) 
\ .   
\end{equation}
$\zeta_f$ is the $CP-$parity of the decay product, and $\Delta M_q$ is the mass difference between the two $B_q$ mass eigenstates. $\phi_M$ is the weak phase of the $B-\bar{B}$ mixing, which is dominated by the SM box diagram with an internal $t$ quark. The NP contribution to this phase is negligible from what we saw in Section~\ref{Constraints}, so we can write: $\phi_M = \text{Arg} (V_{tb}^\ast V_{td}) \simeq \text{Arg}(V_{td})= -\beta$, for $B_d$, and $\phi_M = \text{Arg} (V_{tb}^\ast V_{ts}) \simeq \text{Arg}(V_{ts})= -\beta_s$, for $B_s$.\footnote{$V_{ij}$ are elements of the CKM matrix. More details about $\beta$ and $\beta_s$  can be found in Appendices~\ref{sec:A} and~\ref{sec:CPasym}. Their experimental values, as well as that of $\Delta M_d$ and $\Delta M_s$ are shown in Table~\ref{tab:1}.}. Therefore one can measure $\delta_{dbQ}$ and $\delta_{sbQ}$ ($\text{Arg}(\lambda_{dQ}\lambda_{bQ}^\ast)$ and $\text{Arg}(\lambda_{sQ}\lambda_{bQ}^\ast)$)\footnote{Note that while the CPV phases $\delta_{ij}$ are convention dependent, $\delta_{qq'Q}=\text{Arg}(\lambda_{qQ}\lambda_{q'Q}^\ast)$, the arguments of the Wilson coefficients, are not, in the 1 DF case.} via the asymmetries: 
\begin{equation}
\begin{aligned}
\mathcal{A}_{CP}(B_q \rightarrow Q\bar{Q}) 
& = \frac{\Gamma (B_q^0(t)\rightarrow Q\bar{Q})-\Gamma (\bar{B}_q^0(t)\rightarrow Q\bar{Q})}{\Gamma (B_q^0(t)\rightarrow Q\bar{Q})+\Gamma (\bar{B}_q^0(t)\rightarrow Q\bar{Q})} \ , \\
& = - \sin \left[2( \beta_q+\delta_{qbQ})\right]\sin (\Delta M_q t) 
\ .
\end{aligned}
\label{AsymBQQBar}
\end{equation}
The value of the time $t$ is the duration between the creation of a $B_q^0-\bar{B}_q^0$ pair and the disappearance of either $B_q^0(t)$ or $\bar{B}_q^0(t)$ into missing energy.  Reconstructing the time in invisible decays is the biggest experimental challenge. The phases $\delta_{dbQ}$ and $\delta_{sbQ}$ coincide respectively with $(-\delta')$ and $(\delta-\delta')$ if we use the parameterization of equation~(\ref{Deflambda1DF}). 

 If there are $n_f\geq 2$ dark flavors and still $M_B$ is above $\Lambda_{DQCD}$, but the $\lambda$ matrix is hierarchical such that only one tree diagram involving one dark flavor ($Q$ for $q=d$ and $Q'$ for $q=s$) is dominant over the others then the situation can be mapped onto the $n_f=1$ case. We then have 
\begin{equation}
 \label{chiSp}
 \begin{aligned}
 &\chi_{db}=\sum\limits_{Q,Q'}|\lambda_{dQ}\lambda_{bQ'}^\ast|^2 \simeq |\lambda_{dQ}\lambda_{bQ}^\ast|^2 \ , \\
 &\chi_{sb}=\sum\limits_{Q,Q'}|\lambda_{sQ}\lambda_{bQ'}^\ast|^2 \simeq |\lambda_{sQ'}\lambda_{bQ'}^\ast|^2 \ .
 \end{aligned}
 \end{equation}
Therefore, there will be the same asymmetry as equation~(\ref{AsymBQQBar}).
If $\Lambda_{DQCD} > M_B > 2M_D $, then the decay is $B_{d(s)} \rightarrow \pi^{(')}_{\text{D}}\pi^{(')}_{\text{D}}$ which has the same dominant mode (considering equation~(\ref{chiSp})) $B_{d(s)} \rightarrow Q^{(')} \bar{Q}^{(')}$ at the quark level and again gives the same asymmetry as equation~(\ref{AsymBQQBar}).

\subsection{\boldmath{$ B_d \rightarrow \pi^0(K_{\text{S}}) \slashed{E}$}}
\label{pheno2}
In this section, we consider the scenario $\Lambda_{\text{DQCD}}\geq M_B$ and a number of DFs $n_f\geq 2$ such that the only kinematically allowed dark final states are dark pions. We have calculated the BRs of $( B_d\rightarrow \pi^0 \pi_{\text{D}})$ and $( B_d\rightarrow K_{\text{S}} \pi_{\text{D}})$ in DQCD, where $\pi_{\text{D}}$ is not necessarily flavor diagonal ($\pi_{\text{D}}\equiv Q\bar{Q'}$). Using the factorization approximation \cite{Factorization} on the tree diagram of Figure~\ref{bsQQbar}, we find, after summing over all dark flavors:
\begin{equation}
\begin{aligned}
\label{BRBtoPiOrKsPiD1}
\sum\limits_{Q,Q'}\mathcal{B}( B_d \rightarrow \pi^0(K_{\text{S}})\pi_{\text{D}}) \approx  & \  N_c\frac{\chi_{d(s)b}}{2048\pi M_X^4} 
\frac{[F_0^{B\pi^+(K^0)}(M_D^2)]^2\,F_D^2}{\Gamma_B}
\frac{(M_B^2-M_{\pi (K)}^2)^2}{M_B^3} \times \\
& \times \sqrt{\left[(M_B+M_{\pi (K)})^2-M_D^2\right]\left[(M_B-M_{\pi (K)})^2-M_D^2\right]} \ ,
\end{aligned}
\end{equation}
where $\chi_{d(s)b}$ is defined in equation~\eqref{chi}, $M_{\pi}$ ( $M_K$) is the mass of the $\pi^0$ ($K_S$)~\cite{PDG}, $M_D$ is the mass of the dark pion, and $F_D$ is its decay constant. $F_0^{B\pi^+(K^0)}(q^2)$\footnote{$q$ is the  momentum exchange between $B$ and $\pi^0$ or $K$. Hence, $q^2=M_D^2$.} is a form factor of the hadronic matrix element $\bra{\pi^+(K^0)}\bar{b}\gamma^{\mu}d\ket{B^0}$ \cite{AliEtAl,FormFactors}, which can be calculated by using lattice QCD. From \cite{RBCandUKQCD,FermilabLatticeAndMILC}, we can find that for $q^2\lesssim 10 \ \text{GeV}^2$ (which corresponds to $M_D\lesssim 3$ GeV), we have $F_0^{B\pi^+}(q^2)\approx 0.2$; and from \cite{FermilabLatticeAndMILC2}, we can find that $F_0^{BK^0}(q^2\lesssim 10 \ \text{GeV}^2)\approx 0.3$. We recall that $N_c$ is the number of dark colors. For $F_D\approx M_D\approx 1$ GeV, we have
\begin{equation}
 \label{BR-BtoSPiD}
 \begin{aligned}
 \sum\limits_{Q,Q'}\mathcal{B}( B_d \rightarrow \pi^0\pi_{\text{D}}) \simeq & \ 2.03\times 10^{-3}N_c\left(\frac{F_D}{1\text{  GeV}}\right)^2\left(\frac{1\text{  TeV}}{ M_X}\right)^4 \chi_{db} \ , \\
 \sum\limits_{Q,Q'}\mathcal{B}( B_d \rightarrow K_{\text{S}}\pi_{\text{D}}) \simeq & \ 4.46\times 10^{-3}N_c\left(\frac{F_D}{1\text{  GeV}}\right)^2\left(\frac{1\text{  TeV}}{ M_X}\right)^4 \chi_{sb} \ .
\end{aligned}
\end{equation}

The most recent experimental limits have been established by Belle collaboration \cite{Belle2017}: $\mathcal{B}(B_d \rightarrow \pi^0\nu\bar{\nu})_{\text{exp}}< 0.9\times 10^{-5}$, and $\mathcal{B}(B_d \rightarrow K_{\text{S}}\nu\bar{\nu})_{\text{exp
}}< 1.3\times 10^{-5}$, both at $90\%$ CL, so for those parameters, $\chi_{sb}=1$ and $\chi_{db}=1$ are excluded.
The corresponding SM BRs are $\mathcal{B}(B_d \rightarrow \pi^0\nu\bar{\nu})_{\text{SM}}=(1.2\pm 0.15)\times 10^{-7}$ \cite{HambrockEtAL} and $\mathcal{B}(B_d \rightarrow K_{\text{S}}\nu\bar{\nu})_{\text{SM}}\simeq (2.00\pm 0.25)\times 10^{-6}$~\cite{BurasEtAl},\footnote{This reference gives $\mathcal{B}(B^+ \rightarrow K^+\nu\bar{\nu})_{\text{SM}}\simeq (4.0\pm 0.5)\times 10^{-6}$. Because $\tau_{B^+}\simeq \tau_{B^0}$, isospin implies 
$\mathcal{B}(B^0 \to K^0\nu\bar{\nu})_{\text{SM
}} \simeq \mathcal{B}(B^+ \to K^+\nu\bar{\nu})_{\text{SM}}$. 
Then, using the fact that 
$K_{\text{S}}\propto K^0/\sqrt{2}$ 
we get the quoted value.}
so the SM process with two neutrinos in the final state could have a comparable rate to the one with dark hadrons in the final state, as opposed to the processes considered in Section~\ref{pheno1} where there is no SM background.
As far as we know, there is no projection of the Belle II sensitivity for these modes. Belle II is expected to take roughly 100 times as much data as Belle, and the limits are expected to scale linearly with luminosity. Therefore, we can perform a naive luminosity scaling from the Belle result to estimate that Belle II should be sensitive to a branching ratio $\mathcal{O}(10^{-7})$ for $B_d \rightarrow \pi^0 \nu\bar{\nu}$. For $(B_d \rightarrow K_{\text{S}}\nu\bar{\nu})$, a luminosity extrapolation implies Belle II should be able to collect $\mathcal{O}(10)$ events if the rate is equal to the SM expectation, which implies a $\sim 30$\% precision measurement of the branching ratio.

The NP final state differs from that of the SM as it is two body, so the SM hadron ($\pi^0/K_{\text{S}}$) energy will be mono-chromatic and given by
\begin{equation}
E_{\pi(K)} = \frac{M_B^2 + M_{\pi(K)}^2-M_D^2}{2 M_B}  \, ,
\label{Efinal}
\end{equation}
in the rest frame of the decaying $B$.
This feature would be a smoking gun for new physics, and could also allow experiments to distinguish the dark QCD signal from the SM background. Finally the location of the peak in the $E_{\pi(K)}$ distribution in the $B$ rest frame would provide a measurement of the dark pion mass $M_D$. 

We can now explore specific parameterizations for $\chi_{qb}$.
If there are 2 DFs with $c_{13}=0$, then the sum over DFs of $\Gamma ( B_d \rightarrow \pi^0\pi_{\text{D}})$ is zero. If in addition, $d_1=d_2$, then [see equation~\eqref{xi2DFdeg}] there are no $\Delta F=2$ flavor constraints on $B_d \rightarrow K_{\text{S}}\pi_{\text{D}}$. Finally, the sum over the DFs of the rates of both modes are zero if $s_{13}=0=s_{23}$ (because $\chi_{db}=0=\chi_{sb}$ in this case).
For 3 DFs with a $12-$degeneracy and $s_{13}=0=s_{23}$ [see equation~\eqref{xi3DF12deg}], the $\Delta F=2$ flavor constraints disappear, and $\chi_{db}=\chi_{sb}$. Therefore $\chi_{d(s)b}$ are only constrained by the existing Belle upper limit of this process. Thus, if semi-invisible $B_d$ decays are measured, and $\chi_{db}$ and $\chi_{sb}$ are found to be equal, then this means that $\lambda_1=\lambda_2$ and $s_{13}=0=s_{23}$.

If we are not in these special scenarios, $\chi_{d(s)b}$ will be affected by the $\Delta F=2$ flavor constraints on the $\xi_M$ terms. Although the bounds on the BRs from these constraints are not obvious, we estimate that $\chi_{d(s)b}$ will have upper bounds of order~$\sim 10^{-2}$ at the most (the highest upper bound in Table~\ref{tab:2}). 

The following  equations summarize our results.

\begin{equation}
\begin{aligned}
\label{BR-BtoSPiD2}
& \sum\limits_{Q,Q'}\mathcal{B}( B_d \rightarrow \pi^0(K_{\text{S}})\pi_{\text{D}}) \lesssim \  \mathcal{O}(10^{-5})\left(\frac{F_D}{1\text{  GeV}}\right)^2\left(\frac{1\text{  TeV}}{ M_X}\right)^2 \ , \\
&\bigg[\sum\limits_{Q,Q'}\mathcal{B}( B_d \rightarrow \pi^0\pi_{\text{D}})\bigg]_{(\text{2DF, }c_{13}=0)} =  0 \ , \\
&\bigg[\sum\limits_{Q,Q'}\mathcal{B}( B_d \rightarrow K_{\text{S}}\pi_{\text{D}})\bigg]_{(\text{2DF, }d_1=d_2,\ c_{13}=0)} \\
& \hspace{3.7cm} \simeq \ 4.46\times 10^{-3}N_c\left(\frac{F_D}{1\text{  GeV}}\right)^2\left(\frac{1\text{  TeV}}{ M_X}\right)^4 \chi_{sb}^{\text{\tiny{Generic}}} \ , \\
&\bigg[\sum\limits_{Q,Q'}\mathcal{B}( B_d \rightarrow \pi^0(K_{\text{S}})\pi_{\text{D}})\bigg]_{(2\text{DF},\ s_{13}=0=s_{23})}= 0 \ , \\
&\bigg[\sum\limits_{Q,Q'}\mathcal{B}( B_d \rightarrow \pi^0\pi_{\text{D}})\bigg]_{(3\text{DF, }12-\text{deg, }s_{13}=0=s_{23})} \\
& \hspace{3.7cm} = \ 2.03\times 10^{-3}N_c\left(\frac{F_D}{1\text{  GeV}}\right)^2\left(\frac{1\text{  TeV}}{ M_X}\right)^4 \chi_{db}^{\text{\tiny{Generic}}} \ , \\
&\bigg[\sum\limits_{Q,Q'}\mathcal{B}( B_d \rightarrow K_{\text{S}}\pi_{\text{D}})\bigg]_{(3\text{DF, }12-\text{deg, }s_{13}=0=s_{23})} \\
& \hspace{3.7cm} = \ 4.46\times 10^{-3}N_c\left(\frac{F_D}{1\text{  GeV}}\right)^2\left(\frac{1\text{  TeV}}{ M_X}\right)^4 \chi_{sb}^{\text{\tiny{Generic}}} \ , \\
&\big[3\text{DF, }12-\text{deg, }s_{13}=0=s_{23})\big] \Rightarrow    \chi_{sb}^{\text{\tiny{Generic}}}=\chi_{db}^{\text{\tiny{Generic}}}=(\lambda_0+\lambda_1)^2(\lambda_0-2\lambda_1)^2 \ ,
\end{aligned}
\end{equation}
where $\chi^{\text{\tiny{Generic}}}$ means there are no bounds from $\Delta F = 2$ process, but it is still constrainted by the Belle limit on $B\rightarrow \pi^0 (K_{\text{S}})\nu\nu$.

\paragraph{\boldmath{$CP-$}Asymmetries in \boldmath{$ B_d \rightarrow \pi^0(K_{\text{S}})\slashed{E}$} decays:}
Continuing to the scenario $\Lambda_{\text{DQCD}}\geq M_B$ and $n_f\geq 2$, we now also restrict ourselves to the case where equation~(\ref{chiSp}) applies, i.e.~there is a hierarchy in the matrix $\lambda$ that makes only one tree diagram dominant. Of course, the dark pions for $( B_d \rightarrow \pi^0\pi_{\text{D}})$ and $( B_d \rightarrow K_{\text{S}}\pi_{\text{D}})$ can be different. Let us call them $\pi_{\text{D}}$ and $\pi_{\text{D}}'$ and their constituent dark quarks $Q$ and $Q'$ respectively.
The dark pions are PNGBs, so they are light pseudoscalars, which means they have an odd parity and spin zero. Because all the dark pions are degenerate, we can work in a basis of real fields where all pion states are their own antiparticles. Therefore, they are $CP-$eigenstates with \textit{the same}  $CP-$parity: $\zeta_{\pi_{\text{D}}}=(P_Q) (P_{\bar{Q}})(-1)^S=(-1)^{S+1}=-1$. We know that $\pi^0 \ (K_{\text{S}})$ is $CP-$odd (even),\footnote{We can approximately consider the meson $K_{\text{S}}$ as a $CP-$even eigenstate. This is a good approximation, where we neglect $\epsilon_K$ which is of order $10^{-3}$.}  therefore, the final state of the decay ($ B_d \rightarrow  \pi^0(K_{\text{S}})\pi_{\text{D}}$) is a $CP-$even (odd) eigenstate.\footnote{Since the final total spin is $S=0$, then the $CP-$parity of the 2 body final state is simply the product of the $CP-$parities of the individual (dark) mesons.}

\begin{figure}[hbpt]
 \centering
 \includegraphics[width=150mm,scale=1.0]{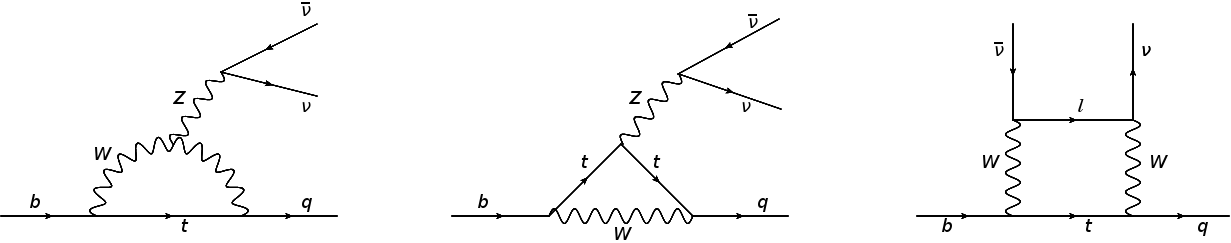}
 \caption{Dominant diagrams, in the SM, for the modes $ B_d\rightarrow \pi^0(K_{\text{S}})\nu\bar{\nu}$ and $B^+\rightarrow \pi^+(K^+)\nu\bar{\nu}$. $q$ is a $d$ ($s$) SM quark. The same diagrams hold for $K\rightarrow \pi\nu\bar{\nu}$ if we replace the $b$ quark by an $s$ quark, and $q\equiv d$.}
 \label{SMb-to-q}
\end{figure}

Assuming that we can experimentally discern the difference between the NP and SM decay products by using the fact that the NP (SM)'s decay is a two (three) body final state and exploiting equation~(\ref{Efinal}), we can separately consider the $CP-$asymmetries in NP and in the SM.
For the mixing induced $CP-$asymmetries in $( B_d \rightarrow \pi^0 (K_{\text{S}}) \nu\nu)$ in the SM, the decays are dominated by two types of diagrams at the quark level \cite{Buras2005,BurasRareDec}, the $Z-$mediated penguins, and the box diagram, all with an internal top quark, as shown in Figure~\ref{SMb-to-q}, with $q\equiv d(s)$.
%
 We notice that there is only one weak phase of decay in each of the modes $( B_d\rightarrow \pi^0(K_{\text{S}})\nu\bar{\nu})$, that is, $\phi_D=\text{Arg}(V_{tb}V_{td(s)}^\ast)\simeq \text{Arg}(V_{td(s)}^\ast)=\beta_{(s)}$, and the weak phase of mixing is $\phi_M=-\beta$. In addition, the decay product is a $CP-$even (odd) eigenstate, so the conditions to get the simple  form of equation~(\ref{mixCPasym}) apply.\footnote{There is a factor of 3 in the decay width from the sum over the neutrino flavors that cancels in the asymmetry.}
\begin{equation}
\begin{aligned}
\label{AsymSM}
\mathcal{A}_{CP}( B_d \rightarrow \pi^0 \nu\bar{\nu})_{\text{SM}} & = + \sin \left(4 \beta\right)\sin (\Delta M_dt) \ , \\
\mathcal{A}_{CP}( B_d \rightarrow K_{\text{S}} \nu\bar{\nu})_{\text{SM}} & = - \sin \left[2( \beta+\beta_s)\right]\sin (\Delta M_dt) \ .
\end{aligned}
\end{equation}
In DQCD,  the inclusive modes  of ($ B_d \rightarrow  \pi^0(K_{\text{S}})\pi_{\text{D}^{(')}}$) are $(b\rightarrow d(s) Q^{(')}\bar{Q}^{(')})$, where we again have a unique phase of decay. Thus the asymmetries can take the simple form equation~(\ref{mixCPasym}), where $\phi_D=\text{Arg}(\lambda_{qQ^{(')}}\lambda_{bQ^{(')}}^\ast)\equiv \delta_{qbQ^{(')}} \ (q=d \ \text{or} \ s)$. The weak phase of mixing is still $\phi_M=-\beta$ for both decays. Thus we have:
\begin{equation}
\begin{aligned}
\mathcal{A}_{CP}( B_d \rightarrow \pi^0\pi_{\text{D}})
& = + \sin \left[2( \beta+\delta_{dbQ})\right]\sin (\Delta M_dt) \ , \\
\mathcal{A}_{CP}( B_d \rightarrow K_{\text{S}}\pi_{\text{D}})
& = - \sin \left[2( \beta+\delta_{sbQ'})\right]\sin (\Delta M_dt) \ .
\end{aligned}
\label{AsymBPiOrKsPiD}
\end{equation}
Measuring such an asymmetry will enable us to measure the NP's CPV phases $\delta_{dbQ}$ and $\delta_{sbQ'}$. As in the case of fully invisible decays, time dependant asymmetries in semi-visible decays have not yet been shown to be feasible, but we hope it will be possible with future detectors.


\subsection{\boldmath{$CP-$}Asymmetry in \boldmath{$B^+\rightarrow \pi^+(K^+)\slashed{E}$}}
\label{pheno3}
We consider again the scenario ($\Lambda_{\text{DQCD}}\geq M_B$) and a number of dark flavors $n_f\geq 2$. From isospin symmetry, the rate of the decay ($ B_d\rightarrow \pi^0(K_{\text{S}})\pi_{\text{D}}$)  is half the rate of ($B^+\rightarrow \pi^+(K^+)\pi_{\text{D}}$). Dividing by the total widths, we find
\begin{equation}
\label{BRchargedB}
\mathcal{B}(B^+\rightarrow \pi^+(K^+)\pi_{\text{D}})\simeq (2.16)\mathcal{B}(B_d\rightarrow \pi^0(K_{\text{S}})\pi_{\text{D}}) \ ,
\end{equation}
so we can quickly deduce the BR of the charged process from equations~(\ref{BR-BtoSPiD}) and (\ref{BR-BtoSPiD2}). 
The most recent experimental limits are \cite{Belle2017}: $\mathcal{B}(B^+ \rightarrow \pi^+\nu\bar{\nu})_{\text{exp}}< 1.4\times 10^{-5}$, and $\mathcal{B}(B^+ \rightarrow K^+\nu\bar{\nu})_{\text{exp
}}< 1.9\times 10^{-5}$, both at $90\%$ CL.\footnote{For a more recent but weaker search, see~\cite{Abudinen:2021emt}.} The corresponding SM BRs are $\mathcal{B}(B^+ \rightarrow \pi^+\nu\bar{\nu})_{\text{SM}}=(2.39\pm 0.30)\times 10^{-7}$ \cite{HambrockEtAL} and $\mathcal{B}(B^+ \rightarrow K^+\nu\bar{\nu})_{\text{SM}}= (4.0\pm 0.5)\times 10^{-6}$ \cite{BurasEtAl}, so they could constitute a background to the NP channels that must be distinguished. 
We can again estimate the reach at Belle II using naive luminosity scaling and find a sensitivity of $\mathcal{O}(10^{-7})$ for $B^+\rightarrow \pi^+\nu\nu$. For $B^+\rightarrow K^+\nu\nu$, Belle II should be able to make a roughly 20\% measurement of this mode if the rate is equal to its SM value. This means that for a large swath of parameter space, the modes ($ B_d\rightarrow \pi^0(K_{\text{S}})\pi_{\text{D}}$) are within the reach  of Belle II. 

For charged mesons, mixing effects are absent and we instead have direct $CP-$violation (DCPV) which has the advantage of being time independent. This type of asymmetry has the disadvantage of being dependent on strong phases and hadronic matrix elements,  as we will see below. For the decays ($B^\pm\rightarrow f^\pm$), where $f^\pm\equiv \pi^\pm\slashed{E} \text{ or } K^\pm\slashed{E}$, the asymmetry for DCPV is defined by 
\begin{equation}
\label{DCPVasym1}
\mathcal{A}_{\text{DCPV}}=\frac{\Gamma (B^+\rightarrow f^+)-\Gamma (B^-\rightarrow f^-)}{ \Gamma (B^+\rightarrow f^+)+\Gamma (B^-\rightarrow f^-)} \ .
\end{equation}
In the case where the amplitude of the decay ($B^+\rightarrow f^+$) is dominated by two diagrams with distinct strong phases ($\delta_1$ and $\delta_2$)  and distinct weak phases ($\phi_1$ and $\phi_2$) , that is,
\begin{equation}
\label{AmpBcharged}
A(B^\pm\rightarrow f^\pm)=A_1e^{i(\delta_1\pm\phi_1)}+A_2e^{i(\delta_2\pm\phi_2)}\ ,
\end{equation}
where $A_i$ are real, the asymmetry equation~(\ref{DCPVasym1}) takes the form \cite{Buras2005}
\begin{equation}
\label{DCPVasym2}
\mathcal{A}_{\text{DCPV}}=\frac{-2A_1A_2\sin{(\delta_1-\delta_2)}\sin{(\phi_1-\phi_2)}}{A_1^2+A_2^2+2A_1A_2\cos{(\delta_1-\delta_2)}\cos{(\phi_1-\phi_2)}} \ .
\end{equation}
We now see how this asymmetry depends on the strong phases, in addition to its dependence on hadronic  uncertainties included in $A_i$. This renders the measurement of the CPV phases $\phi_i$ complicated. 

In the SM, the decays ($B^+\rightarrow \pi^+(K^+)\nu\bar{\nu}$) are dominated by the diagrams of Figure~\ref{SMb-to-q}, where all the weak phases are equal. Hence,
\begin{equation}
\label{DCPVasymSM}
\mathcal{A}_{\text{DCPV}}^{\text{SM}}(B^+\rightarrow \pi^+(K^+)\nu\bar{\nu})\simeq 0 \ ,
\end{equation}
so that a measurement of an asymmetry $\mathcal{A}_{\text{DCPV}}(B^+\rightarrow \pi^+(K^+)\slashed{E})$ is a direct manifestation of new physics.
%
%

In DQCD, the decays $(B^+\rightarrow \pi^+(K^+)\pi_{\text{D}})$ are dominated by the tree diagrams of Figure~\ref{bsQQbar}. As there is a $SU(n_f)$ flavour symmetry, the dark pion states will form a multiplet which is a representation of that symmetry. 
Let us assume that the dominant modes produce a dark pion that is a Cartan generator of the $SU(n_f)$ group. For example, the $T_3$ dark  can be written as follows analogous to the  $\pi^0$ in the SM:
\begin{equation} 
\ket{\pi_{\text{D}}}=\frac{1}{\sqrt{2}}[\ket{Q_1 \bar{Q}_1}-\ket{Q_2 \bar{Q}_2}] \ .
\end{equation} 
Therefore, two tree diagrams dominate those specific decay modes, one replacing $Q$ and $\bar{Q'}$ of Figure~\ref{bsQQbar} with $Q_1$ and $\bar{Q}_1$ respectively, and the other replacing them with $Q_2$ and $\bar{Q}_2$. Thus we can have relevant tree-tree interference,  and we can apply equation~(\ref{DCPVasym2}),\footnote{We just need to change the sign of $A_2$.} where $\phi_1$ and $\phi_2$ are the weak phases of the two interfering trees. We have:
\begin{equation} 
\begin{aligned}
& \{\phi_1-\phi_2\}_{\pi^+\pi_{\text{D}}}=\delta_{dbQ_1}-\delta_{dbQ_2}\equiv \text{Arg}[\lambda_{dQ_1}\lambda_{bQ_1}^\ast]-\text{Arg}[\lambda_{dQ_2}\lambda_{bQ_2}^\ast] \ , \\
& \{\phi_1-\phi_2\}_{K^+\pi_{\text{D}}}=\delta_{sbQ_1}-\delta_{sbQ_2}\equiv \text{Arg}[\lambda_{sQ_1}\lambda_{bQ_1}^\ast]-\text{Arg}[\lambda_{sQ_2}\lambda_{bQ_2}^\ast] \ .
\end{aligned}
\end{equation}
Note that, while the phases $\delta_{qbQ}$ are convention dependent, their differences are not. The measurement of such an asymmetry will not only prove the presence of NP, but also, if the NP is DQCD, it will prove that the dark sector is multi-flavored, contains more than one new $CP-$violating phase, and has at least one dark pion that is a superposition of at least two dark quark$-$antiquark states.


\subsection{\boldmath{$CP-$}Asymmetry in \boldmath{$K \rightarrow \pi \slashed{E}$}}
\label{pheno4}
We now turn to rare kaon decays, specifically the mode ($K_{\text{L}} \rightarrow \pi^0 \slashed{E}$), and the related mode ($K^+ \rightarrow \pi^+ \slashed{E}$). 
In the SM, these modes are dominated by the diagrams of Figure~\ref{SMb-to-q} with a quark $s$ instead of $b$ and $q\equiv d$. These diagrams have the same CPV phase of decay, that is Arg$(V_{td}V_{ts}^\ast)=\beta_s-\beta$. The branching ratios have been calculated to be: $\mathcal{B}(K_{\text{L}} \rightarrow \pi^0 \nu\bar{\nu})_{\text{SM}}=(3.00\pm 0.31)\times 10^{-11}$, and $\mathcal{B}(K^+ \rightarrow \pi^+ \nu\bar{\nu})_{\text{SM
}}=(9.11\pm 0.72)\times 10^{-11} $ \cite{BurasEtAlKtoPi}. These flavor changing neutral currents (FCNCs) are very suppressed in the standard model because of the GIM mechanism, the smallness of the CKM matrix elements, and the loop suppression.
At the experimental level, the KOTO and NA62 collaborations place limits on these processes. The latest upper bound from KOTO on the neutral Kaon process is, at $90\%$ CL  \cite{Ahn:2020opg},
\begin{equation}
\label{KOTOexp}
\mathcal{B}(K_{\text{L}} \rightarrow \pi^0\nu\bar{\nu})_{\text{KOTO}} < 4.9\times 10^{-9} \ .
\end{equation}
The NA62 upper limit for the charged mode is, at $90\%$ CL \cite{CortinaGil:2020vlo},
\begin{equation}
\label{NA62exp}
\mathcal{B}(K^+ \rightarrow \pi^+\nu\bar{\nu})_{\text{NA62}} < 1.78\times 10^{-10} \ .
\end{equation}
%
We consider the equivalent modes ($K_{\text{L}}(K^+) \rightarrow \pi^{0(+)}\pi_{\text{D}}$) in DQCD with 150 MeV $\lesssim M_D \lesssim 350$ MeV. This is below the GeV scale where we have taken our DQCD confinement scale, but this can happen if $n_f \geq 2$ and if the dark quark masses are small.

The (dark) quark level of these decays is dominated by the tree diagrams of Figure~\ref{bsQQbar} with a quark $s$ instead of $b$ and $q\equiv d$. 
The neutral Kaon decay of this type is $CP$-violating because the $K_L$ is $CP$-odd while the final state is $CP$-even. This is not the case for the charged Kaon decay which can proceed in the absence of $CP$-violation. 
We have calculated the branching ratio of the semi-invisible neutral kaon decay by using the factorization approximation. Summing over all dark flavors, we find:
\begin{equation}
\begin{aligned}
\label{BRKtoPiPiD}
\sum\limits_{Q,Q'}\mathcal{B}(K_{\text{L}} \rightarrow \pi^0\pi_{\text{D}}) \approx & \frac{N_c\sum\limits_{Q,Q'}|\text{Im}(\lambda_{dQ}\lambda_{sQ'}^\ast)|^2}{1024\pi M_X^4} 
\frac{F_D^2 [F_0^{K\pi^+}(M_D^2)]^2}{\Gamma_{K_{\text{L}}}}
\frac{(M_K^2-M_{\pi}^2)^2}{M_K^3} \times \\
& \hskip1.5cm \times \sqrt{\left[(M_K+M_{\pi})^2-M_D^2\right]\left[(M_K-M_{\pi})^2-M_D^2\right]} \ ,
\end{aligned}
\end{equation}
where $\Gamma_{K_{\text{L}}}$ is the decay rate of $K_{\text{L}}$, which value is listed in Table~\ref{tab:1} and $F_0^{K\pi^+}(q^2)$ ($q$ being the  momentum exchange between $K_{\text{L}}$ and $\pi^0$) is a form factor of the hadronic matrix element $\bra{\pi^+}\bar{s}\gamma^{\mu}d\ket{K^0}$. We can find from \cite{CarrascoEtal} that $F_0^{K\pi^+}(q^2)\approx 1$. The $CP$-violation is seen in the fact that if the effective coupling $\lambda_{dQ}\lambda_{sQ'}^\ast$ were real, then this decay would not occur. 

The constraint on $\lambda$ from the $CP$-violating $\Delta F = 2$ observable $|\epsilon_K|$ (see Table~\ref{tab:2}), is:
\begin{equation}
\label{xiK1DF}
 N_c \ | \ \text{Im}\Big(\big[\sum\limits_Q(\lambda_{dQ}^\ast\lambda_{sQ})\big]^2\Big) \ | \left(\frac{1\text{TeV}}{ M_X}\right)^2\leq   1.64\times 10^{-6} \ . 
\end{equation}
We notice that the bound is on the modulus of the imaginary part of the square of the sum over the same DF index, whereas equation~(\ref{BRKtoPiPiD}) contains the sum over 2 DF indices of the squared moduli of the imaginary parts. Thus, the two types of constraints (i.e. from $\Delta F = 2$ and $\Delta F = 1$ processes) cannot be compared in a model independent way. Assuming that an upper bound of the same order of magnitude as equation~(\ref{xiK1DF}) applies to the factor of equation~(\ref{BRKtoPiPiD}), we get for $M_D\approx F_D\approx 0.2$ GeV,
\begin{equation}
\label{BRKtoPiPiD2}
\sum\limits_{Q,Q'}\mathcal{B}(K_{\text{L}} \rightarrow \pi^0\pi_{\text{D}})\lesssim  \mathcal{O}(10^{-7})\left(\frac{F_D}{0.2\text{  GeV}}\right)^2\left(\frac{1\text{ TeV}}{M_X} \right)^2 \ ,    
\end{equation}
implying that the $\Delta F = 1$ bound from KOTO is significantly stronger than the one from $|\epsilon_K|$.


The branching ratio of the semi-invisible charged kaon decay, after summing over all dark flavors is:
\begin{equation}
\begin{aligned}
\label{BRK+toPi+PiD}
&\sum\limits_{Q,Q'}\mathcal{B}(K^+ \rightarrow \pi^+\pi_{\text{D}}) \approx  \chi_{ds}\ \frac{N_c}{1024\pi M_X^4} 
\frac{F_D^2 [F_0^{K\pi^+}(M_D^2)]^2}{\Gamma_{K^+}}
\frac{(M_{K^+}^2-M_{\pi^+}^2)^2}{M_{K^+}^3}\times \\
& \hskip4cm \times \sqrt{\left[(M_{K^+}+M_{\pi^+})^2-M_D^2\right]\left[(M_{K^+}-M_{\pi^+})^2-M_D^2\right]} \ ,
\end{aligned}
\end{equation}
where $ \chi_{ds}=\sum\limits_{Q,Q'}|\lambda_{dQ}\lambda_{sQ'}^\ast|^2$, and $\Gamma_{K^+}$, given in Table~\ref{tab:1}, is the decay rate of $K^+$.

It has been shown in~\cite{GNpaper}, that in any lepton flavor conserving model, the following relation (from which the Grossman-Nir (GN) bound resulted) applies:
\begin{equation}
\label{GNrelation}
\mathcal{A}_{CP}^{\text{GN}}=\frac{\mathcal{B}(K_{\text{L}} \rightarrow \pi^0\nu \bar{\nu})}{\mathcal{B}(K^+ \rightarrow \pi^+\nu \bar{\nu})}\simeq (4.3)\sin^2{\theta} \ ,
\end{equation}
where $\theta$ is the CPV phase of decay. As mentioned at the beginning of this section, the phase of decay in the SM is $\theta_{\text{SM}}=\beta_s-\beta $.

We have verified that a  similar relation holds in DQCD. The charged and neutral Kaon semi-invisible decays can be related to one another with a GN-like relation.
Both processes have the same inclusive decay $(s\to d Q \bar{Q}')$ whose CPV phase is $\text{Arg}(\lambda_{dQ}\lambda^\ast_{sQ'})\equiv \delta_{dsQQ'}$. Thus we can define an asymmetry as follows:
\begin{equation}
\label{GNNP}
\mathcal{A}_{CP}^{\text{NP}}=\frac{\mathcal{B}(K_{\text{L}} \rightarrow \pi^0\pi_{\text{D}})}{\mathcal{B}(K^+ \rightarrow \pi^+\pi_{\text{D}})}\simeq (4.3)\sin^2{(\delta_{dsQQ'})} \ .
\end{equation}
We can define the total CPV asymmetry as the sum of GN asymmetries of all contributing decays:
\begin{equation}
\label{GNNPall}
\mathcal{A}^{\text{NP}}_{\text{Total}}\equiv \sum\limits_{Q,Q'}\mathcal{A}_{CP}^{\text{NP}}=\sum\limits_{Q,Q'}\frac{\mathcal{B}(K_{\text{L}} \rightarrow \pi^0\pi_{\text{D}})}{\mathcal{B}(K^+ \rightarrow \pi^+\pi_{\text{D}})}\simeq (4.3)\sum\limits_{Q,Q'}\sin^2{(\delta_{dsQQ'})} \ .
\end{equation}
On the experimental side, what can be measured is the quotient of the sums of BRs (whereas the theoretical asymmetry is the sum of the quotients). From equations~(\ref{BRKtoPiPiD}) and (\ref{BRK+toPi+PiD}) we get:
\begin{equation}
\label{GNexp}
\mathcal{A}_{\text{Exp}}=\frac{\sum\limits_{Q,Q'}\mathcal{B}(K_{\text{L}} \rightarrow \pi^0\pi_{\text{D}})}{\sum\limits_{Q,Q'}\mathcal{B}(K^+ \rightarrow \pi^+\pi_{\text{D}})}\simeq (4.3) \frac{\sum\limits_{Q,Q'}|\text{Im}(\lambda_{dQ}\lambda_{sQ'}^\ast)|^2}{\sum\limits_{Q,Q'}|\lambda_{dQ}\lambda_{sQ'}^\ast|^2} \ .
\end{equation}
Let us now assume that only one inclusive decay mode, $(s\to d Q \bar{Q}')$ is dominant over all the others. This means that the kaon decays dominantly to a pion and a dark pion composed of the specific flavors $Q$ and $\bar{Q}'$. In this case, both the theoretical and experimental asymmetries are given by equation~(\ref{GNNP}).
Therefore, a measurement of the GN asymmetry for $(K \rightarrow \pi\pi_{\text{D}})$ would allow for the measurement of the new CPV phase $\delta_{dsQQ'}$.  As mentioned in Section~\ref{pheno2}, SM decays to neutrinos can potentially be distinguished from those NP decays to dark pions via the energy distribution of the visible SM pion using equation~\eqref{Efinal}. The KOTO experiment is expected to be upgraded to gain significant sensitivity to ($K_L \rightarrow \pi^0$ invisible) decay~\cite{KOTOfuture}, potentially being able to make a 20\% measurement of the SM rate. This could allow for significantly better sensitivity to this type of GN asymmetry. 

Tables~\ref{tab:3} and \ref{tab:4} summarize all the results of this phenomenology section. In these Tables, "BR", "Exp", "Asym", and "DF" stand respectively for, Branching Ratio, Experiment,  Asymmetry, and Dark Flavor. 

\begin{center}
\begin{table}[hbpt]
\tiny
\centering
\begin{tabular}{||c|c|c|c|c||}
\hline \hline
\rule{0pt}{5ex} 
SM mode & BR(SM) & BR(Exp) & BR(Projected Exp) & NP modes   \\
 & & & &  
\\ \hline\hline
 \rule{0pt}{5ex}  
$ B_d \rightarrow \nu\bar{\nu}$ 
  & $1.24\times 10^{-25}$ \cite{BadinPetrov} &  $ < 2.4\times 10^{-5}$ & $ < 1.5\times 10^{-6}$  &  $ B_d \rightarrow Q\bar{Q}$    \\ 
$ B_d \rightarrow \nu\bar{\nu}\nu\bar{\nu}$ & $(1.51\pm 0.28)\times 10^{-16}$ \cite{PetrovEtAl} &  BaBar \cite{BaBar} & Belle II \cite{Belle2} & $ B_d \rightarrow \pi_D\pi_D$  \\
 & & & &  
 \\ \hline
 \rule{0pt}{5ex}  
$B_s \rightarrow \nu\bar{\nu}$  & $ 3.07\times 10^{-24}$ \cite{BadinPetrov} & ---  & < $1.1\times 10^{-5}$ & $B_s \rightarrow Q\bar{Q}$ \\
$B_s \rightarrow \nu\bar{\nu}\nu\bar{\nu}$ & $(5.48\pm 0.89)\times 10^{-15}$ \cite{PetrovEtAl} & & Belle II \cite{Belle2} & $B_s \rightarrow \pi_D\pi_D$ \\
 & & & & 
\\ \hline
\rule{0pt}{5ex}
 $ B_d \rightarrow \pi^0\nu\bar{\nu}$ & $ (1.2\pm 0.15)\times 10^{-7}$ \cite{HambrockEtAL} &  $ < 0.9\times 10^{-5}$  & $\mathcal{O}(10^{-7})$ & $ B_d \rightarrow \pi^0 Q\bar{Q}$   \\ 
 & & Belle \cite{Belle2017} & Belle II & $ B_d \rightarrow \pi^0\pi_{\text{D}}$\\
 & & & &    
 \\ \hline
 \rule{0pt}{5ex}  
  $ B_d \rightarrow K_{\text{S}}\nu\bar{\nu}$ & $ (2.00\pm 0.25)\times 10^{-6}$ \cite{BurasEtAl} &  $ < 1.3\times 10^{-5}$  & $\mathcal{O}(30\%)$  & $ B_d \rightarrow K_{\text{S}} Q\bar{Q}$   \\
  & & Belle \cite{Belle2017} & Belle II &  $ B_d \rightarrow K_{\text{S}}\pi_{\text{D}}$  \\
  & & & & 
\\ \hline
\rule{0pt}{5ex}
 $B^+ \rightarrow \pi^+\nu\bar{\nu}$ & $ (2.39\pm 0.30)\times 10^{-7}$ \cite{HambrockEtAL} &  $ < 1.4\times 10^{-5}$ & $\mathcal{O}(10^{-7})$ & $B^+ \rightarrow \pi^+ Q\bar{Q}$   \\ 
 & & Belle \cite{Belle2017} & Belle II & $B^+ \rightarrow \pi^+\pi_{\text{D}}$  \\
 & & & &   
 \\ \hline
 \rule{0pt}{5ex}  
  $B^+ \rightarrow K^+\nu\bar{\nu}$ & $ (4.00\pm 0.50)\times 10^{-6}$ \cite{BurasEtAl} &  $ < 1.9\times 10^{-5}$  & $\mathcal{O}(20\%)$ & $B^+ \rightarrow K^+ Q\bar{Q}$ \\
  & & Belle \cite{Belle2017} & Belle II & $B^+ \rightarrow K^+\pi_{\text{D}}$ \\
 & & & &   
 \\ \hline 
 \rule{0pt}{5ex}  
  $K_{\text{L}} \rightarrow \pi^0\nu\bar{\nu}$ & $ (3.00\pm 0.31)\times 10^{-11}$ \cite{BurasEtAlKtoPi} &  $  < 4.9\times 10^{-9}$ & $\mathcal{O}(20\%)$ & $K_{\text{L}} \rightarrow \pi^0\pi_{\text{D}}$   \\
  & & KOTO \cite{Ahn:2020opg} &  KOTO Step2~\cite{KOTOfuture} &  \\
 & & & & 
  \\ \hline 
 \rule{0pt}{5ex}  
  $K^+ \rightarrow \pi^+\nu\bar{\nu}$ & $ (9.11\pm 0.72)\times 10^{-11}$ \cite{BurasEtAlKtoPi} &  $  < 1.78\times 10^{-10}$ & --- & $K^+ \rightarrow \pi^+\pi_{\text{D}}$   \\
  & & NA62 \cite{CortinaGil:2020vlo} &   &  \\
 & & & & 
 \\ \hline \hline
\end{tabular}
\caption{Summary of the invisible and semi-invisible decays we studied. See text for more details, and for BRs of specific scenarios. 
All current limits are at $90\%$ CL. The projected branching ratios for semi-invisible $B$ decays (rows 3-6) are extrapolated from the Belle result. Those above the SM value (rows 3,5) 
use linear luminosity scaling, while those with a percentage (rows 4, 6, and 7) denote the expected precision on the measurement assuming an SM-like central value using $1/\sqrt{N}$ estimate for the uncertainty.}
\label{tab:3}
\end{table}
\end{center}

\begin{center}
\begin{table}[hbpt]
\scriptsize
\centering
\begin{tabular}{||c|c|c|c|c||}
\hline \hline
\rule{0pt}{5ex} 
NP mode & Asym Type & SM & \# DF &  Asym(NP)
\\ \hline\hline
 \rule{0pt}{5ex}  
$B_d \rightarrow Q\bar{Q}$ &  Mixing & 
$\simeq 0$ & $1$  &  $ -\sin \left[2( \beta+\delta_{dbQ})\right]\sin (\Delta M_dt)$\\ 
& & & & 
 \\ \hline
 \rule{0pt}{5ex}  
$B_s \rightarrow Q\bar{Q}$  & Mixing & 
$\simeq 0$ & $1$
&   $-\sin \left[2( \beta_s+\delta_{sbQ})\right]\sin (\Delta M_st)$\\
 & & & & 
\\ \hline
\rule{0pt}{5ex}
 $B_d \rightarrow \pi^0\pi_{\text{D}}$  & Mixing &  $ + \sin \left(4 \beta\right)\sin (\Delta M_dt)$ & $\geq 2$  &   $ +\sin \left[2( \beta+\delta_{dbQ})\right]\sin (\Delta M_dt)$\\ 
 & & & & 
 \\ \hline
 \rule{0pt}{5ex}  
 $B_d \rightarrow K_{\text{S}}\pi_{\text{D}}$  &  Mixing &  $ - \sin \left[2( \beta+\beta_s)\right]\sin (\Delta M_dt)$ & $\geq 2$   &  $ - \sin \left[2( \beta+\delta_{sbQ'})\right]\sin (\Delta M_dt)$\\
 & & & & 
\\ \hline
\rule{0pt}{5ex}
 $B^+ \rightarrow \pi^+\pi_{\text{D}}$  & Direct &  $\simeq 0$ & $\geq 2$  &   $\propto\sin(\delta_{dbQ_1}-\delta_{dbQ_2})$\\ 
 & & & & 
 \\ \hline
 \rule{0pt}{5ex}  
 $B^+ \rightarrow K^+\pi_{\text{D}}$  &  Direct &  $\simeq 0$ & $\geq 2$   &   $\propto\sin(\delta_{sbQ_1}-\delta_{sbQ_2})$\\
 & & & & 
 \\ \hline 
 \rule{0pt}{5ex}  
 $K_{\text{L}} \rightarrow \pi^0\pi_{\text{D}}$ &  Grossman-Nir & $(4.3)\sin^2{(\beta_s-\beta)}$ & $\geq 2$   & $(4.3)\sin^2{(\delta_{dsQQ'})}$\\
 & & & & 
 \\ \hline \hline
\end{tabular}
\caption{Summary of the $CP-$asymmetries  studied (see text for details).
We recall that $\delta_{qq'Q}\equiv \text{Arg}(\lambda_{qQ}\lambda_{q'Q}^\ast)$ and $\delta_{dsQQ'}\equiv \text{Arg}(\lambda_{dQ}\lambda_{sQ'}^\ast)$. The fourth column shows the number of dark flavours for which the  expressions in the last column apply.}
\label{tab:4}
\end{table}
\end{center}
%
\section{Combination of \boldmath $\Delta F=1$ and $\Delta F=2$ Constraints}
\label{Constraints2}
We will study in this section the constraints on the DQCD model from the experimental measurements of the BRs of the (semi-)invisible decays considered in the previous section and listed in Table~\ref{tab:3}. The $\Delta F=1$ constraints apply to the $\chi_{qq'}$ terms - defined in equation~(\ref{chi}) -  while the $\Delta F=2$ constraints studied in Section~\ref{Constraints} apply to the $\xi_M$ terms - defined in equation~(\ref{xi}). These last, summarised in Table~\ref{tab:2}, are valid for any dark quark/pion mass, and any value of the decay constants of the dark pions. The $\Delta F=1$ constraints have on-shell dark pions in the final state, so they will only be kinematically allowed for certain values of $M_D$. The matrix element also depends on $F_D$ which can have a significant effect on the constraint.  For simplicity we will take $M_D = F_D$ in our calculations of the $\Delta F=1$ constraints. 

In this study, we will consider two dark pion mass regimes: first, we will take $M_D$ and $F_D$ at the GeV order, where only (semi-)invisible $B$ meson decays contribute ($K$ decays to dark pions being kinematically forbidden); and second, we will take $M_D$ and $F_D$ at the 0.1 GeV order, where $K$ decays contribute. We will evaluate the $\Delta F=1$ constraints in the three scenarios of one, two, and three DFs, then combine them to the $\Delta F=2$ constraints and plot the results. We will consider the real Yukawa matrix case as well as scenarios of maximum CPV.  We recall that in the case of 1 DF, $\chi_{ds},\chi_{db}$ and $\chi_{sb}$ coincide respectively with $|\xi_K|^2,|\xi_{B_d}|^2$ and $|\xi_{B_s}|^2$.  In order to obtain our allowed parameter space in the cases of 2 and 3 DFs, we randomly generated all the parameters and subjected each point of the parameter space to the $\Delta F=1$ and  $\Delta F=2$  flavor constraints. 
\subsection{0.5 GeV\boldmath $\lesssim M_D\approx F_D \lesssim 4$ GeV}
\label{BoundsHighMdark}
\begin{figure}[hbpt]
 \centering
\begin{subfigure}{0.45\textwidth}
    \includegraphics[width=\textwidth]{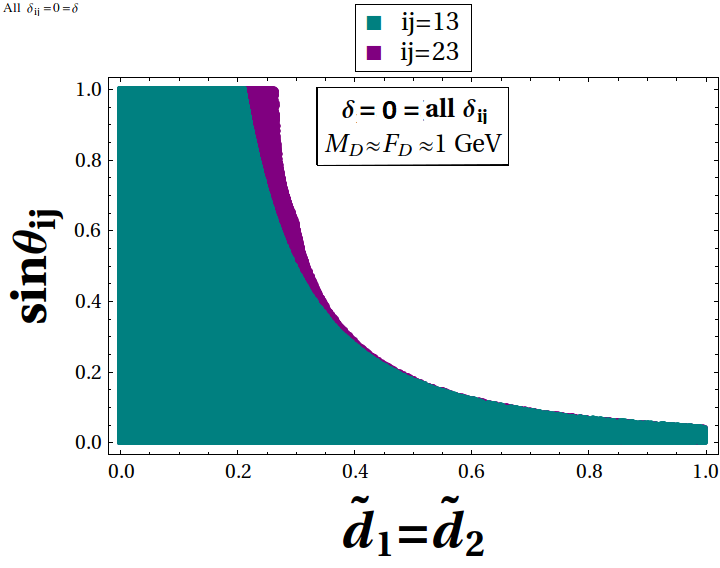}
    \end{subfigure}
\hspace{1cm}
\begin{subfigure}{0.45\textwidth}
    \includegraphics[width=\textwidth]{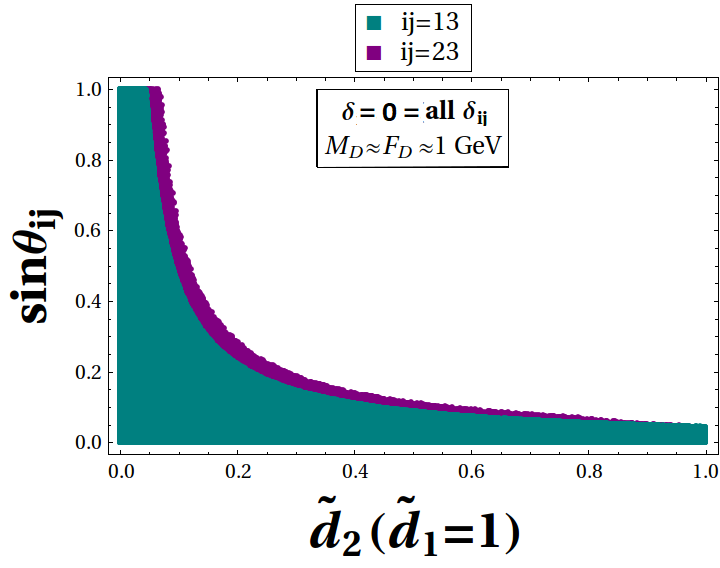}
    \end{subfigure}

\begin{subfigure}{0.45\textwidth}\vspace{1cm}
    \includegraphics[width=\textwidth]{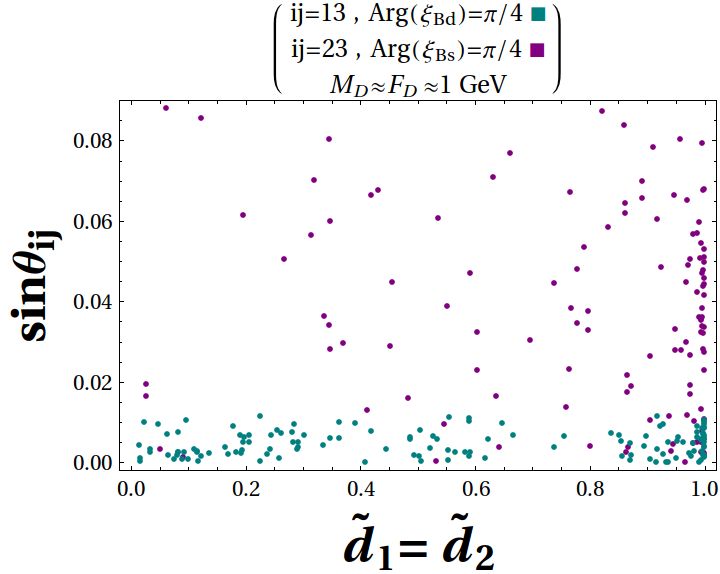}
    \end{subfigure}
\hspace{1cm}
\begin{subfigure}{0.45\textwidth}\vspace{1cm}
    \includegraphics[width=\textwidth]{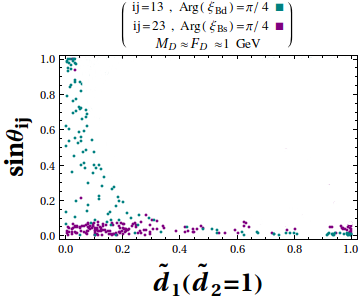}
    \end{subfigure}
\caption{Combined $\Delta F=2$ and $\Delta F=1$ Constraints, for $F_D\approx M_D\approx 1$ GeV and $M_X=1$ TeV, in the case of 2 DFs. The upper panels represent the case of a real Yukawa matrix, and the lower ones represent the complex case in a scenario of maximum CPV. The very small allowed region in the lower left panel is zoomed in.} 
\label{Figallbounds2DF}
\end{figure}

\begin{figure}[hbpt]
 \centering
%
%
%
%
%
%
%


  \begin{subfigure}{0.45\textwidth}\vspace{1cm}
    \includegraphics[width=\textwidth]{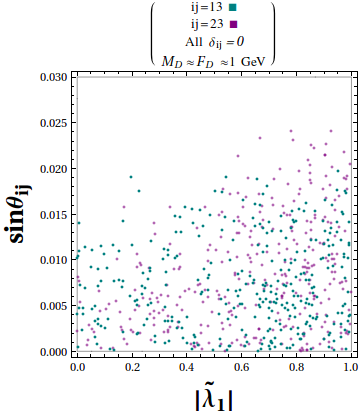}
    \end{subfigure}
\hspace{1cm}
  \begin{subfigure}{0.45\textwidth}\vspace{1cm}
    \includegraphics[width=\textwidth]{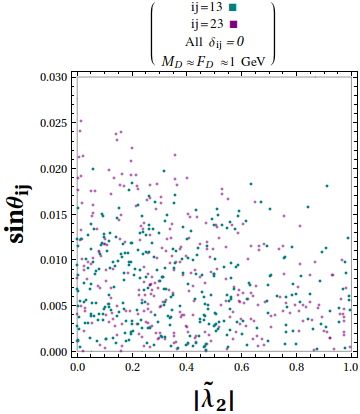}
    \end{subfigure}
%
%
%
%
\caption{Combined $\Delta F=2$ and $\Delta F=1$ Constraints, for $F_D\approx M_D\approx 1$ GeV and $M_X=1$ TeV, in the case of 3 DFs and a real Yukawa matrix. In these plots the very small allowed regions are zoomed in.}
\label{Figallbounds3DF}
\end{figure}

In this regime the $K$ decays of the last column of Table~\ref{tab:3} do not apply, therefore there are no bounds on $\chi_{ds}$. The strongest bound on $\chi_{db}$ ($\chi_{sb}$) is from ($B^+ \rightarrow \pi^+(K^+)\pi_{\text{D}}$). For $F_D\approx M_D\approx 1$ GeV, we have
\begin{equation}
 \label{BRconstraints1}
 \begin{aligned}
& \left[N_c\left(\frac{F_D}{1\text{  GeV}}\right)^2\left(\frac{1\text{  TeV}}{ M_X}\right)^4 \chi_{db}\right]_{_{_{[M_D\approx 1 \text{ GeV}]}}} \ \leq 3.19\times 10^{-3} \ , \\
& \left[N_c\left(\frac{F_D}{1\text{  GeV}}\right)^2\left(\frac{1\text{  TeV}}{ M_X}\right)^4 \chi_{sb}\right]_{_{_{[M_D\approx 1 \text{ GeV}]}}} \ \leq 1.97\times 10^{-3} \ .
\end{aligned}
\end{equation}
Combining these constraints to those of Table~\ref{tab:2}, we plot Figures~\ref{Figallbounds2DF} and \ref{Figallbounds3DF} for 2 DFs and 3 DFs respectively. 
In the 1 DF scenario, the $\Delta F=1$ constraints do not affect $\chi_{ds}$. They are stronger  (weaker) than the $\Delta F=2$ constraints for $\chi_{sb}$ ($\chi_{db}$)$\equiv |\xi_{B_s}|^2$ ($|\xi_{B_d}|^2)$. We have considered the same scenarios as those of Figure~\ref{Fig1DF}, that is, the case where the Yukawa matrix $\lambda$ is real, and the case where it is complex, with maximum CPV. The red and blue regions of Figure~\ref{Fig1DF} remain the same, but the green regions shrink more than those of all the panels of this Figure.

In the 2 DF scenario, $s_{12}$ remains generic in the real Yukawa case. In the complex case, the orange regions of all the panels of Figure~\ref{FIGsupMAXCPV2F} remain the same. This is why we only show $s_{13}$ and $s_{23}$ in Figure~\ref{Figallbounds2DF}, where the plots for $\tilde{d}_2=1$ are about the same as those for $\tilde{d}_1=1$ in the real Yukawa case (the 2 upper panels). For the complex Yukawa case (the 2 lower panels), we have considered scenarios of maximum CPV, where the Wilson coefficients are imaginary (like in Section~\ref{sec:2DF}). We observe that the parameter spaces are strongly reduced when we add the  $\Delta F=1$ constraints, for example, when $\tilde{d}_1=\tilde{d}_2$, $s_{23}$ ($s_{13}$) does not exceed 0.09 (0.015).

The 3 DF scenario with a real Yukawa matrix is presented in Figure~\ref{Figallbounds3DF}. $s_{12}$ is not shown because the yellow allowed region of the upper panel of Figure~\ref{Fig3DF} is unchanged. 
We observe that $s_{13}$ and $s_{23}$ do not exceed 0.02 and 0.03 respectively. In the case of a complex Yukawa coupling, we considered the same maximum CPV scenarios as those of the lower right panel of Figure~\ref{Fig3DF}, and zero points were generated out of 1 billion scans! This result suggests that these scenarios may be excluded when the dark pions' mass and decay constant are $\sim $ 1 GeV.

\vspace{2cm}

\vspace{1cm}

\subsection{0.15 GeV\boldmath $\lesssim M_D\approx F_D \lesssim 0.3$ GeV}
\label{BoundsLowMdark}
\begin{figure}[hbpt]
 \centering
\begin{subfigure}{0.3\textwidth}\vspace{1cm}
    \includegraphics[width=\textwidth]{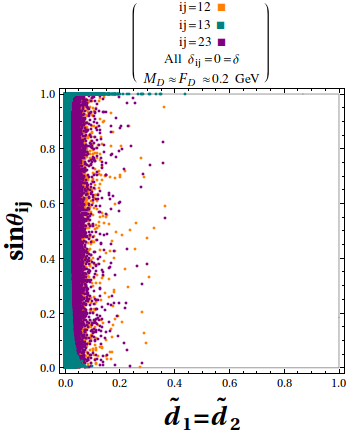}
    \end{subfigure}
\hspace{0.45cm}
\begin{subfigure}{0.3\textwidth}\vspace{1cm}
    \includegraphics[width=\textwidth]{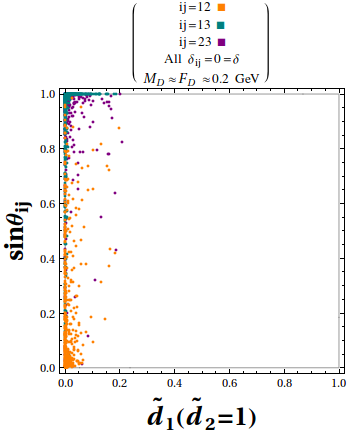}
    \end{subfigure}
\hspace{0.45cm}
\begin{subfigure}{0.3\textwidth}\vspace{1cm}
    \includegraphics[width=\textwidth]{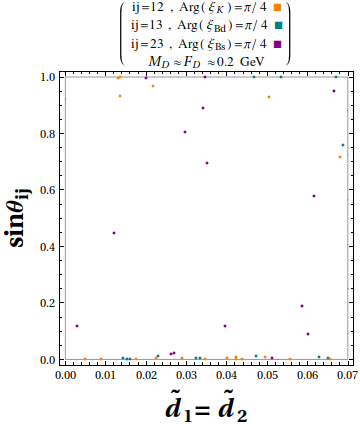}
    \end{subfigure}    
\caption{Combined $\Delta F=2$ and $\Delta F=1$ Constraints, for $F_D\approx M_D\approx 0.2$ GeV and $M_X=1$ TeV, in the case of 2 DFs. The left and middle panels represent the case of a real Yukawa matrix, and the right one, in which we zoomed in the very small allowed region, represents the complex case in a scenario of maximum CPV.} 
\label{Figallbounds2DF2}
\end{figure}

Here we do the same analysis we did in Section~\ref{BoundsHighMdark} in a lighter DM regime where all the decays of the last column of Table~\ref{tab:3} apply. The strongest constraint on $\chi_{ds}$ is from ($K^+ \rightarrow \pi^+\pi_{\text{D}}$), and again, the strongest bound on $\chi_{db}$ ($\chi_{sb}$) is from ($B^+ \rightarrow \pi^+(K^+)\pi_{\text{D}}$). For $F_D\approx M_D\approx 0.2$ GeV:
\begin{equation}
 \label{BRconstraints1}
 \begin{aligned}
& \left[N_c\left(\frac{F_D}{0.2\text{  GeV}}\right)^2\left(\frac{1\text{  TeV}}{ M_X}\right)^4 \chi_{db}\right]_{_{_{[M_D\approx 0.2 \text{ GeV}]}}} \ \leq 0.077 \ , \\
& \left[N_c\left(\frac{F_D}{0.2\text{  GeV}}\right)^2\left(\frac{1\text{  TeV}}{ M_X}\right)^4 \chi_{sb}\right]_{_{_{[M_D\approx 0.2 \text{ GeV}]}}} \ \leq 0.048 \ , \\
& \left[N_c\left(\frac{F_D}{0.2\text{  GeV}}\right)^2\left(\frac{1\text{  TeV}}{ M_X}\right)^4 \chi_{ds}\right]_{_{_{[M_D\approx 0.2 \text{ GeV}]}}} \ \leq 1.035\times 10^{-8} \ .
\end{aligned}
\end{equation}
Figure~\ref{Figallbounds2DF2} shows the combined flavor constraints in the 2 DFs case. 
In the 1 DF scenario, the $\Delta F=1$ constraints are weaker than the $\Delta F=2$ ones for $\chi_{db}$  and $\chi_{sb}$. Therefore, the blue and green regions of Figure~\ref{Fig1DF} remain the same. On the other hand, the constraint from NA62 is the strongest on $\chi_{ds}$, this restricts the parameter space, in the $(\tilde{\lambda}_d,\tilde{\lambda}_s)$ plane to almost a red cross in the real Yukawa case as well as in the same cases of maximum CPV as those of Figure~\ref{Fig1DF}.

In the 2 DF scenario and a real Yukawa matrix, when $\tilde{d}_1=1$, only a few points were generated out of a billion scans. These points have $\tilde{d}_2 \lesssim 0.2$, $s_{12} \gtrsim 0.9$, $s_{13} \gtrsim 0.9$, and $s_{23} \lesssim 0.2$.
In the case where $\tilde{d}_1=\tilde{d}_2$ or $\tilde{d}_2=1$,  we have $\tilde{d}_1 \lesssim 0.4$, as we observe in the left and middle panels of Figure~\ref{Figallbounds2DF2}.
If the Yukawa matrix is complex and $\tilde{d}_2=1$, zero points were generated in the planes $(\tilde{d}_1,s_{ij})$, out of a billion scans, in the same maximum CPV scenarios as those considered in the previous subsection. The case where $\tilde{d}_1=\tilde{d}_2$ is represented in the right panel of Figure~\ref{Figallbounds2DF2} where only a few points were allowed out of one billion scans and all have $\tilde{d}_1 \lesssim 0.07$ .

In the 3 DF scenario, zero points were allowed in the planes $(\tilde{\Delta}_{ij},s_{ij})$ out of 1 to 3 billion scans, either in the real Yukawa matrix case or in the maximum CPV cases considered in the previous subsection. Its likely that the constraint from NA62 excludes DQCD with 3 flavors, $M_X=1$ TeV and $M_D=F_D=200$ MeV. 
%
%
%
%
%


\section{Summary and Conclusion}
\label{sec:conclusions}

Dark QCD models that confine at the GeV scale with a TeV scale mediator charged under both QCD and dark QCD are a well motivated paradigm for asymmetric dark matter~\cite{BaiSchwaller014}. This setup, however, requires a flavourful coupling between the mediator and the SM quarks of the form given in equation~\eqref{IntLag}. This coupling generically mediates flavour and $CP$-violating processes at one-loop, and in this work we have quantitatively studied a broad range of such processes. We extend the analysis of~\cite{RennerSchwaller018} and consider $CP$-violation for the first time. We also consider all possible number of dark flavours and enumerate the number of $CP$-conserving and violating parameters for all cases. The parameter counting for $n_f > 3$ turns out to be equivalent to that of $n_f=3$ as long as the dark quarks are mass degenerate.

In most BSM theories with TeV scale new physics and $\mathcal{O}(1)$ $CP$-violating phases, the strongest bounds come from null searches for electric dipole moments. We have shown, however, that even though this model can have $\mathcal{O}(1)$ $CP$-violating phases, the contributions to EDMs cancel at one and two loops, and the constraints turn out to be quite weak. If there are multiple flavours of scalars, then there may be stronger constraints, and future more sensitive EDM experiments may have sensitivity to the vanilla model. 

We have studied constraints from $\Delta F = 2$ processes including meson mixing and $CP$-violation. We have found that generic $\mathcal{O}(1)$ couplings are excluded and we performed a detailed numerical exploration of the allowed parameter space. We have found some slices of parameters where the constraints can be weak. 

We have also explored $\Delta F = 1$ processes from rare meson decays to dark sector states. For the parameter space we consider, the lightest dark sector states have long lifetimes and can be treated as missing energy. Therefore rare decays of mesons to missing energy can be used to place bounds that are complementary to those from $\Delta F = 2$ processes with different scaling with the dark sector parameters. Some of these processes such as $B \rightarrow$ invisible have essentially no SM background and would be a smoking gun for new physics. Many measurements will be performed in the near future at Belle II and KOTO that could have sensitivity to decays to invisible dark sector states, and we have shown how these can be distinguished from SM backgrounds when they exist.

Finally, we explore novel $CP$-violating observables that can be used to probe the $CP$-violation in the dark sector. While some of these measurements are quite speculative, they could shed light on new sources of $CP$-violation in the universe. 


\acknowledgments 

We thank Prateek Agrawal, Monika Blanke, Kevin Graham, Yuval Grossman, Maxim Pospelov, Alberto Tonero, and Yongcheng Wu for useful discussions. This work is supported in part by the Natural Sciences and Engineering Research Council (NSERC) of Canada, and the work of WB is supported in part by the M. Hildred Blewett Fellowship of the American Physical Society, and the Arthur B. McDonald Canadian Astroparticle Physics Research Institute.


\appendix
\section{A Brief Review of Meson Mixing and  \boldmath{$CP$} Violation}
\label{sec:A}

\subsection{The \boldmath{$K-$}System}
Let us first start with the neutral kaon system~\cite{Buras2005,Chau1983,BurasGuadagnoliIsidori2010,NirCPV}. The flavour eignestates are given by $\ket{K^0}$ and $\ket{\bar{K}^0}$,
 but these differ from the mass eigenstates $K$-long and $K$-short,  $\ket{K_{\text{S}}^0}$ and $\ket{K_{\text{L}}^0}$.
%
%
 %
The time evolution of the kaon states in the flavour basis is given by the Schr\"{o}dinger equation:
\begin{equation}
\label{KtimeEvol}
i\frac{d}{dt} \ket{\psi}= \hat{H} \ket{\psi} \ .
\end{equation}
The Hamiltonian of weak interactions, $\hat{H}$, is given by  
\begin{equation}
\label{H}
\hat{H}=
\begin{pmatrix}
M-i\Gamma /2 & M_{12}-i\Gamma_{12} /2 \\
M_{12}^\ast-i\Gamma_{12}^\ast /2 & M-i\Gamma /2
\end{pmatrix} \ ,
\end{equation}
where $M$ and $\Gamma$ are respectively the mass and the width of $K^0$. $M_{12}$ is the transition matrix element between $K^0$ and $\bar{K}^0$ of the Hamiltonian of mixing, when the intermediate states are off-shell (virtual), and $\Gamma_{12}$ is the matrix element when the intermediate states go on-shell (physical). Both $M_{12}$ and $\Gamma_{12}$ are complex because their coefficients are complex. The calculation of the eigenstates of $\hat{H}$ gives
\begin{equation}
 \label{KlKs}
 \ket{K_{S,L}^0}=p\ket{K^0} \pm q\ket{\bar{K}^0}  \ ,
\end{equation}
with 
\begin{equation}\begin{aligned}
\label{q/p def}
\sqrt{\frac{M_{12}^\ast-i\Gamma_{12}^\ast /2}{M_{12}-i\Gamma_{12} /2}}= & -\frac{q}{p} \ , \\
|p|^2+|q|^2= & 1\ .
\end{aligned}
\end{equation}
The physical states of weak interaction, $\ket{K_{\text{S}}^0}$ and $\ket{K_{\text{L}}^0}$, oscillate in time between the flavor eigenstates $K^0$ and $\bar{K^0}$. 

We can also use $\hat{H}$ to compute the masses and widths of the physical states:
\begin{equation}
\begin{aligned}
M_{L,S}= & M \pm \text{Re}[Q] \ , \ \ \ \ \ \Gamma_{L,S}=\Gamma \mp 2\text{Im}[Q] \ , \\
Q = & \sqrt{(M_{12}-i\Gamma_{12} /2 )(M_{12}^\ast-i\Gamma_{12}^\ast /2)} \ ,
\end{aligned}
\label{MandGamma}
\end{equation}
and from here, we can get the mass and width differences:
\begin{equation}
 \label{DeltaMandGamma}
 \Delta M_K = M_L - M_S = 2\text{Re}[Q] \ , \ \ \ \ \ \ \ \Delta\Gamma_K = \Gamma_L - \Gamma_S = -4\text{Im}[Q].
\end{equation}

The flavour eignestates can also be rearranged into $CP$ eignestates
\begin{equation}
\begin{aligned}
 \ket{K_1^0}= & \frac{1}{\sqrt{2}}\left[\ket{K^0}-\ket{\bar{K^0}} \right] \ , \\
 \ket{K_2^0}= & \frac{1}{\sqrt{2}}\left[\ket{K^0}+\ket{\bar{K^0}} \right] \ .
\end{aligned}
\label{K1K2vsKKbar}
\end{equation}
with $CP\ket{K_{1,2}^0}=\pm\ket{K_{1,2}^0}$. As the weak interactions violate $CP$, the $CP$ basis differs from the mass basis:
\begin{equation}
\begin{aligned}
 \ket{K_{\text{S}}^0}= & \frac{1}{\sqrt{1+|\bar{\epsilon}|^2}}\left[\ket{K_1^0}+\bar{\epsilon}\ket{K_2^0} \right] \ , \\
 \ket{K_{\text{L}}^0}= & \frac{1}{\sqrt{1+|\bar{\epsilon}|^2}}\left[\bar{\epsilon}\ket{K_1^0}+\ket{K_2^0} \right] \ ,
\end{aligned}
\label{K1K2}
\end{equation}
where $\bar{\epsilon}$ parameterizes how much the physical states deviate from the $CP$ eigenstates.

From equations~(\ref{KlKs}), (\ref{K1K2vsKKbar}), and (\ref{K1K2}), we can find another way the weak interaction  eigenstates are related to the flavor eigenstates:
\begin{equation}
\label{KfalvKmass}
\ket{K_{S,L}^0}=\frac{1}{\sqrt{2(1+|\bar{\epsilon}|^2)}}\left[(1+\bar{\epsilon})\ket{K^0} \mp (1+\bar{\epsilon})\ket{\bar{K}^0}  \right]
  \ ,
\end{equation}
and how $\bar{\epsilon}$ is related to the $p$ and $q$ parameters, which in turn are related to the Hamiltonian off diagonal elements (equation~(\ref{q/p def})):
\begin{equation}
\label{q/p/epsilonbar}
-\frac{q}{p} =\frac{1-\bar{\epsilon}}{1+\bar{\epsilon}}= \sqrt{\frac{M_{12}^\ast-i\Gamma_{12}^\ast /2}{M_{12}-i\Gamma_{12} /2}} \ .
\end{equation}
The deviation of ($q/p$) from unity parameterizes the amount of $CP$ violation. We will return to this quantity in the B$-$system section.

The parameter $\bar{\epsilon}$ depends on the phase convention of the flavor states. In order to construct a phase independant parameter, we note that from equation~(\ref{K1K2}) the $K_{\text{S}}$ ($K_{\text{L}}$) is mostly $CP-$even (odd), therefore it will mostly decay to 2 (3) pions. This also explains the significant difference in their lifetimes as the decay to three pions is kinematically suppressed. The decay of $K_{\text{L}}$ to 2 pions is $CP-$violating and can be used to construct a phase independent measure of $CP$ violation. 
 $\epsilon_K$ is defined as  the ratio of the amplitude of the decay ($K_{\text{L}}\rightarrow \pi\pi$)  to the amplitude of the decay ($K_{\text{S}}\rightarrow \pi\pi$) when the final state's isospin is zero, namely
\begin{equation}
 \label{e/ebar}
 \epsilon_K \equiv \frac{A(K_{\text{L}}\rightarrow (\pi\pi)_{I=0})}{A(K_{\text{S}}\rightarrow (\pi\pi)_{I=0})} \ \  \Rightarrow  \ \ \text{Re}(\epsilon_K) = \text{Re}(\bar{\epsilon}) \ .
\end{equation}
Thereby, the phase convention dependence is cancelled in $\epsilon_K$. 

The simplest way $\epsilon_K$ can be related to the Hamiltonian off diagonal elements is as follows~\cite{BurasGuadagnoliIsidori2010} 
\begin{equation}
\label{epsilonDef1}
\epsilon_K  = \frac{\kappa_{\epsilon}e^{i\phi_\epsilon}}{\sqrt{2}\Delta  M_K} \text{Im}[M_{12}^{\rm SD}] \ ,
\end{equation}
where $\phi_\epsilon = (43.51 \pm 0.05)^o$  and the phenomenological factor, $\kappa_{\epsilon} = 0.94 \pm 0.02$~\cite{BurasGirrbach014}, summarizes the corrections to $\epsilon_K$ due to long distance (LD) effects~\cite{BurasGuadagnoliIsidori2010}. $M_{12}^{SD}$ is the short distance (SD) contribution to $M_{12}$ defined in \ref{H}. This SD contribution is calculated from the box diagrams of the $K^0-\bar{K}^0$ mixing.\footnote{The box diagrams in DQCD are those of Figure~\ref{fig:Boxes}. They are similar in the SM, but replacing $X$ by the $W$ boson and $Q$ by the up-type quarks.} Equation~(\ref{epsilonDef1})  clearly shows how $\epsilon_K$ represents $CP$ violation due to mixing ($\epsilon_K$ vanishes if $M_{12}^{\rm SD}=0$).
The SD matrix element is given by
\begin{equation}
\label{MvsH}
M_{12}^{\rm SD}=\frac{1}{2m_K}\bra{\bar{K}^0}{H_{\Delta S=2}^{\rm Box}}\ket{K^0}^\ast .
\end{equation}
Here, $H_{\Delta S=2}^{\rm Box}$ is the effective Hamiltonian calculated from the box diagrams of the $K^0-\bar{K}^0$ mixing. Finally, for the case of the $K$ mesons only, since $\epsilon_K$ is very small, $\mathcal{O}(10^{-3})$, the imaginary parts of $M_{12}$ and $\Gamma_{12}$ are much smaller than their real parts, 
thus, from equation~(\ref{DeltaMandGamma}), we have the approximation
\begin{equation}
\label{DeltaMK}
\Delta M_K \simeq 2\text{Re}\left[ M_{12} \right]  \ ,
\end{equation}
where $M_{12}$ contains both SD and LD contributions. 


\subsection{The \boldmath{$B-$}Systems}
We now explore the basic meson mixing formalism for $B_d$ and $B_s$. The equations are the same as equations~(\ref{H})$-$(\ref{q/p/epsilonbar}) developed for $K$ mesons, except that the mass eigenstates are called $B_{\text{L}}$  and $B_{\text{H}}$ for the light and heavy state, respectively.
Since $\Gamma_{12}\ll M_{12}$  for the $B$-system~\cite{BurasEtAlBBbar}, we have, from equation~(\ref{DeltaMandGamma})~\cite{Buras2005},
\begin{equation}
\label{DeltaMB}
\Delta M_B = M_H - M_L \approx 2\mid M_{12}^B\mid  \ .
\end{equation}
Like equation~(\ref{MvsH}), we have for the B$-$system 
\begin{equation}
\label{MvsHB}
M_{12}^B=\frac{1}{2m_B}\bra{\bar{B}^0}{H_{\Delta B=2}}\ket{B^0}^\ast 
\end{equation}
Because the mass of the $B$ mesons is significantly above the QCD scale, the LD effects are negligible in the calculation of $\Delta M_B$~\cite{Buras1998}. 

Now expanding equation~(\ref{q/p def}), into powers of $|\Gamma_{12}/M_{12}|$, we can keep the leading term only, since $|\Gamma_{12}/M_{12}|$ is estimated to be of order $10^{-4}$~\cite{Buras2005,NirCPV}, giving 
\begin{equation}
\label{q/pB}
\left(\frac{q}{p}\right)_B  \approx  \frac{M_{12}^\ast}{\mid M_{12}\mid} \ \ \ \mysimeqq \ \ \ \frac{2M_{12}^\ast}{\Delta M_B} \ .
\end{equation}
Thus $(q/p)_B$ is a pure phase to an excellent approximation, which means
\begin{equation}
\label{q/pPhase}
\left( \frac{q}{p} \right)_B = e^{2i\phi_M}\simeq \ \ \ \frac{2M_{12}^\ast}{\Delta M_B} \ ,
\end{equation}
where $\phi_M$ is a CPV phase due to mixing. By convention, $\phi_M=-\beta$ for $B^0_d$, and $\phi_M=-\beta^0_s$ for $B_s$. 
\begin{equation}
\label{beta}
\left( \frac{q}{p} \right)_{B_d} = e^{-2i\beta} \approx \frac{2M_{12}^\ast(B_d)}{\Delta M_d} \ ,
\end{equation}
\begin{equation}
 \label{betas}
 \left( \frac{q}{p} \right)_{B_s} = e^{-2i\beta_s} \approx \frac{2M_{12}^\ast(B_s)}{\Delta M_s} \ .
\end{equation}
%
Within the Standard Model, the contribution to the amplitudes of the $B-\bar{B}$ mixing are dominated by the box diagrams with two $W$ propagators and 2 internal $t$ quarks, so the SM weak phase of mixing comes dominantly from the arguments of the CKM matrix elements. For $B_d^0$ this comes from $V_{td}$:
\begin{equation}
\label{ARGq/pBd}
\text{Arg}\left( \frac{q}{p}(\text{SM}) \right)_{B_d} = 2\text{Arg}(V_{td})=-2\beta_{\text{SM}} \ ,
\end{equation}
and for $B_s^0$ this comes from $V_{ts}$: 
\begin{equation}
\label{ARGq/pBs}
\text{Arg}\left( \frac{q}{p}(\text{SM}) \right)_{B_s} = 2\text{Arg}(V_{ts})=-2(\beta_s)_{\text{SM}} 
 \ .
\end{equation}
Both $\beta_{\text{SM}}$ and $(\beta_s)_{\text{SM}}$ are related to the unique phase of the CKM matrix. 
Their predicted SM values are reported in Table~\ref{tab:1} from UTfit collaboration~\cite{Bona2016Utfit2}. This same reference provides the experimental measurements of these angles, reported in Table~\ref{tab:1} as well.

\section{Mixing Induced  \boldmath{$CP-$}Asymmetry}
\label{sec:CPasym}

In what follows, we will briefly introduce a very interesting class of $CP-$asymmetries which result from $CP-$violation in the interference between mixing and decay \cite{Buras2005,NirFPandCPV,NirCPV,CPasymExp} which can be defined for neutral meson decays. Let $B^0$ be any neutral meson; knowing that it oscillates in time between  the flavour eigenstates $B^0$ and $\bar{B}^0$, we define $B^0(t)$ as the meson at a time $t$, starting as a $B^0$ at $t=0$. Similarly, $\bar{B}^0(t)$ is defined such as $\bar{B}^0(0)=\bar{B}^0$, the $CP-$conjugate of $B^0$. In a decay $B^0\rightarrow f$, where $f$ is a $CP-$eigenstate, 
the time-dependent mixing induced $CP-$asymmetry is defined as:
\begin{equation}
\label{CPasym1}
\mathcal{A}_{CP}(t)=\frac{\Gamma (B^0(t)\rightarrow f)-\Gamma (\bar{B}^0(t)\rightarrow f)}{\Gamma (B^0(t)\rightarrow f)+\Gamma (\bar{B}^0(t)\rightarrow f)} \ .
\end{equation}
This quantity can be separated into two terms, a decay only contribution with coefficient $C_f$, and a term due to the interference with coefficient $S_f$:
\begin{equation}
\label{CPasym2}
\mathcal{A}_{CP}(t)=C_f\cos (\Delta M_Bt)-S_f\sin (\Delta M_Bt)  \ ,
\end{equation}
where $\Delta M_B$ is the mass difference between the two
$B^0$ mass eigenstates. The expressions of the decay and mixing contributions can be calculated and found to be \cite{Buras2005,NirFPandCPV,NirCPV}:
\begin{equation}
\begin{aligned}
C_f = & \frac{1-|\xi_f|^2}{1+|\xi_f|^2} \ ,\\
S_f = & \frac{-2\text{Im}(\xi_f)}{1+|\xi_f|^2} \ , \\
\xi_f = & \frac{q}{p}\frac{A(\bar{B}^0\rightarrow f)}{A(B^0\rightarrow f)}
\ ,
\end{aligned}
\label{CandS}
\end{equation}
where $A$ is the decay amplitude of the flavour eigenstate at $t=0$, and $p$ and $q$ are the mixing parameters defined in Appendix~\ref{sec:A}. 

In general, $A(B^0\rightarrow f)$ has different terms coming from different Feynman diagrams. For example, if only two diagrams contribute to the decay then 
\begin{equation}
 \frac{A(\bar{B}^0\rightarrow f)}{A(B^0\rightarrow  f)}=-\ \zeta_f\ \frac{A_1e^{i(\delta_1-\phi_1)}+A_2e^{i(\delta_2-\phi_2)}}{A_1e^{i(\delta_1+\phi_1)}+A_2e^{i(\delta_2+\phi_2)}}
\ ,
 \label{Abar/A}
\end{equation}
where $\phi_i$ are the CPV (or the weak) phases, $\delta_i$ are the strong phases,\footnote{The strong phases are complex phases that are not present in the Lagrangian. They appear in decay amplitudes when intermediate states  go on-shell via $CP-$conserving processes, so they do not change sign under $CP$. They are called ``strong phases'' because they are usually due to strong interactions.} $\zeta_f$ is the $CP-$parity of the decay product $f$, and the $A_i$ contain the hadronic matrix elements.  If $\phi_1\neq \phi_2$, and the two diagrams contribute with similar strengths, then the asymmetry will suffer from hadronic uncertainties due to the presence of the strong phases $\delta_i$, and the hadronic matrix elements in $A_i$. However, a very interesting scenario happens when the decay is either dominated by one diagram, or all diagrams have the same CPV phase ($\phi_1=\phi_2\equiv \phi_D$ in the example). In these cases, the strong phases as well as the hadronic matrix elements cancel completely from $A(\bar{B}^0\rightarrow f)/A(B^0\rightarrow f)$, and we are left with:
\begin{equation}
\label{AAbar}
\left[\frac{A(\bar{B}^0\rightarrow f)}{A(B^0\rightarrow f)}\right]_{\text{sp}}=-\zeta_fe^{-i2\phi_D}
\ ,
\end{equation}
where we added the index ``$D$'' to the weak phase $\phi$ to express its decay origin, and we added the index ``sp'' to show that we are in the special case where either one diagram dominates the decay or all diagrams have the same CPV phase. 
In this special case, where equation~(\ref{AAbar}) applies, since $(q/p)$ is almost a pure phase for both $B^0_d$ and $B^0_s$ (equations~(\ref{beta}, \ref{betas})), then we have $|\xi_f|^2=1$ and thus, $C_f=0$ and $S_f=-\zeta_f\sin(2\phi_D-2\phi_M)$. Thus we get from equations~(\ref{CPasym2}) and~(\ref{CandS}):
\begin{equation}
\label{CPasym3}
[\mathcal{A}_{CP}(t)]_{\text{sp}}= \zeta_f\sin(2\phi_D-2\phi_M)\sin (\Delta M_Bt) 
\end{equation}
We note that  the difference between two weak phases, $(\phi_D-\phi_M)$, 
is convention-independent. 

The decays $(B_d\rightarrow \psi K_{\text{S}})$ and $(B_s\rightarrow \psi \phi)$,  have the same inclusive mode $b\rightarrow s c \bar{c}$. In the SM, it is dominated by the tree diagram with an internal $W$ boson. In NP, our DQCD model's contribution starts only at the 2-loop level, so it is negligible. Therefore, we are in the special case of one dominant diagram, where equation~(\ref{CPasym3}) applies. For the considered decays, $\phi_D$ is the weak phase of the SM tree diagram, that is, $\phi_D=\text{Arg}(V_{cs}V_{cb}^\ast)\simeq 0$.  So the asymmetries in these decays \textbf{depend only on the weak phase of the mixing $\phi_M$},  which is $(-\beta)$ for $B_d$ and $(-\beta_s)$ for $B_s$. Therefore
\begin{equation}
\label{ACP}
\begin{aligned}
\mathcal{A}_{CP}(B_d\rightarrow \psi K_{\text{S}})&=\zeta_{\psi K_{\text{S}}}\sin (2\beta) \sin [(\Delta M_d)t] \\
\mathcal{A}_{CP}(B_s\rightarrow \psi \phi)&=\zeta_{\psi \phi}\sin (2\beta_s) \sin [(\Delta M_s)t] \ ,
\end{aligned}
\end{equation}
 where $\Delta M_{d(s)}$ is the mass difference between the two $B_{d(s)}$ states, and the $CP-$parity $\zeta_{\psi K_{\text{S}}}=- 1$. The final state $(\psi \phi)$ is a mixture of $CP-$even and $CP-$odd states, nevertheless, its $CP-$parity can be resolved experimentally~\cite{LinnPhD}.
 

\section{Calculation of The \boldmath{$\Delta F = 2$} Constraints}
\label{sec:B}

In this section we show the explicit calculation of the $\Delta F = 2$ constraints on the DQCD scenario that are shown in Table~\ref{tab:2}. In the calculations of the new physics (NP) contributions to the flavor observables, the off-diagonal elements of the mixing mass matrix (see equations~\eqref{MvsH} and~\eqref{MvsHB}) are involved. For all three mesons ($M=K, \ B_d, \ B_s$), they can be written as 
\begin{equation}
\label{M12all}
M_{12}^{\rm NP} = \frac{1}{ 2 m_M} \bra{\bar{M}^0} H_{\rm eff}^{\rm NP}(\Delta F=2) \ket{M^0} ^\ast = \frac{N_c}{384{\pi}^2M_X^2} m_M F_M^2 \hat{B}_M \eta_M (\xi_M^\ast )^2 \ ,
\end{equation}
where $m_M$ is the experimental value of the meson's mass, $F_M$ is its decay constant, $\hat{B}_M$ is its bag parameter, and $\eta_M$ includes any RG and QCD corrections, and DQCD non-perturbative uncertainties, which, as mentioned in Section~\ref{Constraints}, we take to be $1\pm 0.5$.

Table~\ref{tab:1} contains all the experimental input data we use as well as the derived $\Delta F = 2$ constraints, along with the corresponding references. These constraints will impose limits on the parameters of our model, whether the real ones, or the $CP-$violating phases, that we will abbreviate to ``CPV phases.'' The $C_M$ parameters in Table~\ref{tab:1} are defined by  
\begin{equation}
\label{Cdef}
 C_M = \Delta M_M^{\rm exp}/\Delta M_M^{\rm SM} \ , \ M=K,B_d,B_s \ , 
\end{equation}
that is, the ratio of the experimental mass difference to the prediction from the SM. 
\begin{center}
\begin{table}[hbpt]
\centering
\begin{tabular}{|| c || c ||}
\hline\hline
\rule{0pt}{5ex} 
Experimental values~\cite{PDG}  &  Lattice QCD values \\
 & 
 \\ \hline
\rule{0pt}{5ex}  
 $m_K= 497.611 \pm 0.013\text{ MeV}$  &  $F_K=155.8\pm 1.7\text{ MeV}$~\cite{QCDlatticeUT}  \\
  & \\
 $m_{B_d}=5279.62 \pm 0.15\text{ MeV}$  &  $\hat{B}_K=0.7625\pm 0.0097$~\cite{FLAG016}  \\
  & \\
 $m_{B_s}=5366.82 \pm 0.22\text{ MeV}$  &  $F_{B_d}\sqrt{\hat{B}_{B_d}}=219\pm14\text{ MeV}$~\cite{FLAG016}  \\
  & \\
 $\Delta M_K=(3.484 \pm 0.006)\times 10^{-12}\text{ MeV}$  &  $F_{B_s}\sqrt{\hat{B}_{B_s}}=270\pm 16\text{ MeV}$~\cite{FLAG016}   \\ 
  & \\
 $\Delta M_d=(3.354 \pm 0.022)\times 10^{-10}\text{ MeV}$  & $F_{B_d}=192.0\pm4.3\text{ MeV}$~\cite{FLAG016} \\
   & \\
$\Delta M_s=(1.1688\pm 0.0014)\times 10^{-8}\text{ MeV}$  & $F_{B_s}=228.4\pm 3.7\text{ MeV}$~\cite{FLAG016}\\
& \\
  \cline{2-2} 
  \rule{0pt}{5ex}
$\Gamma_{K_{\text{L}}} = (1.286\pm 0.005)\times 10^{-17}$ GeV &  $\Delta F =2$ constraints \\
& \\
  \cline{2-2} 
  \rule{0pt}{5ex}  
$\Gamma_{K^+} = (5.315\pm 0.009)\times 10^{-17}$ GeV & $C_K\in [0.51,2.07]$~\cite{UTFit2008} \\
& \\
$\Gamma_{B_s}=(4.358\pm 0.014)\times 10^{-13}$ GeV  & $C_{B_d}\in [0.77,1.35]$~\cite{UTfit2016} \\
&\\
$\Gamma_{B_d}=(4.329\pm 0.011)\times 10^{-13}$ GeV  & $C_{B_s}\in [0.87,1.30]$~\cite{UTfit2016} \\
& \\
 $|\epsilon_K|=(2.228 \pm 0.011) \times 10^{-3}$ & $10^3|\epsilon_K|^{\rm SM} \in [1.76,2.33]$~\cite{BurasGirrbach014}  \\
  & \\ 
 $\sin 2\beta_d\in [0.634,0.726]$~\cite{Bona2016Utfit2} & $\sin 2\beta_d^{\rm SM}\in [0.665,0.785]$~\cite{Bona2016Utfit2}   \\
  & \\
 $\beta_s [^o]\in [-0.91,2.85]$~\cite{Bona2016Utfit2} &  $\beta_s^{\rm SM} [^o] \in [0.97,1.13]$~\cite{Bona2016Utfit2}  \\
  & \\ 
\hline\hline
\end{tabular}
\caption{Summary of quantities and constraints used. All values are taken at 95\% confidence level (CL). }
\label{tab:1}
\end{table}
\end{center}

\subsection{Meson Mass Differences}
\label{sec:CPCcons}
We show in this section how we get the constraints on the $\xi_M$ terms, due to the mass difference data, for the $K$, $B$, and $B_{s}$ mesons. 



The $CP$ conserving constraint, at the 2$\sigma$ level, from the $K-\bar K$ mixing is~\cite{UTFit2008}
\begin{equation}
 \label{CdeltaM5}
 0.51  \leq  \  C_K \equiv \Delta M_K^{\rm exp}/\Delta M_K^{\rm SM}   \  \leq 2.07 \ ,
 \end{equation}
from which we can derive the allowed interval for the new physics contribution:
\begin{equation}
\label{DeltaMNPdef}
\Delta M_K^{\rm NP} = \Delta M_K^{\rm exp} - \Delta M_K^{\rm SM} \pm \sqrt{\sigma_{\rm exp}^2 + \sigma_{SM}^2 } \ .
\end{equation}
From equation~(\ref{DeltaMK}), we have the relation  
\begin{equation} 
\label{DeltaMKNP}
 \Delta M_K^{\rm NP} = 2 {\rm Re} (M_{12}^{\rm NP}) \ , 
 \end{equation}
then we use equation~(\ref{M12all}) for $M_{12}^{\rm NP}$. Using the values given in Table~\ref{tab:1}, we can constrain $\Delta M_K^{NP}$, and obtain:
\begin{equation}
 \label{xiK6}
-6.93 \times 10^{-4} \leq 
N_c\text{Re}\left[( {\xi}_K^*)^2 \right] \left(\frac{1\text{ TeV}}{M_X} \right)^2 \leq  3.77 \times 10^{-4},
\end{equation}
where ${\xi}_K$ is defined in equation~(\ref{xi}), and $N_c$ is the number of dark colors. 


%
Moving to the $B_d$ system,
the $B_d-\bar B_d$ mixing mass difference constraint, at the 2$\sigma$ level, is~\cite{UTfit2016}           
\begin{equation}
 \label{11}
0.77 \leq C_{B_d} \leq  1.35 \ .
\end{equation}
From equation~(\ref{DeltaMB}), the relation between the new physics contribution to $\Delta M_{B_{d(s)}}$ and $M_{12}^{NP}$, the NP's contribution to $M_{12}$, is given by:
\begin{equation} 
 \label{DeltaMBNP}
 \Delta M_{B_{d(s)}}^{\rm NP} = 2 \mid M_{12}^{\rm NP}\mid \,=\, \mid \bra{\bar{B}_{d(s)}^0}H_{\rm eff}^{\rm NP}(\Delta B=2)\ket{B_{d(s)}^0} \mid / m_{B_{d(s)}}.
\end{equation}
Using the same strategy as in the $K-$system, we find
\begin{equation}
 \label{13}
N_c \mid{\xi}_{B_d}^2\mid \left( \frac{1\text{ TeV}}{M_X} \right)^2 \leq 6.55 \times 10^{-4}.
 \end{equation}
%
Similarly, for the $B_s$ system the bound is given by
\begin{equation}
 \label{11}
0.87 \leq C_{B_s} \leq  1.3 \ .
\end{equation}
which then leads to
\begin{equation}
 \label{13s}
 N_c\mid{\xi}_{B_s}^2\mid \left( \frac{1\text{ TeV}}{M_X} \right)^2 \leq 13.15 \times 10^{-3}.
 \end{equation}


\subsection{The  \boldmath{$\Delta F=2 \ CP-$}Violating Constraints}
\label{sec:CPVcons}

Now we briefly show how we get the constraints from the $CP$ violating  observables in the $K-\bar K$, $B_d-\bar B_d$, and $B_s-\bar B_s$ mixings.

We start with the $K-\bar K$ mixing, where we consider the $CP$ violating observable $\mid\epsilon_K\mid$ which is the modulus of  equation~(\ref{epsilonDef1}). From the SM prediction at 95\% CL~\cite{BurasGirrbach014}
\begin{equation} 
\label{epsilonSM3} 
1.76  \times 10^{-3} \leq |\epsilon_K^{\rm SM}| \leq 2.33 \times 10^{-3} \ ,
\end{equation}
and the experimental value~\cite{PDG} is $|\epsilon_K^{\rm exp}| = 2.228(11) \times 10^{-3}$. We evaluate the constraint on the new physics contribution to $|\epsilon_K|$ at 95\% CL is
\begin{equation} 
   \mid\epsilon_K\mid_{\rm NP} \  \leq  \ 0.75 \times 10^{-3 }.
\label{20}
\end{equation}
Taking the modulus of $\epsilon_K$ in equation~(\ref{epsilonDef1}) and using equation~(\ref{M12all}), we find
\begin{equation}
 \begin{aligned}
 \mid\epsilon_K\mid_{\rm NP} & \approx \frac{\kappa_{\epsilon}}{\sqrt{2}(\Delta  M_K)_{\rm exp}}\mid \text{Im}[M_{12}^{\rm NP}]\mid \\
 & = \frac{\kappa_{\epsilon}m_K F_K^2 \hat{B}_K \eta_K}{384\sqrt{2}{\pi}^2(\Delta  M_K)_{\rm exp}M_X^2}   N_c\mid \text{Im}\left[(\xi_K^\ast )^2 \right]\mid \ , \\
 & \ \ \kappa_{\epsilon} = 0.94 \pm 0.02 \ .
 \end{aligned}\label{epsilon}
\end{equation}
With the data of Table~\ref{tab:1} we then get
\begin{equation} \label{12CPVxiConstraint}
 N_c\mid \text{Im}\left[ ( {\xi}_K^*)^2 \right] \mid \left( \frac{1\text{ TeV}}{M_X} \right) ^2   \leq  1.64 \times 10^{-6}.                      
\end{equation}

Let us now move to the constraints from the $CP-$violating  observables in the $B-$ system, that is, $\beta$ and $\beta_s$. These parameters can be extracted from global fits~\cite{HFAG} and are also measured directly from the mixing-induced $CP$  
asymmetries (defined in the previous section) in the decays of $B$ mesons. This allows one to place a constraint on new physics contributing to these processes. 
 For example, the asymmetry in $(B_d\rightarrow \psi K_{\text{S}})$ measures $\beta$ and the asymmetry in $(B_s\rightarrow \psi\phi)$ measures $\beta_s$, as shown in equation \eqref{ACP} . These decay modes provide ``clean'' measurements of those phases,  because the hadronic matrix elements and the strong phases drop out from their $CP-$asymmetries, so the measurements of the CPV phases are free from hadronic uncertainties. 
 From equations~(\ref{beta})-(\ref{ARGq/pBs}), we can write the relations 
\begin{equation}
\label{beta2}
S_{\psi K_{\text{S}}} = \sin 2\beta_{\text{exp}} \approx -\frac{2}{\Delta M_d} \text{Im}\left[ M_{12}^\ast ({\rm NP})+M_{12}^\ast ({\rm SM})\right]_{B_d} \ ,
\end{equation}
\begin{equation}
 \label{betas2}
 S_{\psi \phi} = -\zeta_{\psi \phi}\sin 2(\beta_s)_{\text{exp}} \approx  -\frac{2}{\Delta M_s} \text{Im}\left[ M_{12}^\ast ({\rm NP})+M_{12}^\ast ({\rm SM})\right]_{B_s} \ ,
\end{equation}
and the SM terms in equations~(\ref{beta2}) and (\ref{betas2}) represent respectively $\sin 2\beta_{\text{SM}}$ and $\sin 2(\beta_s)_{\text{SM}}$.  
 Thus from equations~(\ref{beta2}) and~(\ref{M12all}), we can easily find 
\begin{equation}
\begin{aligned}
\sin [2\beta_{\text{exp}}] - \sin [2\beta_{\text{SM}}] = & -\frac{2}{\Delta M_d} \text{Im}\left[ M_{12}^\ast ({\rm NP})\right] \\
= & -\frac{1}{m_{B_d}\Delta M_d} \text{Im}\left[ \bra{ B_d^0}H_{\rm eff}^{\rm NP}(\Delta B=2)\ket{B_d^0}^\ast \right] \\
= & -
\frac{ m_{B_d} F_{B_d}^2 \hat{B}_{B_d} \eta_{B_d}}{192{\pi}^2\Delta M_dM_X^2} N_c\text{Im}\left[(\xi_{B_d}^\ast )^2\right] \ ,
\end{aligned}\label{betaNP}
\end{equation}
where we take the experimental values for $m_{B_d}$ and $\Delta M_d$. Using the numerical values of Table~\ref{tab:1}, we find the constraint
\begin{equation} 
\label{13CPVxiConstraint}
 -0.86\times 10^{-4} \leq N_c\text{Im}\left[ ( {\xi}_{B_d}^*)^2 \right] \left( \frac{1\text{ TeV}}{M_X} \right) ^2   \leq 3.12 \times 10^{-4}.                               
 \end{equation}

We apply the same procedure for the $B_s-\bar B_s$ mixing, and we find the following constraint from the difference between the experimental and the SM values of the $CP-$violating observable $\beta_s$:
\begin{equation} 
\label{23CPVxiConstraint}
  -3.54\times 10^{-3} \leq N_c\text{Im}\left[ ( {\xi}_{B_s}^*)^2 \right] \left( \frac{1\text{ TeV}}{M_X} \right) ^2   \leq  3.88 \times 10^{-3}.                                 
\end{equation}

All the constraints are summarized in Table~\ref{tab:2} in Section~\ref{Constraints}.
From these results we can infer model independent bounds on any new-physics model, contributing to flavor observables, and which $\Delta F=2$ effective Hamiltonian, at some scale $\Lambda$, contains (V+A)(V+A) (or (V-A)(V-A), since strong interactions conserve $CP$) operators in the form:
\begin{equation} 
\label{HeffNP}  
H_{\rm eff}^{\rm NP}(\Delta F=2) = \frac{c_{qq'}^{NP}}{\Lambda^2} \left[ (\bar{q}\gamma_{\mu}P_{R(L)}q')(\bar{q}\gamma^{\mu}P_{R(L)}q')\right]  + {\rm h.c.} .
\end{equation}
The scale $M_X$ can be mapped to the scale suppressing the dimension 6 operator, for example $c_{ds}^\text{NP}= N_c{\xi}_K^2 /(128{\pi}^2)$.

Before closing this section, we compare our results to the work in~\cite{IsidoriEtAL010} which performed a similar analysis. 
At a scale $\Lambda \approx M_X\approx 1\text{ TeV}$, equation~(\ref{xiK6}) yields:
\begin{equation}
\label{CdeltaMK7}
 \text{Re}[c_{ds}^{\rm NP}] \left( \Lambda = 1 \text{TeV}\right) \leq 2.98 \times 10^{-7},
 \end{equation}
which is at the same order of magnitude, but three times stronger than that of reference \cite{IsidoriEtAL010}.
In addition, the $CP-$violating constraint we find in equation~(\ref{12CPVxiConstraint}) is also about three times stronger than the constraint of reference~\cite{IsidoriEtAL010}. We believe these disagreements originate from the fact that this reference took the SM leading contribution to $K-\bar K$ mixing amplitude with only the top quark running inside the loop~\cite{Isidori015,NirFPandCPV}. This is only true for the $B-$ system, while for the $K-$system, the dominant contribution to the real part of the amplitude is from an internal $c$ quark, and the dominant contribution to the imaginary part is from both $c$ and $t$ internal quarks (See section 10.2 and 10.4 of reference~\cite{Buras1998}). 

The constraints we find in equations~(\ref{13}), (\ref{13s}), and~(\ref{13CPVxiConstraint}), are all about one order of magnitude stronger than the ones of reference~\cite{IsidoriEtAL010}. We believe this is due to the improvement in the experimental data, which have significantly changed for the $B-$system. We find a result that is the same order of magnitude as the one in~\cite{IsidoriEtAL010} when using the same input data.



\bibliography{Paper1Bibliography} 
\bibliographystyle{JHEP}

\end{document}